\documentclass[prd,reprint,onecolumn,superscriptaddress,tightenlines,nofootinbib,eqsecnum
]{revtex4-2}
\usepackage[hhmmss]{datetime}
\usepackage[english]{babel}
\usepackage{amssymb,amsmath,amsfonts,graphicx,mathrsfs,color,bm}
\usepackage{graphicx}% Include figure files
\usepackage{dcolumn}% Align table columns on decimal point
\usepackage{hyperref}% add hypertext capabilities
\hypersetup{colorlinks=true,linkcolor=magenta,anchorcolor=green,citecolor=cyan,filecolor=black,menucolor=black,urlcolor=blue}
\usepackage[table]{xcolor}
\allowdisplaybreaks
\usepackage{epstopdf}
\usepackage[utf8]{inputenc} 
\usepackage{pgf}
\usepackage{float}
\usepackage{subcaption}

\def\switch{[1 \leftrightarrow 2]}

\newcommand{\p}{\partial}
\newcommand{\ud}{\mathrm{d}}

\newcommand{\bdm}{\begin{displaymath}}
\newcommand{\edm}{\end{displaymath}}

\newcommand{\calF}{\mathcal{F}}

\newcommand{\be}{\begin{equation}}
\newcommand{\ee}{\end{equation}}
\newcommand{\bse}{\begin{subequations}}
\newcommand{\ese}{\end{subequations}}
\newcommand{\bea}{\begin{eqnarray}}
\newcommand{\eea}{\end{eqnarray}}

\newcommand{\di}{\mathrm{i}}
\newcommand{\e}{\mathrm{e}}
\newcommand\calO{{\mathcal{O}}}
\newcommand{\dd}{\mathrm{d}}

\newcommand{\nn}{\nonumber}

\newcommand{\ADM}{\mathcal{M}}

\newcommand{\etidal}{\epsilon_\text{tidal}}

\newcommand{\Meudon}{\affiliation{Laboratoire d'étude de l'Univers et des phénomènes eXtrêmes (LUX), Observatoire de Paris, Université PSL, Sorbonne Université, CNRS, 92190 Meudon, France}}

\newcommand{\UPC}{\affiliation{Université Paris Cité, 75013 Paris, France}}

\newcommand{\IBS}{\affiliation{Cosmology, Gravity, and Astroparticle Physics Group, Center for Theoretical Physics of the Universe, Institute for Basic Science (IBS), Daejeon, 34126, Korea}}

\begin{document}

\title{Tidal effects in the total flux and waveform in massless scalar-tensor theories to, respectively, relative 2PN and 1.5PN orders}

\author{Eve Dones}
\email{edones@ibs.re.kr}
\Meudon
\UPC
\IBS

\author{Laura Bernard}
\email{laura.bernard@obspm.fr} 
\Meudon

\date{\today}% It is always \today, today,
             %  but any date may be explicitly specified

\begin{abstract}
Within scalar-tensor (ST) theories, neutron stars in binary systems experience tidal deformations caused by both their companion and the scalar field. These deformations are strongly correlated to the star’s internal structure and composition. Accurately modeling their imprint on the emitted gravitational waves will be essential for interpreting the high-precision data expected from future detectors and for disentangling potential signatures of modified gravity from those arising due to the properties of neutron star matter. Using the post-Newtonian multipolar-post-Minkowskian formalism adapted to ST theories, and working within the adiabatic approximation, we compute the tidal corrections to the total energy flux, accounting for both gravitational and scalar radiation, and to the waveform phasing, at the
next-to-next-to-leading order (NNLO). This corresponds to second post-Newtonian (2PN) order beyond the leading-order dipolar tidal contribution. At this accuracy, three independent types of tidal deformability (scalar, tensorial, and mixed scalar-tensorial) contribute to the signal. We also derive the full waveform amplitude modes, including gravitational and scalar modes,  as well as the memory (m=0) ones, to the $\text{N}^{1.5}\text{LO}$ (i.e. to relative 1.5PN order).
\end{abstract}

%\keywords{,}

%\notoc
%\toccontinuoustrue

\maketitle

%\flushbottom

%------------------------------------------------------------------------
%------------------------------------------------------------------------
\section{Introduction} 
%------------------------------------------------------------------------
%------------------------------------------------------------------------

%------------------------------------------------------------------------
\subsection{Context}
\label{subsec:context}
%------------------------------------------------------------------------

With the advent of gravitational wave astronomy, there has been a renewal of interest in the modeling of gravitational waves in alternative theories of gravity. Such efforts have concerned both analytical treatment of the early inspiral as well as numerical ones to describe the merger of compact binaries. In the latter case, the first simulations were done in the context of binary neutron stars in scalar-tensor theories, illustrating the phenomenon of dynamical scalarization~\cite{Barausse:2012da,Palenzuela:2013hsa}. More recently, advances in the mathematical understanding of gravity theories beyond general relativity (GR)~\cite{Kovacs:2020ywu} have allowed to perform full numerical simulations in some specific theories belonging to the Horndeski or Einstein-scalar-Gauss-Bonnet classes~\cite{East:2020hgw,Cayuso:2023xbc,Corman:2024cdr}.
\newline

On the analytical side, significant progress has also been made in modeling inspiral waveforms within ST theories. Most of these works have been performed in the framework of the post-Newtonian multipolar post-Minkowskian (PN-MPM) formalism~\cite{Blanchet:2013haa}, a perturbative approach based on an expansion in small velocities and for weak gravitational fields.
The dynamics of compact binaries is currently known up to 3PN order~\cite{Mirshekari:2013vb,Bernard:2018hta,Bernard:2018ivi} and the radiative sector has been tackled in different groups. Using the direct integration of the relaxed Einstein equations (DIRE) formalism, Lang derived the tensorial waveform to 2PN order, the scalar waveform to $1.5$PN order and the total energy flux to 1PN order, all beyond the standard quadrupole radiation in GR, and for binaries on generic orbits~\cite{Lang:2013fna,Lang:2014osa}. In Ref.~\cite{Sennett:2016klh}, these results were specialized to binaries on quasi-circular orbits. The scalar waveform amplitude modes were derived for the first time up to $1.5$PN order in Ref. ~\cite{Bernard:2022noq}, along with the $1.5$PN total energy flux, using the PN-MPM formalism. More recently, the results of Ref.~\cite{Bernard:2022noq} have been extended to systems on quasi-elliptical orbits~\cite{Trestini:2024zpi,Trestini:2024mfs}. 
\newline

 The effects of tidal deformation of neutron stars in binary inspirals have also been investigated within the PN-MPM framework and in the adiabatic approximation. In GR, such corrections have been computed in Refs.~\cite{Henry:2019xhg,Henry:2020ski,Dones:2024odv,Henry:2025uta,Henry:2026bqh} for comparable mass binaries. 
 %A few preliminary works within the gravitational self-force formalism have looked at tidal effects in the case of IMRIs or EMRIs with a highly deformable secondary~\cite{Steinhoff:2012rw,Xu:2022fyp}. 
 In ST theories, leading-order dipolar tidal corrections were included in both the dynamics and radiation in Ref.~\cite{Bernard:2019yfz}. In Ref.~\cite{Creci:2023cfx} was emphasized that, in ST theories, three independent tidal deformabilities are needed to fully characterize the response to external tidal fields: scalar, tensorial, and mixed scalar-tensorial deformabilities.  All three types of deformation are encountered at the NNLO. Hence, the first step to obtain gravitational waveforms that include these three types of effects consists in deriving the dynamics at that same accuracy. This was the subject of our first paper~\cite{Bernard:2023eul}, where we cross-checked our results with the alternative but equivalent post-Newtonian effective field theory (EFT) framework~\cite{Goldberger:2004jt}. In parallel, the leading-order dipolar and quadrupolar scalar effects, as well as the leading-order quadrupolar tensorial and scalar-tensorial effects have been studied in both the dynamical and radiative sectors in Ref.~\cite{Creci:2024wfu}. 
\newline

In the present work, we extend our previous results by including tidal corrections to the total energy flux, accounting for both gravitational and scalar radiation, and to the GW phasing at the NNLO. We further derive the full waveform, consistently incorporating gravitational and scalar modes, up to $\text{N}^{1.5}\text{LO}$ (i.e. relative 1.5PN order). In addition, we compute for the first time the memory ${\rm m}=0$ modes up to the NLO, which are necessary to achieve the desired accuracy in the gravitational part of the waveform.\newline

The paper is organized as follows. In Section~\ref{subsec:massless_ST}, we recall the gravitational and matter parts of the action used to describe tidal effects within scalar-tensor theories and the resulting field equations. In Section~\ref{sec:dynamics_circ},  we specialize the conservative dynamics derived in~\cite{Bernard:2023eul} to the case of circular orbits.  Section~\ref{sec:mPMPN} presents how the PN-MPM formalism is used to compute the total energy flux and waveform. In Section~\ref{sec:intermediate_quantities}, we detail the intermediate computations required for deriving these quantities.  Section~\ref{sec:fluxes_phases} presents the final expressions for the fluxes and GW phasing, the latter being given both in the time and Fourier domains. In Section~\ref{sec:amplitudemodes}, we provide the gravitational and scalar waveform amplitude modes in the standard PN-expanded form. Sources moments for circular orbits are listed in Appendix~\ref{App:source_moments_circ}. In Appendix~\ref{eq:comparison_Creci},  we discuss the comparison of our results with those of~\cite{Creci:2024wfu}. The expressions for the energy fluxes, the GW phasing, and the waveform modes, including tidal effects, are provided in a \textit{Mathematica} file directly available in the Supplemental Material~\cite{SuppMaterial}. Finally, some intermediate PN results for the conservative dynamics on circular orbits, dissipative equations of motion and center-of-mass position, and source moments, will be available to interested readers upon request.

%------------------------------------------------------------------------
\subsection{Notations}
\label{subsec:notations}
%------------------------------------------------------------------------

In this section, we present the notations that will be used throughout the article. Some quantities are related to ST theories and their generalized PPN parameters, while others are linked to compact objects and binary systems.
\begin{itemize}	
\item[-] We consider two compact objects of masses $m_{1}$ and $m_{2}$.  The total mass is $m=m_1+m_2$, the reduced mass is $\mu = m_1 m_2 / m$ and the symmetric mass ratio is $\nu= \mu/ m \in \ ]0,1/4]$. The total mass $m$ should not be confused with the order $\mathrm{m}$ associated to the waveform modes. Morever, we assume that $m_1 \geq m_2$ and define the relative mass difference as $\delta = (m_1-m_2)/m \in [0,1[ $. The symmetric mass ratio and the relative mass difference are linked by the relation $\delta^2 = 1-4\nu$. 
\item[-]  Both objects are endowed with tidal deformations parametrized by a set of scalar-type, tensorial-type and scalar-tensorial-type tidal deformabilities. For the purposes of this work, we restrict our attention to the dipolar and quadrupolar scalar-type deformabilities, denoted by $\{\lambda_A(\phi),\mu_A(\phi)\}$, the quadrupolar tensorial-type deformability $c_A(\phi)$, and the quadrupolar scalar-tensorial-type deformability $\nu_A(\phi)$. All these quantities are introduced in the tidal action~\eqref{eq:Stidal}. The deformabilities are Taylor expanded around the asymptotic value of the scalar field $\phi_0$. The expansion coefficients relevant for this work are related to dimensionless parameters via the relations
\begin{align}
    k_{A,\lambda}^{(n)}\equiv \lambda_A^{(n)}\cdot\frac{\tilde{G}\alpha }{c^2R_A^3}\cdot\left(\frac{\tilde{G}\alpha M_A}{c^2R_A}\right)^n\,,\qquad
    k_{A,\mu}^{(0)}\equiv \mu_A^{(0)}\cdot\frac{\tilde{G}\alpha }{c^2R_A^5}\,,\qquad
    k_{A,\nu}^{(0)}\equiv \nu_A^{(0)}\cdot\frac{\tilde{G}\alpha }{R_A^5}\,,\qquad
    k_{A,c}^{(0)}\equiv c_A^{(0)}\cdot\frac{\tilde{G}\alpha }{R_A^5}\,,
\end{align}
where $R_A$ is the typical radius of body A. Here, the superscript (n) labels the order in the expansion of the tidal deformabilities in powers of the scalar perturbation $\psi$, as defined in Eqs.~\eqref{tiddeform}. In this work, for the coefficients $\lambda_A^{(n)}$, we will only consider the cases $n=0,1,2$. 
%$R_A$ is the typical radius of body $A$. 
Using the fact that compact objects have a compactness $\mathcal{C}_A = \frac{\tilde{G}\alpha m_A}{c^2R_A}\sim 1$, we find that the tidal deformabilities are at least 3PN quantities. In particular, we have
\begin{align}\label{eq:dimless_TLN}
\frac{\lambda_a^{(0)}}{c^2} \sim \mathcal{O}\left(\frac{1}{c^6}\right)=\mathcal{O}\left(\etidal \right)\,,\qquad \frac{\mu_a^{(0)}}{c^2}\sim\nu_a^{(0)}\sim c_a^{(0)}\sim\mathcal{O}\left(\frac{1}{c^{10}}\right)=\mathcal{O}\left(\frac{\etidal}{c^4} \right)\,.
\end{align}
We note that scalar-type tidal deformabilities are accompanied by an explicit factor of $1/c^2$. This comes from the fact that the leading-order scalar tidal corrections (whether dipolar or quadrupolar) always enter the dynamics and the radiation suppressed by this factor. 
\item[-]There are two ways of thinking the PN power counting when dealing with tidal eﬀects: the absolute and the relative ones. Throughout this paper, we use the relative way. In this scheme, the PN order of a given contribution is defined with respect to the leading-order term of the same type (point-particle or tidal). Hence, when we say that a quantity is required up to $\mathrm{N}^n\mathrm{LO}$ (with $\mathrm{N}^0\mathrm{LO}\equiv\mathrm{LO}$ corresponding to the leading order, $\mathrm{N}^1\mathrm{LO}\equiv\mathrm{NLO}$ the next-to-leading order, etc.), it is meant that both the point-particle and the tidal contributions are derived up to $n$PN orders beyond their respective leading orders. This convention is summarized by the remainder notation
\begin{equation}
\mathcal{O}\!\left(\frac{1}{c^{2n}}\,;\,\frac{\epsilon_{\rm tidal}}{c^{2n}}\right),
\end{equation}
which indicates that terms of order $1/c^{2n}$ relative to the leading point-particle contribution, as well as terms of order $\epsilon_{\rm tidal}/c^{2n}$ relative to the leading tidal contribution, are neglected.

\item[-] We denote by $\bm{y}_A(t)$ the two ordinary coordinate trajectories in a harmonic coordinate system $\left\{t,\mathbf{x}\right\}$, by $\bm{v}_A(t)=\ud\bm{y}_A/\ud t$ the two ordinary velocities and by $\bm{a}_A(t)=\ud\bm{v}_A/\ud t$ the two ordinary accelerations.  The ordinary separation vector is given $\bm{x}=\bm{y}_1-\bm{y}_2$, the orbital radius by $r_{12}=\lvert\bm{x}\vert$, and the unit vector by $\bm{n}_{12}=\bm{x}/r_{12}$; ordinary scalar products are denoted by parentheses, \textit{e.g.} $\left(n_{12}v_{1}\right)=\bm{n}_{12}\cdot\bm{v}_{1}$, while 3-dimensional Dirac function is denoted $\delta^{(3)}(\mathbf{x})$, and its value at the position $\bm{y}_A$ is written $\delta_A \equiv \delta^{(3)} (\mathbf{x} - \bm{y}_A) $. We denote by $L=i_1\cdots i_\ell$ a multi-index with $\ell$ spatial indices; $\nabla_L = \nabla_{i_1}\cdots\nabla_{i_\ell}$ and so on;  similarly,  $n_{L} = n_{i_1}\cdots n_{i_\ell}$.
\item[-] To express quantities in the center-of-mass (CoM) frame, the “12” label is typically dropped. We pose $v^2=(vv)=\bm{v}\cdot\bm{v}$ and $\dot{r}=(nv)=\bm{n}\cdot\bm{v}$. When working in the CoM frame, it is convenient to define the combinations of the tidal deformabilities to reduce the length of the expressions, namely : 
\begin{align}\label{eq:tidal_deformabilities_CoM}
\lambda^{(n)}_{\pm} = & \,  \frac{m_2}{m_1} \bar{\delta}_2 \, \lambda^{(n)}_1 \pm \frac{m_1}{m_2} \bar{\delta}_1 \, \lambda^{(n)}_2  \,, \nn \\
\Lambda^{(n)}_{\pm} = & \,  \frac{m_2}{m_1} \frac{\zeta(2+\bar{\gamma})}{(1-\zeta)}\bar{\delta}_2 \,(1-2s_2) \, \lambda^{(n)}_1 \pm \frac{m_1}{m_2} \frac{\zeta(2+\bar{\gamma})}{(1-\zeta)}\bar{\delta}_1 \,(1-2s_1)\, \lambda^{(n)}_2  \,, \nn \\
\mu^{(n)}_{\pm} = & \, \frac{m_2}{m_1} \bar{\delta}_2 \, \mu^{(n)}_1 \pm \frac{m_1}{m_2} \bar{\delta}_1 \, \mu^{(n)}_2  \,, \nn \\
c^{(n)}_{\pm} = & \, \frac{m_2}{m_1} \left[ 4\bar{\delta}_2 + (2+\bar{\gamma})^2\left(\frac{1-\zeta}{\zeta}+2(1-2s_2) \right) \right] c^{(n)}_1 \pm \frac{m_1}{m_2} \left[ 4\bar{\delta}_1 + (2+\bar{\gamma})^2\left(\frac{1-\zeta}{\zeta}+2(1-2s_1) \right) \right]c^{(n)}_2  \,, \nn \\
{\nu}^{(n)}_{\pm} = & \, \frac{m_2}{m_1} \left[ 4\bar{\delta}_2 + (2+\bar{\gamma})^2(1-2s_2) \right]\nu^{(n)}_1 \pm \frac{m_1}{m_2} \left[ 4\bar{\delta}_1 + (2+\bar{\gamma})^2(1-2s_1) \right]\nu^{(n)}_2 \,.
\end{align}
Note that these combinations of the tidal deformabilities present some differences compared to those used in~\cite{Bernard:2023eul}. 
\item[-]  In the case of quasi-circular orbits, we introduce $\bm{\ell}$ the vector normal to the orbital plane and $\boldsymbol{\lambda}$ the vector tangent to the orbit such that $\boldsymbol{\lambda}\cdot \boldsymbol{v} \ge 0 $. The vectors are oriented such that $(\mathbf{n}, \boldsymbol{\lambda}, \bm{\ell})$ be an orthonormal tetrad. When focusing on quasi-circular orbits, we conveniently define a normalized version of Eqs.~\eqref{eq:tidal_deformabilities_CoM} in order to shorten our expressions, namely :
\begin{align}\label{eq:tidal_deformabilities_Circ}
& \tilde{\lambda}^{(n)}_{\pm} = \frac{\alpha \tilde{G}}{c^2}\left(\frac{c^2}{\alpha \tilde{G}m}\right)^3 {\lambda}^{(n)}_{\pm} \,, \qquad  \tilde{\Lambda}^{(n)}_{\pm} = \frac{\alpha \tilde{G}}{c^2}\left(\frac{c^2}{\alpha \tilde{G}m}\right)^3 {\Lambda}^{(n)}_{\pm} \,, \qquad \tilde{\mu}^{(n)}_{\pm} = \frac{\alpha \tilde{G}}{c^2}\left(\frac{c^2}{\alpha \tilde{G}m}\right)^5 {\mu}^{(n)}_{\pm}\nn \\
& \tilde{c}^{(n)}_{\pm} = \alpha \tilde{G}\left(\frac{c^2}{\alpha \tilde{G}m}\right)^5 {c}^{(n)}_{\pm} \,, \qquad  \;\;\; \tilde{\nu}^{(n)}_{\pm}= \alpha \tilde{G}\left(\frac{c^2}{\alpha \tilde{G}m}\right)^5 {\nu}^{(n)}_{\pm}  \,.
\end{align}
\item[-]  We denote $(R, N_i)$ the orbital separation and unit vector in a radiative coordinate system. 
\item[-] To present our results, following~\cite{Bernard:2018hta}, we introduce a number of ST and post-Newtonian parameters. The ST parameters are defined based on the value $\phi_0$ of the scalar field $\phi$ at spatial infinity, on the Brans-Dicke-like scalar function $\omega(\phi)$ and on the mass-functions $m_A(\phi)$. We pose $\varphi\equiv\phi/\phi_{0}$. The post-Newtonian parameters naturally extend and generalize the usual PPN parameters to the case of a general ST theory~\cite{Will:1972zz,Will:2018bme}. All these parameters are given and summarized in the following Table~\ref{tab:ST_parameters}.
\item[-] Finally, expressions abbreviated with ($\cdots$) indicate that terms are too lengthy to be displayed in full; dependencies on the quadrupolar scalar, tensorial or scalar-tensorial tidal deformabilities are shown explicitely where they appear; the complete expressions are provided in the Supplemental Material~\cite{SuppMaterial}. 
\hspace{0.5cm}\begin{small}
\begin{center}
\begin{tabular}{|c||cc|}
	\hline
	%\hline 
	& \multicolumn{2}{|c|}{\textbf{ST parameters}} \\[2pt]
	\hline &&\\[-10pt]
	general & \multicolumn{2}{c|}{$\omega_0=\omega(\phi_0),\qquad\omega_0'=\left.\frac{\ud\omega}{\ud\phi}\right\vert_{\phi=\phi_0}, \qquad\omega_0''=\left.\frac{\ud^2\omega}{\ud\phi^2}\right\vert_{\phi=\phi_0},\qquad\varphi = \frac{\phi}{\phi_{0}},\qquad\tilde{g}_{\mu\nu}=\varphi\,g_{\mu\nu},$} \\[12pt]
	& \multicolumn{2}{|c|}{$\tilde{G} = \frac{G(4+2\omega_{0})}{\phi_{0}(3+2\omega_{0})},\qquad \zeta = \frac{1}{4+2\omega_{0}},$} \\[8pt]
	& \multicolumn{2}{|c|}{$\lambda_{1} = \frac{\zeta^{2}}{(1-\zeta)}\left.\frac{\ud\omega}{\ud\varphi}\right\vert_{\varphi=1},\qquad \lambda_{2} = \frac{\zeta^{3}}{(1-\zeta)}\left.\frac{\ud^{2}\omega}{\ud\varphi^{2}}\right\vert_{\varphi=1}, \qquad \lambda_{3} = \frac{\zeta^{4}}{(1-\zeta)}\left.\frac{\ud^{3}\omega}{\ud\varphi^{3}}\right\vert_{\varphi=1}.$} \\[8pt]
	& \multicolumn{2}{|c|}{$\switch$ switches the particle's labels (note the index on the $\lambda_i$'s in not a particle label)} \\[7pt]
	\hline &&\\[-7pt]
	~sensitivities~ & \multicolumn{2}{|c|}{$s_A = \left.\frac{\ud \ln{m_A(\phi)}}{\ud\ln{\phi}}\right\vert_{\phi=\phi_0},\qquad s_A^{(k)} = \left.\frac{\ud^{k+1}\ln{m_A(\phi)}}{\ud(\ln{\phi})^{k+1}}\right\vert_{\phi=\phi_0},\qquad(A=1,2)$} \\[9pt]
    & \multicolumn{2}{|c|}{$s'_A = s_A^{(1)},\qquad s''_A = s_A^{(2)},\qquad s'''_A = s_A^{(3)},$} \\[5pt]
	& \multicolumn{2}{|c|}{$\mathcal{S}_+ = \frac{1-s_1 - s_2}{\sqrt{\alpha}}\,,\qquad \mathcal{S}_- = \frac{s_2 - s_1}{\sqrt{\alpha}}.$} \\[7pt]	\hline\hline 
	Order & \multicolumn{2}{|c|}{\textbf{PN parameters}} \\[2pt]
	\hline &&\\[-10pt]
	N & \multicolumn{2}{|c|}{$\alpha= 1-\zeta+\zeta\left(1-2s_{1}\right)\left(1-2s_{2}\right)$}   \\[5pt]
	\hline &&\\[-10pt]
	1PN & $\overline{\gamma} = -\frac{2\zeta}{\alpha}\left(1-2s_{1}\right)\left(1-2s_{2}\right),$ & Degeneracy \\[5pt]
	& ~~$\overline{\beta}_{1} = \frac{\zeta}{\alpha^{2}}\left(1-2s_{2}\right)^{2}\left(\lambda_{1}\left(1-2s_{1}\right)+2\zeta s'_{1}\right),$~~~~&  $\alpha(2+\overline{\gamma})=2(1-\zeta)$ \\[5pt]
	& $\overline{\beta}_{2} = \frac{\zeta}{\alpha^{2}}\left(1-2s_{1}\right)^{2}\left(\lambda_{1}\left(1-2s_{2}\right)+2\zeta s'_{2}\right),$~~~~& \\[5pt]
	&  $\overline{\beta}_+ = \frac{\overline{\beta}_1+\overline{\beta}_2}{2}, \qquad \overline{\beta}_- = \frac{\overline{\beta}_1-\overline{\beta}_2}{2}.$ &  \\[5pt]
	\hline &\\[-10pt]
	2PN & $\overline{\delta}_{1} = \frac{\zeta\left(1-\zeta\right)}{\alpha^{2}}\left(1-2s_{1}\right)^{2}\,,\qquad \overline{\delta}_{2} = \frac{\zeta\left(1-\zeta\right)}{\alpha^{2}}\left(1-2s_{2}\right)^{2},$ & Degeneracy \\[5pt]
	&  $\overline{\delta}_+ = \frac{\overline{\delta}_1+\overline{\delta}_2}{2}, \qquad \overline{\delta}_- = \frac{\overline{\delta}_1-\overline{\delta}_2}{2},$ &  $16\overline{\delta}_{1}\overline{\delta}_{2} = \overline{\gamma}^{2}(2+\overline{\gamma})^{2}$\\[5pt]
	& $~~\overline{\chi}_{1} = \frac{\zeta}{\alpha^{3}}\left(1-2s_{2}\right)^{3}\left[\left(\lambda_{2}-4\lambda_{1}^{2}+\zeta\lambda_{1}\right)\left(1-2s_{1}\right)-6\zeta\lambda_{1}s'_{1}+2\zeta^{2}s''_{1}\right],~~$  &  \\[5pt]
	& $\overline{\chi}_{2} = \frac{\zeta}{\alpha^{3}}\left(1-2s_{1}\right)^{3}\left[\left(\lambda_{2}-4\lambda_{1}^{2}+\zeta\lambda_{1}\right)\left(1-2s_{2}\right)-6\zeta\lambda_{1}s'_{2}+2\zeta^{2}s''_{2}\right],$ &  \\[5pt]
	&  $\overline{\chi}_+ = \frac{\overline{\chi}_1+\overline{\chi}_2}{2}, \qquad \overline{\chi}_- = \frac{\overline{\chi}_1-\overline{\chi}_2}{2}.$ &  \\[5pt]
\hline
\end{tabular}
\captionof{table}{Parameters for the general ST theory and our notation for PN parameters. \label{tab:ST_parameters}}
\end{center}
\end{small}

\end{itemize}

%------------------------------------------------------------------------
%------------------------------------------------------------------------
\section{Massless scalar-tensor theories beyond GR}
\label{subsec:massless_ST}
%------------------------------------------------------------------------
%------------------------------------------------------------------------

%------------------------------------------------------------------------
\subsection{The action}
\label{subsec:action}
%------------------------------------------------------------------------

In this section, we summarize the framework in which we have studied the tidal effects on the dynamics of an inspiralling compact binary system in massless scalar-tensor theories of gravity~\cite{Bernard:2023eul}. The starting point is the generalized version of Brans-Dicke theories where a scalar field $\phi$ is minimally coupled to the metric $g_{\mu \nu}$. The theory is characterized by the coupling function $\omega(\phi)$. In order to decouple scalar and tensor degrees of freedom, we perform a conformal transformation to go from the physical metric to the conformal one, defined by
\be\label{eq:conformal_metric}
\tilde{g}_{\mu\nu}=\varphi\,g_{\mu\nu}\,, \qquad \varphi=\frac{\phi}{\phi_0}\,.
\ee
In particuliar, this conformal transformation implies that $g^{\mu\nu}=\varphi\,\tilde{g}^{\mu\nu}$ and $\sqrt{-\tilde{g}}=\varphi^2 \sqrt{-g}$.  After some reshuffling, the gravitational action can be displayed in the Einstein frame as 
\be\label{eq:STactionEF}
S^\text{GF}_{\mathrm{ST}} = \frac{c^{3}\phi_{0}}{16\pi G} \int\ud^{4}x\,\sqrt{-\tilde{g}}\left[ \tilde{R} -\frac{1}{2}\tilde{g}_{\mu\nu}\tilde{\Gamma}^{\mu}\tilde{\Gamma}^{\nu} - \frac{3+2\omega(\phi)}{2\varphi^{2}}\tilde{g}^{\alpha\beta}\p_{\alpha}\varphi\p_{\beta}\varphi\right] +S_{\mathrm{m}}\left(\mathfrak{m},g_{\alpha\beta}\right)\,,
\ee
where $\tilde{R}$ is the Ricci scalar related to the conformal metric. A gauge fixing term has been introduced, $\propto\tilde{g}_{\mu\nu}\tilde{\Gamma}^{\mu}\tilde{\Gamma}^{\nu}$ with $\tilde{\Gamma}^{\nu}\equiv\tilde{g}^{\rho\sigma}\tilde{\Gamma}^{\nu}_{\rho\sigma}$, to enforce working in the harmonic gauge,  defined by $\tilde{\Gamma}^{\mu}=0$. 

As for the matter part, we follow an EFT approach and modify the skeletonized matter action associated to point-particles~\cite{1975ApJ...196L..59E} by adding non-minimal worldline couplings involving higher-order derivatives of the field(s)~\cite{Damour:1995kt,Damour:1998jk,Goldberger:2004jt}. In GR, it was shown that the tidal action can be constructed solely from two sets of tidal multipole moments: the electric-type $G_L^{A}$ and the magnetic-type $H_L^{A}$. These are built from the Riemann tensor $R_{\mu \rho \nu \sigma}$ and its covariant derivatives contracted with the 4-velocity $u_A^{\mu}= \ud y_A/{\ud\tau_A}$, and are evaluated along the worldline of body $A$~\cite{Bini:2012gu}. At the lowest PN orders, it is enough to consider quadratic (kinetic-like) couplings in these tidal moments~\cite{Bini:2012gu,Henry:2019xhg}. In ST theories, additional non-minimal couplings must be included, involving the scalar field $\varphi$, its covariant derivatives and the 4-velocity, also evaluated along the worldline of body $A$. At the lowest PN orders, it is sufficient to consider quadratic couplings in the scalar tidal field $\mathcal{E}_L \sim \nabla_L^{\perp} \varphi$, where $\nabla_{\mu}^{\perp}\equiv\left(\delta_{\mu}^{\nu}+u_{\mu}u^{\nu}\right)\nabla_{\nu}$ with $\nabla^{\perp}_L=\nabla^{\perp}_{\mu_1}\cdots\nabla^{\perp}_{\mu_{\ell}}$, as well as mixed quadratic couplings between the scalar tidal field and the gravitational (electric-type and magnetic-type) tidal moments. 
For the purpose of our work, we find that the matter action, decomposed as $S_{\rm m}=S_{\rm pp}+S_{\rm tidal}$, reduces to~\cite{Bernard:2023eul,Creci:2023cfx,Creci:2024wfu}
\begin{subequations}\label{eq:Smatter+tidal}
\begin{align}\label{eq:Smatter}
    S_{\rm pp} &= -c\sum_{A=1,2}\,\int\ud \tau_A\,m_A\left(\phi\right)\,,\\ \nn
    S_{\rm tidal} &= -\frac{c}{2}\,\sum_{A=1,2}\,\int\ud \tau_A\,\biggl\{\lambda_A\left(\phi\right) \left(\nabla_{\alpha}^{\perp}\varphi\right)_A\,\left(\nabla^{\alpha}_{\perp}\varphi \right)_A  + \frac{1}{2} \mu_A\left(\phi\right) \left(\nabla_{\alpha\beta}^{\perp}\varphi\right)_A\,\left(\nabla^{\alpha\beta}_{\perp}\varphi \right)_A \\
& \hspace{3cm} + \nu_A\left(\phi\right) \left(\nabla_{\alpha\beta}^{\perp}\varphi\right)_A\,(G^{\alpha\beta})_A - \frac{1}{2c^2} c_A(\phi)  \,  (G_{\alpha\beta})_A \, (G^{\alpha\beta})_A \; \biggr\}  \label{eq:Stidal}\,,
\end{align}
\end{subequations}
where $\ud\tau_A/c\equiv \ud t\,\sqrt{-(g_{\mu\nu})_A v_A^{\mu}v_A^{\nu}/c^2}$ is the proper time of particle $A$ along its worldline $y_A^{\mu}$ . The object masses are functions of the scalar field to take into account their internal self-gravity~\cite{1975ApJ...196L..59E}. The action is also made of the mass quadrupole tidal moment defined as 
\begin{align}\label{eq:tensor_field}
(G_{\mu \nu})_A= - c^2\,{\left(R_{\mu \rho \nu \sigma}\right)}_A \, u_A^{\rho} \, u_A^{\sigma}\,.
\end{align}
where the 4-velocity $u_A^{\mu}$ is normalized to $g_{\mu \nu}^{a} u_A^{\mu} u_A^{\nu}=-1$. The tidal action~\eqref{eq:Stidal} is sufficient to capture the impact of tidal deformations on the dynamics and radiation of a binary system up to the NNLO. Indeed, using Eqs.~\eqref{tiddeform} and~\eqref{eq:dimless_TLN}, one finds that the first term of~\eqref{eq:Stidal} generates the leading tidal effects together with the subleading contributions up to NNLO, while the three subsequent terms contribute purely at the NNLO. Higher-order terms in the action that naturally arise in the EFT construction, such as non-quadratic couplings between the scalar tidal field and the gravitational tidal moments, enter necessarily at NNNLO or beyond, which is beyond the scope of this work.  Hence, the results presented in this work depend only on the four tidal deformability parameters $(\lambda_A,\mu_A,\nu_A,c_A)$. 

%------------------------------------------------------------------------
\subsection{The field equations}
\label{subsec:field_eqns}
%------------------------------------------------------------------------

For the treatment of the radiative sector, it is necessary to consider and solve the full Einstein field equations including the tidal contributions. Varying the gauge-fixed action ~\eqref{eq:STactionEF}, supplemented by the matter and tidal actions ~\eqref{eq:Smatter+tidal}, w.r.t the conformal metric $\tilde{g}_{\mu \nu}$ and the scalar field $\varphi$, yields to the following field equations :

\begin{subequations}\label{eq:rEFE_g_tilde}
\begin{align}
\tilde{G}_{\mu\nu} + \frac{3+2 \omega(\phi)}{2 \varphi^2}\left( \frac{1}{2} \tilde{g}^{\alpha \beta}\tilde{g}_{\mu\nu} - \delta^{\alpha}_{\mu} \delta^{\beta}_{\nu}\right) \p_{\alpha} \varphi \p_{\beta} \varphi &= \frac{8 \pi G}{c^4 \phi_0} \left( \tilde{T}_{\mu \nu}^{\rm pp} +  \tilde{T}_{\mu \nu}^{\rm tidal} \right)\,,\\ 
\tilde{g}^{\alpha \beta} \tilde{\nabla}_{\alpha}\tilde{\nabla}_{\beta} \varphi + \frac{1}{2} \frac{d}{d \varphi} \left[\ln{\left(\frac{3+2 \omega(\phi)}{\varphi ^2} \right)} \right]\tilde{g}^{\alpha \beta} \p_{\alpha} \varphi \p_{\beta} \varphi &= \frac{8 \pi G}{c^4 \phi_0} \frac{\varphi}{3+2 \omega(\phi)} \left( \Delta \tilde{S}_{\rm pp} + \Delta \tilde{S}_{\rm tidal}\right)\label{eq:rEFE_g_tilde_2}\,,
\end{align}
\end{subequations}
where we have introduced  the four-dimensional Kronecker delta $\delta_{\nu}^{\mu}$, the conformal Einstein tensor $\tilde{G}_{\mu\nu}$ and the conformal stress-energy tensor for point-particles $\tilde{T}^{\mu\nu}_{\rm pp} \equiv \frac{2}{\sqrt{-\tilde{g}}}\frac{\delta S_{\mathrm{pp}}}{\delta \tilde{g}_{\mu\nu}}$ . Note that $\tilde{T}^{\mu\nu}_{\rm pp}$ can be related to the physical (Jordan-frame) stress-energy tensor  $T^{\mu\nu}_{\rm pp} \equiv \frac{2}{\sqrt{-g}}\frac{\delta S_{\mathrm{pp}}}{\delta g_{\mu\nu}}$ by the relation $T^{\mu\nu}_{\rm pp}= \varphi^3 \tilde{T}^{\mu\nu}_{\rm pp}$ where 
\begin{align}\label{eq:Tmunupp}
    \tilde{T}^{\mu\nu}_{\rm pp}\left(t,\mathbf{x}\right) = \sum_{A=1,2}\,\dfrac{m_A\left(\phi\right)v_A^{\mu}v_A^{\nu}}{\sqrt{-\tilde{g}_{\mu\nu}v_A^{\mu}v_A^{\nu}/c^2}}\frac{1}{\sqrt{{\varphi}}}\frac{\delta_A^{(3)} (\mathbf{x}-\mathbf{y_A}(t))}{\sqrt{-\tilde{g}}}\,.
\end{align}
We denote by $\tilde{T}^{\mu\nu}_{\rm tidal} \equiv \frac{2}{\sqrt{-\tilde{g}}}\frac{\delta S_{\mathrm{tidal}}}{\delta \tilde{g}_{\mu\nu}}$ the tidal correction to the conformal stress-energy tensor, and by $\tilde{T}$ its trace taken with respect to the conformal metric $\tilde{g}_{\mu\nu}$. For convenience, we have also defined $ \Delta \tilde{S}_{\rm pp} = \frac{-2 \varphi}{\sqrt{-\tilde{g}}}\frac{\delta S_{\rm pp}}{\delta \varphi}$ and $ \Delta \tilde{S}_{\rm tidal} = \frac{-2 \varphi}{\sqrt{-\tilde{g}}}\frac{\delta S_{\rm tidal}}{\delta \varphi}$.   In particular, the term $\Delta \tilde{S}_{\rm pp}$ is linked to the trace of the conformal stress-energy tensor by the relation
\begin{align}
\Delta \tilde{S}_{\rm pp} = \left.-2 \varphi \frac{\partial \tilde{T}}{\partial \varphi}\right|_{\tilde{g}_{\mu \nu}}
\end{align}
where the partial derivative is taken at fixed Einstein-frame metric $\tilde{g}_{\mu \nu}$. This term accounts for both the explicit dependence of $\tilde{T}^{\mu\nu}$ on $\varphi$, and the implicit dependence arising from the scalar-field dependence of the masses of self-gravitating bodies, treated following Eardley’s approach.\\

Introducing the gothic metric $\tilde{\mathfrak{g}}^{\mu\nu} \equiv \sqrt{-\tilde{g}}\,\tilde{g}^{\mu\nu}$, the gothic metric deviation $h^{\mu\nu}\equiv \tilde{\mathfrak{g}}^{\mu\nu} -\eta^{\mu\nu}$  and the scalar perturbation $\psi\equiv\varphi-1$,  and imposing the harmonic gauge condition $\partial_{\mu}h^{\mu\nu}=0$,  the Einstein field equations~\eqref{eq:rEFE_g_tilde} can be written in the form of inhomogeneous flat d'Alembertian equations, namely
\begin{subequations}\label{eq:rEFE}
\begin{align}
&\Box_{\eta}\,h^{\mu\nu} = \frac{16\pi G}{c^{4}\phi_0}\tau^{\mu\nu}\,,\label{eq:EFE_V1_1}\\
& \Box_{\eta}\,\psi = \frac{8\pi G}{c^{4} \phi_0}\tau_{s}\label{EFE_V1_2}\,,
\end{align}
\end{subequations}
where $\Box_{\eta} = \eta^{\mu \nu} \p_{\mu} \p_{\nu}$ and where the tensor and scalar pseudo-tensors both read
\begin{subequations}\label{tau}
\begin{align}\label{taumunu}
\tau^{\mu\nu} &=  \vert \tilde{g}\vert (\tilde{T}^{\mu\nu}_{\rm pp}+ \tilde{T}^{\mu\nu}_{\rm tidal}) +\frac{c^{4} \phi_0}{16\pi G} \Lambda^{\mu\nu} \,,\\
\label{taus} \tau_{s} &= \frac{  \varphi }{(3+2\omega(\phi))}\sqrt{-\tilde{g}}\left(\Delta \tilde{S}_{\rm pp} + \Delta \tilde{S}_{\rm tidal}\right) +\frac{c^{4} \phi_0}{8\pi G}\Lambda_s\,.
\end{align}
\end{subequations}
The non-linearities are encoded by the scalar source term~\cite{Bernard:2018hta}
\be\label{eq:lambdas} 
\Lambda_s = -h^{\alpha\beta}\p_{\alpha}\p_{\beta}\psi-\p_{\alpha}\psi\p_{\beta}h^{\alpha\beta} +\left(\frac{1}{\varphi}-\frac{\phi_{0}\omega'(\phi)}{3+2\omega(\phi)}\right)\tilde{\mathfrak{g}}^{\alpha\beta}\p_{\alpha}\psi\p_{\beta}\psi\,.
\ee 
and by the gravitational one $\Lambda^{\mu\nu}$, which can in turn be decomposed as $\Lambda^{\mu\nu} = \Lambda^{\mu\nu}_{\mathrm{LL}}+\Lambda_{\mathrm{H}}^{\mu\nu}+\Lambda^{\mu\nu}_{\mathrm{GF}}+\Lambda_{\phi}^{\mu\nu}$ where $\Lambda^{\mu\nu}_{\mathrm{LL}}$ is the Landau-Lifshitz pseudo-energy tensor~\cite{Landau:1975pou}, $\Lambda^{\mu\nu}_{\mathrm{H}}$ comes from our use of the flat version of the d'Alembertian operator in Eqs.~\eqref{eq:rEFE}, $\Lambda^{\mu\nu}_{\mathrm{GF}}$ is due to the gauge-fixing term in the action and $\Lambda_{\phi}^{\mu\nu}$ is sourced by the scalar field. Note that $\Lambda_{\mathrm{GR}}^{\mu\nu}=\Lambda^{\mu\nu}_{\mathrm{LL}}+\Lambda_{\mathrm{H}}^{\mu\nu}+\Lambda^{\mu\nu}_{\mathrm{GF}}$ will take the same form as in GR. The expressions of these source terms are given by~\cite{,Bernard:2018hta}
\begin{subequations}\label{eq:exprLambda}
\begin{align}\label{eq:LambdaLL}
& \Lambda_{\mathrm{LL}}^{\alpha\beta} =\ \frac{1}{2}\tilde{\mathfrak{g}}^{\alpha\beta}\tilde{\mathfrak{g}}_{\mu\nu}\partial_{\lambda}h^{\mu\gamma}\partial_{\gamma}h^{\nu\lambda}-\tilde{\mathfrak{g}}^{\alpha\mu}\tilde{\mathfrak{g}}_{\nu\gamma}\partial_{\lambda}h^{\beta\gamma}\partial_{\mu}h^{\nu\lambda} -\tilde{\mathfrak{g}}^{\beta\mu}\tilde{\mathfrak{g}}_{\nu\gamma}\partial_{\lambda}h^{\alpha\gamma}\partial_{\mu}h^{\nu\lambda} \nn&\\
&\qquad\quad +\tilde{\mathfrak{g}}_{\mu\nu}\tilde{\mathfrak{g}}^{\lambda\gamma}\partial_{\lambda}h^{\alpha\mu}\partial_{\gamma}h^{\beta\nu} +\frac{1}{8}\left(2\tilde{\mathfrak{g}}^{\alpha\mu}\tilde{\mathfrak{g}}^{\beta\nu}-\tilde{\mathfrak{g}}^{\alpha\beta}\tilde{\mathfrak{g}}^{\mu\nu}\right)\left(2\tilde{\mathfrak{g}}_{\lambda\gamma}\tilde{\mathfrak{g}}_{\tau\pi}-\tilde{\mathfrak{g}}_{\gamma\tau}\tilde{\mathfrak{g}}_{\lambda\pi}\right)\partial_{\mu}h^{\lambda\pi}\partial_{\nu}h^{\gamma\tau} \,, \\
\label{eq:LambdaH}
&\Lambda_{\mathrm{H}}^{\alpha\beta} = -h^{\mu\nu}\partial_{\mu}\partial_{\nu}h^{\alpha\beta} +\partial_{\mu}h^{\alpha\nu}\partial_{\nu}h^{\beta\mu} \,,\\
\label{eq:Lambdagf}
& \Lambda_{\mathrm{GF}}^{\alpha\beta} = - \partial_{\lambda}h^{\lambda\alpha} \partial_{\sigma}h^{\sigma\beta} - \partial_{\lambda}h^{\lambda\rho}\partial_{\rho}h^{\alpha\beta} -\frac{1}{2}\tilde{\mathfrak{g}}^{\alpha\beta}\tilde{\mathfrak{g}}_{\rho\sigma} \partial_{\lambda}h^{\lambda\rho} \partial_{\gamma}h^{\gamma\sigma} +2\tilde{\mathfrak{g}}_{\rho\sigma}\tilde{\mathfrak{g}}^{\lambda(\alpha} \partial_{\lambda}h^{\beta)\rho}\partial_{\gamma}h^{\gamma\sigma} \,,\\
\label{eq:Lambdas}
&\Lambda_{\phi}^{\mu\nu} = \frac{3+2\omega(\phi)}{\varphi^2}\left(\tilde{\mathfrak{g}}^{\mu\alpha}\tilde{\mathfrak{g}}^{\nu\beta} -\frac{1}{2}\tilde{\mathfrak{g}}^{\mu\nu}\tilde{\mathfrak{g}}^{\alpha\beta}\right)\p_{\alpha}\psi\p_{\beta}\psi\,.
\end{align}
\end{subequations}
The gauge-fixing term~\eqref{eq:Lambdagf} contains $\p_{\nu}h^{\mu\nu}$ terms which are not zero off-shell, \textit{i.e.} when the accelerations are not replaced by the equations of motion. However in this work, all quantities will be directly computed on-shell, implying that this term is zero.

As for any general matter system in harmonic coordinates, the metric and scalar field perturbations can be parameterized by some PN potentials, $\left(V,\,V^i,\,\hat{W}^{ij}\right)$ for the metric~\cite{Blanchet:2013haa}, and $\left(\psi_{(0)},\,\psi_{(1)}\right)$ for the scalar field~\cite{Bernard:2018hta, Bernard:2023eul}, as
\begin{subequations}
\label{metric_decomp}
\begin{align}
 h^{00ii} = & - \frac{4}{c^2}V- \frac{8}{c^4} V^2 + \mathcal{O}\left(\frac{1}{c^6}\right) \,, \\
h^{0i} = &- \frac{4}{c^3}V_i+ \mathcal{O}\left(\frac{1}{c^5}\right) \,, \\
h^{ij} = & - \frac{4}{c^4}(\hat{W}_{ij}-\frac{1}{2} \delta_{ij} \hat{W})+\mathcal{O}\left(\frac{1}{c^6}\right) \,, \\
\psi = & -\frac{2}{c^2}\psi_{(0)}+\frac{2}{c^4}\left(1-\frac{\phi_0 \omega_0^{'}}{3+2 \omega_0}\right)\psi_{(0)}^2 \nn \\
& \quad +\frac{1}{c^6}\left[-\frac{4}{3}\Biggl(1-\frac{4\phi_0 \omega_0^{'}}{3+2 \omega_0}-\frac{\phi_0^2(-4(\omega_0^{'})^2+ (3+2 \omega_0)\omega_0^{''})}{(3+2 \omega_0)^2}\right)\psi_{(0)}^3 +\psi_{(1)}\Biggr] +\mathcal{O}\left(\frac{1}{c^8}\right) \,, 
\end{align}
\end{subequations}
with $h^{00ii}\equiv h^{00}+h^{ii}$. These potentials satisfy the following flat spacetime wave equations
\begin{subequations}
\label{eq:PotentialWaveEq}
\begin{align}
\Box V = & -4 \pi G \sigma \,, \\
\Box V_i = & -4 \pi G \sigma_i \,, \\
\Box \hat{W}_{ij} = & -4 \pi G \left( \sigma_{ij}-\delta_{ij}\sigma_{kk}\right) -\p_i V\p_iV - (3+2 \omega_0)\p_i \psi_{(0)} \p_j \psi_{(0)} \,, \\
\Box \psi_{(0)} = & -4 \pi G \sigma_s \,, \\
\Box \psi_{(1)} = & -16 \pi G \sigma_s \hat{W} -4 \biggl( 4 V_i\p_{ti}\psi_{(0)} + 2 \hat{W}_{ij}\p_{ij}\psi_{(0)}+2V \p_t^2 \psi_{(0)} \nn \\
& \quad + 2 \bigl( \p_t V_i + \p_j \hat{W}_{ij} -\frac{1}{2} \p_i \hat{W}\bigr)\p_i \psi_{(0)} \biggr) \label{PotentialWaveEq_4}\,,
\end{align}
\end{subequations}
which were obtained by injecting the ansatz~\eqref{metric_decomp} into the field equations~\eqref{eq:rEFE}. Their sources, corresponding to the right hand-sides of Eqs.~\eqref{eq:PotentialWaveEq}, contain both the potentials themselves, and the source densities $\sigma,\sigma_i,\sigma_{ij}$ and $\sigma_s$, defined by 
\begin{subequations}
\begin{align}
	\sigma & = \frac{\tilde{T}^{00}_{\rm pp}+ \tilde{T}^{00}_{\rm tidal} + \tilde{T}^{ii}_{\rm pp}+ \tilde{T}^{ii}_{\rm tidal}}{c^2}\,,\qquad	\sigma_i = \frac{\tilde{T}^{0i}_{\rm pp}+ \tilde{T}^{0i}_{\rm tidal}}{c} \,,\qquad	\sigma_{ij} = \tilde{T}^{ij}_{\rm pp}+\tilde{T}^{ij}_{\rm tidal} \,,\\[5pt]
\sigma_s & =  \frac{1}{c^2}\frac{1}{\sqrt{\left(3+2\omega_0\right)(3+2\omega(\phi))}} \sqrt{-\tilde{g}} \left(\Delta \tilde{S}_{\rm pp} + \Delta \tilde{S}_{\rm tidal}\right) \,.\label{eq:sigmas}	
\end{align}
\end{subequations}
These source densities have compact support, as they are constructed from the full conformal stress-energy tensor, including tidal contributions, and from the terms $\Delta \tilde{S}_{\rm pp}$ and $ \Delta \tilde{S}_{\rm tidal}$. All involve a dependence on the three-dimensional Dirac delta distribution $\delta_A^{(3)}\left(\mathbf{x}-\mathbf{y}_A(t)\right)$, which is confined to the worldline $y_A^i(t)$ of each body.  We expand the masses $m_A\left(\phi\right)$ around the asymptotic value of the scalar field at infinity, $\phi_0$. One finds~\cite{Mirshekari:2013vb}
\begin{align}
m_A\left(\phi\right)=m_A \Biggl[ 1 + s_A \psi + \frac{1}{2} \Bigl( s_A^2 + s_A' - s_A\Bigr) \psi^2 + \frac{1}{6}\Bigl(s_A'' + 3 s_A's_A -3s_A' + s_A^3-3s_A^2 +2s_A\Bigr) \psi^3 + \mathcal{O}\left(\psi^4\right)\Biggr]\,.
\end{align}
where we introduce the dimensionless sensitivities and higher-order sensitivities
\begin{align}
s_A = \left.\frac{\ud \ln{m_A(\phi)}}{\ud\ln{\phi}}\right\vert_{\phi=\phi_0},\qquad s_A' = \left.\frac{\ud^{2}\ln{m_A(\phi)}}{\ud(\ln{\phi})^{2}}\right\vert_{\phi=\phi_0},\qquad s_A'' = \left.\frac{\ud^{3}\ln{m_A(\phi)}}{\ud(\ln{\phi})^{3}}\right\vert_{\phi=\phi_0}\,.
\end{align}
The sensitivities are not free parameters, as they are determined by the underlying scalar-tensor theory and by the internal structure of the compact object~\cite{Damour:1993hw}. In particular, they depend on the scalar coupling function $\omega(\phi)$ characterizing the theory, on the mass and equation of state of the neutron star, and on the asymptotic value $\phi_0$ of the scalar field. For a given choice of theory and equation of state, the sensitivities can be computed by strong field numerical computations~\cite{Doneva:2022ewd}. Typical values are around $s_{\text{NS}}\sim 0.2$ for neutron stars, while stationary black holes satisfy $s_{\text{BH}}\sim 1/2$, consistently with the absence of scalar hair in Brans-Dicke~\cite{Hawking:1972qk} and in more general classes of scalar-tensor theories~\cite{Sotiriou:2011dz}. Similarly, we expand the tidal deformability parameters around the asymptotic value $\phi_0$, in powers of the scalar perturbation
\begin{align}\label{tiddeform}
& \lambda_A(\phi) = \sum_{n=0}^{\infty} \frac{\lambda_A^{(n)}}{n!} \,\phi_0^n \, {\psi}^n\,, \qquad \mu_A(\phi) = \sum_{n=0}^{\infty} \frac{\mu_A^{(n)}}{n!} \,\phi_0^n \, {\psi}^n\,, \qquad \nu_A(\phi) = \sum_{n=0}^{\infty} \frac{\nu_A^{(n)}}{n!} \,\phi_0^n \, {\psi}^n\,,\qquad c_A(\phi) = \sum_{n=0}^{\infty} \frac{c_A^{(n)}}{n!} \,\phi_0^n \, {\psi}^n\,.
\end{align}
As for the sensitivities, the coefficients of these expansions $\lambda_A^{(n)},\mu_A^{(n)}, \nu_A(\phi)$ and $c_A^{(n)}$ are not independent parameters: for a given scalar–tensor theory, neutron-star equation of state, and background scalar field, they are uniquely determined by strong-field calculations of tidally deformed compact objects~\cite{Creci:2023cfx}.

%------------------------------------------------------------------------
%------------------------------------------------------------------------
\section{NNLO tidal effects on the dynamics for circular orbits}
\label{sec:dynamics_circ}
%------------------------------------------------------------------------
%------------------------------------------------------------------------

The dynamics of compact binary systems in ST theories is currently known up to 3PN order in the point-particle approximation~\cite{Bernard:2018ivi,Bernard:2018hta}. In Ref.~\cite{Bernard:2023eul}, building up on the initial work~\cite{Bernard:2019yfz}, we extended these results by computing the NNLO tidal corrections to the equations of motion and to the conserved quantities, notably the energy and angular momentum. In this section, we reduce these results for the dynamical sector to the specific case of circular orbits. Complete and explicit expressions will be made available upon request.

We begin by considering the NNLO conservative equations of motion, as presented in Ref.~\cite{Bernard:2023eul}.  Because we restrict here to circular orbits in the conservative sector, we can set ${\bf n}\cdot{\bf v} = 0$ and ${\bf v}^2=\omega^2 r^2$.  The relative acceleration becomes purely radial and can be written as 
\begin{align}
\label{eq:EoM_radial}
a^i= - \omega ^2 x^i
\end{align}
where $\omega$ is the orbital frequency. This frequency can be expressed either in terms of the orbital separation r or in terms of the PN parameter $\gamma=\tilde{G} \alpha m/(c^2 r)$. We decompose the frequency as $\omega^2=\omega^2_{\text{pp}}+\omega^2_{\text{tidal}}$ and find for the tidal part
\begin{align}
\label{eq:omega2_v2}
\omega^2_{\text{tidal}} = & \frac{8 c^6 \gamma^6}{\alpha^2 \tilde{G}^2 m^2 (-1 + \zeta) \bar{\gamma}} \Bigl\{\zeta \bar{\gamma} \tilde{\lambda}_{+}^{(0)} + \gamma \Bigl[\frac{11}{4} \zeta \bar{\gamma} \nu \tilde{\lambda}_{+}^{(0)} + \frac{1}{16} \delta (80 \bar{\beta}^+ \zeta \tilde{\lambda}_{-}^{(0)} + 12 \zeta \bar{\gamma} \tilde{\lambda}_{-}^{(0)} + 15 \zeta \bar{\gamma}^2 \tilde{\lambda}_{-}^{(0)} - 25 \zeta \bar{\gamma} \tilde{\Lambda}_{-}^{(0)} \nn \\
& + 20 \bar{\gamma} \lambda_1 \tilde{\Lambda}_{-}^{(0)} - 10 \zeta \bar{\gamma} \phi_{0}{} \tilde{\Lambda}_{-}^{(1)} - 80 \bar{\beta}^- \zeta \tilde{\lambda}_{+}^{(0)}) + \frac{1}{16} (-80 \bar{\beta}^- \zeta \tilde{\lambda}_{-}^{(0)} + 80 \bar{\beta}^+ \zeta \tilde{\lambda}_{+}^{(0)} - 92 \zeta \bar{\gamma} \tilde{\lambda}_{+}^{(0)} - 31 \zeta \bar{\gamma}^2 \tilde{\lambda}_{+}^{(0)} \nn \\
& + 25 \zeta \bar{\gamma} \tilde{\Lambda}_{+}^{(0)} - 20 \bar{\gamma} \lambda_1 \tilde{\Lambda}_{+}^{(0)} + 10 \zeta \bar{\gamma} \phi_{0}{} \tilde{\Lambda}_{+}^{(1)})\Bigr] + \gamma^2 \Bigl[- \frac{9}{32} \zeta \bar{\gamma} \bigl(\tilde{c}_{+}^{(0)} - 4 (4 \tilde{\mu}_{+}^{(0)} + \tilde{\nu}_{+}^{(0)})\bigr) + (\cdots)\Bigr] \, + \,  \calO\left(\frac{\etidal}{c^6}\right)\Bigr\} \,,
\end{align}
We recall that the combinations of tidal deformability parameters are defined in Eqs.~\eqref{eq:tidal_deformabilities_CoM} and \eqref{eq:tidal_deformabilities_Circ}. The notation  $(\cdots)$ indicates that terms are too lengthy to be displayed in full. Still, we choose to give explicitely the dependencies on the quadrupolar scalar, tensorial or scalar-tensorial tidal deformabilities (resp. $\mu_+^{(0)}$, $c_+^{(0)}$ and $\nu_+^{(0)}$, see Section~\ref{subsec:notations}). This convention is adopted throughout the article.    The complete expressions are provided in the Supplemental Material~\cite{SuppMaterial}. 

Next, using the relation between the orbital frequency and the PN parameter
\begin{align}
\label{eq:x_param}
x=\left(\frac{\tilde{G}\alpha m \omega}{c^3}\right)^{2/3}\,,
\end{align}
we invert the expression for $\omega^2$ to obtain $\gamma = \gamma_{\text{pp}} +\gamma_{\text{tidal}}$ as a function of $x$. We find
\begin{align}
\label{eq:gamma_x}
\gamma_{\text{tidal}} = & - \frac{8 x^4}{3 (-1 + \zeta) \bar{\gamma}} \Bigl\{\zeta \bar{\gamma} \tilde{\lambda}_{+}^{(0)} + x \Bigl[\frac{1}{12} \zeta \bar{\gamma} \nu \tilde{\lambda}_{+}^{(0)} + \frac{1}{48} \delta (240 \bar{\beta}^+ \zeta \tilde{\lambda}_{-}^{(0)} + 36 \zeta \bar{\gamma} \tilde{\lambda}_{-}^{(0)} + 45 \zeta \bar{\gamma}^2 \tilde{\lambda}_{-}^{(0)} - 75 \zeta \bar{\gamma} \tilde{\Lambda}_{-}^{(0)} \nn \\
& + 60 \bar{\gamma} \lambda_1 \tilde{\Lambda}_{-}^{(0)} - 30 \zeta \bar{\gamma} \phi_{0}{} \tilde{\Lambda}_{-}^{(1)} - 240 \bar{\beta}^- \zeta \tilde{\lambda}_{+}^{(0)} - 256 \bar{\beta}^- \zeta \bar{\gamma} \tilde{\lambda}_{+}^{(0)}) + \frac{1}{48} (-240 \bar{\beta}^- \zeta \tilde{\lambda}_{-}^{(0)} + 240 \bar{\beta}^+ \zeta \tilde{\lambda}_{+}^{(0)} \nn \\
& + 108 \zeta \bar{\gamma} \tilde{\lambda}_{+}^{(0)} + 256 \bar{\beta}^+ \zeta \bar{\gamma} \tilde{\lambda}_{+}^{(0)} + 35 \zeta \bar{\gamma}^2 \tilde{\lambda}_{+}^{(0)} + 75 \zeta \bar{\gamma} \tilde{\Lambda}_{+}^{(0)} - 60 \bar{\gamma} \lambda_1 \tilde{\Lambda}_{+}^{(0)} + 30 \zeta \bar{\gamma} \phi_{0}{} \tilde{\Lambda}_{+}^{(1)})\Bigr]\nn \\
& + x^2 \Bigl[- \frac{9}{32} \zeta \bar{\gamma} \bigl(\tilde{c}_{+}^{(0)} - 4 (4 \tilde{\mu}_{+}^{(0)} + \tilde{\nu}_{+}^{(0)})\bigr) + (\cdots) \Bigr] + \calO\left(\frac{\etidal}{c^6}\right)  \Bigr\}\,.
\end{align}
Using this result for $\gamma(x)$, we can now express the conserved center-of-mass energy $E = E_{\text{pp}}+E_{\text{tidal}}$ and angular momentum  $J^i = J^i_{\text{pp}}+J^i_{\text{tidal}}$ for circular orbits in terms of the gauge-invariant parameter $x$. The NNLO tidal correction to the energy is given by
\begin{align}
\label{eq:E_circ}
E_{\text{tidal}} =& \frac{10 c^2 m \nu x^4}{3 (-1 + \zeta) \bar{\gamma}} \Bigl\{\zeta \bar{\gamma} \tilde{\lambda}_{+}^{(0)} + x \Bigl[- \frac{7}{6} \zeta \bar{\gamma} \nu \tilde{\lambda}_{+}^{(0)} + \frac{7}{60} \delta (48 \bar{\beta}^+ \zeta \tilde{\lambda}_{-}^{(0)} + 9 \zeta \bar{\gamma} \tilde{\lambda}_{-}^{(0)} + 9 \zeta \bar{\gamma}^2 \tilde{\lambda}_{-}^{(0)} - 15 \zeta \bar{\gamma} \tilde{\Lambda}_{-}^{(0)} + 12 \bar{\gamma} \lambda_1 \tilde{\Lambda}_{-}^{(0)} \nn \\
& - 6 \zeta \bar{\gamma} \phi_{0}{} \tilde{\Lambda}_{-}^{(1)} - 48 \bar{\beta}^- \zeta \tilde{\lambda}_{+}^{(0)} - 32 \bar{\beta}^- \zeta \bar{\gamma} \tilde{\lambda}_{+}^{(0)}) -  \frac{7}{60} (48 \bar{\beta}^- \zeta \tilde{\lambda}_{-}^{(0)} - 48 \bar{\beta}^+ \zeta \tilde{\lambda}_{+}^{(0)}  - 15 \zeta \bar{\gamma} \tilde{\lambda}_{+}^{(0)} - 32 \bar{\beta}^+ \zeta \bar{\gamma} \tilde{\lambda}_{+}^{(0)} \nn \\
& - 7 \zeta \bar{\gamma}^2 \tilde{\lambda}_{+}^{(0)}  - 15 \zeta \bar{\gamma} \tilde{\Lambda}_{+}^{(0)} + 12 \bar{\gamma} \lambda_1 \tilde{\Lambda}_{+}^{(0)} - 6 \zeta \bar{\gamma} \phi_{0}{} \tilde{\Lambda}_{+}^{(1)})\Bigr] + x^2 \Bigl[- \frac{27}{80} \zeta \bar{\gamma} \bigl(\tilde{c}_{+}^{(0)} - 4 (4 \tilde{\mu}_{+}^{(0)} + \tilde{\nu}_{+}^{(0)})\bigr) + (\cdots) \Bigr] \nn \\
& + \calO\left(\frac{\etidal}{c^6}\right)\Bigr\}\,, 
\end{align}
and the corresponding correction to the angular momentum reads
\begin{align}
\label{eq:Ji_circ}
J_{\text{tidal}}^i &= \frac{16 \alpha \tilde{G} m^2 \nu  x^{5/2}}{3 c (-1 + \zeta) \bar{\gamma}}\ell^{i}  \Bigl\{\zeta \bar{\gamma} \tilde{\lambda}_{+}^{(0)} + x \Bigl[- \frac{25}{24} \zeta \bar{\gamma} \nu \tilde{\lambda}_{+}^{(0)} + \frac{5}{48} \delta (48 \bar{\beta}^+ \zeta \tilde{\lambda}_{-}^{(0)} + 9 \zeta \bar{\gamma} \tilde{\lambda}_{-}^{(0)} + 9 \zeta \bar{\gamma}^2 \tilde{\lambda}_{-}^{(0)} - 15 \zeta \bar{\gamma} \tilde{\Lambda}_{-}^{(0)} \nn \\
&+ 12 \bar{\gamma} \lambda_1 \tilde{\Lambda}_{-}^{(0)} - 6 \zeta \bar{\gamma} \phi_{0}{} \tilde{\Lambda}_{-}^{(1)} - 48 \bar{\beta}^- \zeta \tilde{\lambda}_{+}^{(0)} - 32 \bar{\beta}^- \zeta \bar{\gamma} \tilde{\lambda}_{+}^{(0)}) -  \frac{5}{48} (48 \bar{\beta}^- \zeta \tilde{\lambda}_{-}^{(0)} - 48 \bar{\beta}^+ \zeta \tilde{\lambda}_{+}^{(0)} - 15 \zeta \bar{\gamma} \tilde{\lambda}_{+}^{(0)} \nn \\
&- 32 \bar{\beta}^+ \zeta \bar{\gamma} \tilde{\lambda}_{+}^{(0)} - 7 \zeta \bar{\gamma}^2 \tilde{\lambda}_{+}^{(0)} - 15 \zeta \bar{\gamma} \tilde{\Lambda}_{+}^{(0)} + 12 \bar{\gamma} \lambda_1 \tilde{\Lambda}_{+}^{(0)} - 6 \zeta \bar{\gamma} \phi_{0}{} \tilde{\Lambda}_{+}^{(1)})\Bigr] +x^2 \Bigl[- \frac{9}{32} \zeta \bar{\gamma}  \bigl(\tilde{c}_{+}^{(0)} - 4 (4 \tilde{\mu}_{+}^{(0)} + \tilde{\nu}_{+}^{(0)})\bigr) \nn \\
&+ (\cdots) \Bigr] + \calO\left(\frac{\etidal}{c^6}\right)\Bigr\}\,. 
\end{align}
The point-particle parts $\omega^2_{\text{pp}}$, $\gamma_{\text{pp}}$, $E_{\text{pp}}$ and $J^i_{\text{pp}}$ can respectively be found up to 3PN order in Eqs. (5.2), (B1), (5.4) and (5.6) of Ref.~\cite{Bernard:2018ivi}, where they are said to be instantaneous. This is due to the fact that at 3PN order, the conservative sector gets corrected by nonlocal tail terms. However, since we only need the point-particle parts up to 2PN order, we will not consider the tail effects in the dynamical sector in this work.

%------------------------------------------------------------------------
%------------------------------------------------------------------------
\section{Waveform generation formalism}
\label{sec:mPMPN}
%------------------------------------------------------------------------
%------------------------------------------------------------------------

%------------------------------------------------------------------------
\subsection{Asymptotic waveforms and fluxes}
\label{sec:asymptotic_waveform}
%------------------------------------------------------------------------

It is well known in GR that the harmonic coordinate system $(t,\bm{x})$ is ill-suited when studying the asymptotic behaviour at spatial infinity of the MPM solution, \textit{i.e.} the solution constructed outside an isolated system using the MPM expansion scheme~\cite{Blanchet:1985sp}. In this coordinate system, the general structure of the MPM solution at future null infinity (\textit{i.e.} when $r \rightarrow + \infty$ with $ u = t-r/c=\text{const}$) involves powers of logarithms of $r$, due to the harmonic gauge. However, one can get rid of these logarithms by going to a radiative coordinate system $(\mathrm{T},\mathrm{\textbf{X}})$, for which the MPM solution at future null infinity (\textit{i.e.} when $R \rightarrow + \infty$ with $ U = T-R/c=\text{const}$) admits a (Bondi-type) expansion in simple powers of $1/R$~\cite{BBM62,Sachs62}. By radiative coordinate system, we mean a coordinate system whose retarded time coordinate $U$ is or becomes a null coordinate at future null infinity, \textit{i.e.} it satisfies $g^{\mu\nu}_{\text{rad}}\,\partial_{\mu}U \,\partial_{\nu}U=0$. For a systematic treatment relating harmonic to radiative coordinates the MPM expansion, we refer for \textit{e.g.} to~\cite{Blanchet:1986dk,Trestini:2022tot}. At low PN orders, a simple linear gauge transformation is enough to remove the logarithmic terms. This is achieved with the gauge vector  
\begin{align}\label{eq:gauge_vector}
\xi^{\mu} = \frac{2\ADM}{c^2}\eta^{0\mu}\text{ln}\left(\frac{r}{c \, b_0}\right)
\end{align}
where $\ADM$ denotes the ADM mass, and $b_0$ is an arbitrary time scale. The retarded time $U=T-R/c$ in radiative coordinates is then related to the retarded time $u=t-r/c$ in harmonic coordinates by
\begin{align}\label{eq:retarded_time_rad}
U = u - \frac{2 G \ADM}{c^3}\text{ln}\left(\frac{r}{c\, b_0}\right) + \mathcal{O}(G^2)
\end{align}

In this radiative coordinate system $(\mathrm{T},\mathrm{\textbf{X}})$, the transverse-traceless (TT) projection of the metric can be uniquely decomposed, at leading order in $1/R$, in terms of two sets of symmetric-trace-free (STF) radiative moments, the mass-type $\mathcal{U}_L$ and the current-type $\mathcal{V}_L$, as~\cite{RevModPhys.52.299}
\begin{equation}\label{eq:hijTT}
h_{ij}^\text{TT} = - \frac{4G}{c^2 R} \perp^\text{TT}_{ijab} \sum_{\ell=2}^{+\infty} \frac{1}{c^\ell \ell!}\Big( N_{L-2}\,\mathcal{U}_{ab L-2}(U) - \frac{2\ell}{c(\ell+1)} N_{c L-2} \varepsilon_{cd(a}\mathcal{V}_{b)dL-2}(U)\Big) + \mathcal{O}\Big(\frac{1}{R^2}\Big)\,,   
\end{equation}
where $\perp^\text{TT}_{ijab}\equiv\frac{1}{2}(\perp_{ia}\perp_{jb}+\perp_{ja}\perp_{ib}-\perp_{ij}\perp_{ab})$ with $\perp_{ij}=\delta_{ij}-N_iN_j$ being the TT projection operator. We recall that \textit{e.g.} $N_{L-2}=N_{i_1}\cdots N_{i_{\ell-2}}$. Similarly, the scalar waveform can also be decomposed, at leading order in $1/R$, in terms of another set of STF scalar radiative moments $\mathcal{U}_L^s$, as~\cite{Bernard:2022noq}
\begin{equation}\label{eq:scalar_waveform}
\psi = - \frac{2G}{c^2 R}\sum_{\ell=0}^{+\infty} \frac{1}{c^\ell \ell!} N_L \mathcal{U}_L^s(U) + \mathcal{O}\Big(\frac{1}{R^2}\Big)\,.  
\end{equation}
From the asymptotic gravitational and scalar waveforms~\eqref{eq:hijTT} and~\eqref{eq:scalar_waveform}, we can compute the associated gravitational and scalar energy fluxes, respectively denoted $\mathcal{F}^g$ and $\mathcal{F}^s$, and defined by~\cite{Bernard:2022noq}
\begin{subequations}\label{eq:fluxes}
\begin{align}
\mathcal{F}^g & =   \frac{c^3 R^2 \phi_0}{32 \pi G} \int \!\dd\Omega \left(\frac{\partial h_{ij}^\text{TT}}{\partial U}\right)^2 \nn \\
& = \sum_{\ell=2}^{+\infty} \frac{G \phi_0}{c^{2\ell+1}}\Bigg( \frac{(\ell+1)(\ell+2)}{(\ell-1) \ell \ell! (2\ell+1)!! }  \overset{(1)}{\mathcal{U}}_L\overset{(1)}{\mathcal{U}}_L + \frac{4 \ell (\ell+2)}{c^2 (\ell-1) (\ell+1)! (2\ell+1)!!} \overset{(1)}{\mathcal{V}}_L\overset{(1)}{\mathcal{V}}_L\Bigg) \,,\\
%%%%%%%%%%%%%%%%%%%%%%%%%%%%%%%%%%%%%%%%%%%%%%%%%%
\mathcal{F}^s &= \frac{c^3 R^2 \phi_0 (3+2\omega_0) }{16\pi G} \int\!\dd\Omega \left( \frac{\partial \Psi}{\partial U}\right)^2  = \sum_{\ell=0}^{+\infty} \frac{G \phi_0 (3+2\omega_0)}{c^{2\ell+1} \ell! (2\ell+1)!!}\overset{(1)}{\mathcal{U}^s_L} \; \overset{(1)}{\mathcal{U}^s_L} \label{eq:fluxes_scalar}\,.
\end{align}
\end{subequations}
In this project, since we aim at computing the tidal corrections to the total energy flux at the NNLO, the above expressions reduce to,
\bse\label{eq:fluxes_1PN}
	\begin{align}
	\mathcal{F}^g &=\frac{G \phi_0}{ c^5} \left\{ \frac{1}{5} \, \overset{(1)}{\mathcal{U}}_{ij} \, \overset{(1)}{\mathcal{U}}_{ij} + \frac{1}{c^2} \, \left[\frac{1}{189} \,  \overset{(1)}{\mathcal{U}}_{ijk} \, \overset{(1)}{\mathcal{U}}_{ijk} + \frac{16}{45} \, \overset{(1)}{\mathcal{V}}_{ij} \, \overset{(1)}{\mathcal{V}}_{ij}\right] + \mathcal{O}\left( \frac{1}{c^3} ; \frac{\epsilon_{\text{tidal}}}{c^3} \right)\right\} \label{eq:flux_grav} \,,\\
	\mathcal{F}^s &=\frac{G \phi_0 (3 +2 \omega_0)}{c^3} \left\{c^2 \,\overset{(1)}{\mathcal{U}^s} \; \overset{(1)}{\mathcal{U}^s} + \frac{1}{3} \, \overset{(1)}{\mathcal{U}^s_i} \; \overset{(1)}{\mathcal{U}^s_i}+ \frac{1}{30 c^2}\, \overset{(1)}{\mathcal{U}^s_{ij} } \; \overset{(1)}{\mathcal{U}^s_{ij} } + \frac{1}{630 c^4} \, \overset{(1)}{\mathcal{U}^s_{ijk}} \; \overset{(1)}{\mathcal{U}^s_{ijk}} + \mathcal{O}\left( \frac{1}{c^5} ; \frac{\epsilon_{\text{tidal}}}{c^5} \right) \right\} \label{eq:flux_scal} \,.
	\end{align}
\ese
In Table~\ref{tab:radiative_moments_for_flux}, we summarize the radiative moments needed to derive the NNLO flux, along with the orders at which they must be obtained to reach that accuracy. 
\begin{table*}[h]
    \centering
    \renewcommand{\arraystretch}{2}
    \begin{tabular}{|c||c|c|c|c||c|c|c|}
        \hline
       Moments & $\quad  {\mathcal{U}}^s \quad $ & $\quad{\mathcal{U}^s_i}\quad$ & $\quad{\mathcal{U}^s_{ij}}\quad$ & $\quad{\mathcal{U}^s_{ijk}}\quad$ & $\quad{\mathcal{U}_{ij}}\quad$ & $\quad{\mathcal{V}_{ij}}\quad$ & $\quad{\mathcal{U}_{ijk}}\quad$ \\
        \hline
       Required precision & NNLO & NNLO & NLO & LO & NLO & LO & LO \\
        \hline
    \end{tabular}
    \caption{Required radiative moments to derive the total energy flux to NNLO.}
    \label{tab:radiative_moments_for_flux}
\end{table*}

\noindent  Note that both $\mathcal{U}^s$ and $\mathcal{U}^s_i$ are required at the same order. This is because the scalar monopole at leading order (both in the point-particle and tidal contributions) is constant, so its time variation, which is required in the flux, is a next-to-leading order effect.  Next, we define the usual polarizations of the gravitational waveform as 
\begin{subequations}
\begin{align}
h_+& \equiv \frac{1}{2}(P_i P_j - Q_i Q_j) h_{ij}^{TT}\,,\\
h_\times & \equiv \frac{1}{2}(P_i Q_j + Q_i P_j) h_{ij}^{TT}\,,
\end{align}
\end{subequations}
using an orthonormal triad $(\bm{N},\bm{P},\bm{Q})$, where $\bm{N}$ is the direction of propagation of the GW, and $\bm{P}$ and $\bm{Q}$ are two unit polarization vectors. The complex gravitational field $h = h_+ - \di h_\times$ can itself be decomposed on the basis of spin-weighted spherical harmonics of weight $-2$,
\begin{subequations}
\be 
h = h_+ - \di h_- = \sum_{\ell=2}^{+\infty} \sum_{m=-\ell}^\ell h^{\ell \mathrm{m}} \,Y^{\ell \mathrm{m}}_{-2}(\Theta,\Phi)\,,
\ee
where the two angles $(\Theta,\Phi)$ characterize the direction $\bm{N}$. Similarly, the pure spin-0 scalar field can also be decomposed on the basis of standard (spin-0) spherical harmonics,
\be 
\psi =\sum_{\ell=0}^{+\infty} \sum_{m=-\ell}^\ell \psi^{\ell \mathrm{m}}\  Y^{\ell \mathrm{m}}(\Theta,\Phi) \label{eq:psi_spheri_harmonic}\,.
\ee
\end{subequations}
Using the orthogonality properties of $Y^{\ell \mathrm{m}}_{-2}$ and $Y^{\ell \mathrm{m}}$, we extract the waveform amplitudes modes characterized by the two numbers $(\ell,\mathrm{m})$ by the surface integrals 
\begin{subequations}
\begin{align}
\label{eq:hlm_psilm_integrals}
h^{\ell \mathrm{m}} &= \int \dd\Omega \, h \,  ({Y^{\ell \mathrm{m}}_{-2}})^{*} \,,\\
\psi^{\ell \mathrm{m}} &= \int \dd\Omega \, \psi \,  (Y^{\ell \mathrm{m}})^{*}\,.
\end{align}
\end{subequations}
where $\dd\Omega = \sin(\Theta)\dd\Theta\dd\Phi$ stands for the solid angle element and the asterisk $(\cdot)^{*}$ denotes the complex conjugation. After injecting the asymptotic waveform decompositions~\eqref{eq:hijTT}-\eqref{eq:scalar_waveform} in terms of the radiative moments and performing the surface integrals, we obtain an expression for the gravitational and scalar amplitude modes directly in terms of the spherical harmonic moments $(\mathcal{U}^{\ell \mathrm{m}},\mathcal{V}^{\ell \mathrm{m}},\mathcal{U}_s^{\ell \mathrm{m}})$~\cite{Faye:2012we,Bernard:2022noq}
\begin{subequations}
\begin{align}
\label{eq:hlm_psilm_UlmVlmUslm}
h^{\ell \mathrm{m}} &= - \frac{G}{\sqrt{2} R c^{\ell+2}}\Big[ \mathcal{U}^{\ell \mathrm{m}} - \frac{\di}{c} \mathcal{V}^{\ell \mathrm{m}}\Big]\,,\\
\psi^{\ell \mathrm{m}} &= \frac{G}{R c^{\ell+2}} \mathcal{U}_s^{\ell \mathrm{m}}\,.
\end{align}
\end{subequations}
In the case of planar orbital motion (where spins are either null or aligned/anti-aligned), symmetry considerations impose that $\mathcal{V}^{\ell \mathrm{m}}$ is zero when $\ell+\mathrm{m}$ is even and $\mathcal{U}^{\ell \mathrm{m}}$ is zero when $\ell+\mathrm{m}$ is odd. These spherical harmonic moments are related to the STF radiative moments via the relations~\cite{Faye:2012we,Bernard:2022noq}
\begin{subequations}
\begin{align}
\mathcal{U}^{\ell \mathrm{m}} &= \frac{4}{\ell!}\sqrt{\frac{(\ell+1)(\ell+2)}{2\ell(\ell-1)}}\alpha_L^{\ell \mathrm{m}} \mathcal{U}_L\,,\\
\mathcal{V}^{\ell \mathrm{m}} &= -\frac{8}{\ell!}\sqrt{\frac{\ell(\ell+2)}{2(\ell+1)(\ell-1)}}\alpha_L^{\ell \mathrm{m}} \mathcal{V}_L\,,\\
\mathcal{U}_s^{\ell \mathrm{m}} &= -\frac{2}{\ell!} \alpha_L^{\ell \mathrm{m}} \mathcal{U}^s_L\,,
\end{align}
\end{subequations}
where we have introduced $\alpha_L^{\ell \mathrm{m}}$ as~\cite{Henry:2021cek}
\begin{equation}\label{eq:alphalmL_def}
\alpha_L^{\ell \mathrm{m}} = \frac{ \sqrt{4 \pi}\,(-\sqrt{2})^\mathrm{m}\,\ell !}{\sqrt{(2\ell+1) (\ell+\mathrm{m})!\,(\ell-\mathrm{m})!}}\,{\mathfrak{m}_0^{*}}^{\langle M}\ell_\text{\textcolor{white}{0}}^{L-M\rangle}\,,
\end{equation}
with $(\bm{n}_0,\bm{\lambda}_0,\bm{\ell})$ being a fixed orthonormal basis and $\mathfrak{m}_0$ defined as $\mathfrak{m}_0=(\bm{n}_0+i\bm{\lambda}_0)/\sqrt{2}$. Here, the unit vector normal to the orbital plane $\bm{\ell}=\bm{\ell}_0$ is constant due to the planar nature of the motion. 

In GR, gravitational modes $h_{\ell \rm{m}}$ can be expressed as the product of a complex amplitude for the $(\ell, \rm{m})$ mode and a complex exponential in the orbital phase, $\e^{-\di \mathrm{m} \phi}$. However, when computing the amplitude of the dominant $(2,2)$ mode, a dependence on the time scale $b_0$ associated with the transformation to radiative coordinates remains, starting at 1.5PN order. This dependency arises from the tail effects. To resolve this, a phase shift was introduced in Refs.~\cite{Blanchet:1993ec,Wiseman:1993aj} of the form
\begin{align}
\psi = \phi -\frac{2G\mathcal{M}\omega}{c^3}\rm{ln}\left(\frac{\omega}{\omega_0}\right)\,
\end{align}
where $\mathcal{M} = m +E/c^2$ is the ADM mass, and $\omega_0 = \e^{11/12-\gamma_E}/(4b_0)$. This phase shift effectively removes the $b_0$-dependence not only in the amplitude of the (2,2) mode, but also in the amplitude of all $(\ell, \rm{m})$ modes, as later observed in the literature. It was argued in Refs.~\cite{Blanchet:1996pi,Arun:2004ff} that this phase shift affects the flux and waveform only from the 4PN order. 

In ST theories, the GW phase $\psi$ (or half phase of the dominant (2,2) mode) similarly differs from the orbital phase $\phi$ due to the propagation of tails in the wave zone. The phase redefinition in ST theories resembles the one found in GR, and takes the form:
\begin{align}\label{eq:phase_redefinition}
\psi \equiv \phi - \frac{2 \tilde{G} \mathcal{M}(1-\zeta)\omega}{c^3} \ln\left(\frac{\omega}{\omega_0}\right) \,,
\end{align} 
where the ADM mass is given by $\mathcal{M} = m +(E_{\text{pp}}+E_{\text{tidal}})/c^2$. We can now factorize out in all modes $h^{\ell \mathrm{m}}$ and $\psi^{\ell \mathrm{m}}$ the appropriate phase factor $\e^{\di \psi}$, so that the dominant gravitational (2,2) mode and the dominant scalar (1,1) mode starts with a one at the Newtonian order, i.e. $\hat{H}^{22} = 1 + \mathcal{O}(x)$ and $\hat{\Psi}^{11} = 1 + \mathcal{O}({x})$. These modes are given by~\cite{Bernard:2022noq} \footnote{Recall that $h^{\ell, -{\rm m}} = (-)^\ell {(h^{\ell \mathrm{m}})}^{*}$ and $\psi^{\ell, -{\rm m}} = (-)^\ell (\psi^{\ell \mathrm{m}})^{*}$. } 
\begin{subequations}
\begin{align}
h^{\ell \mathrm{m}} &= \frac{2 \tilde{G}(1-\zeta) m \nu x}{R c^2}  \sqrt{\frac{16\pi}{5}} \,\hat{H}^{\ell \mathrm{m}} \e^{-\di \mathrm{m} \psi } \,,\\
\psi^{\ell \mathrm{m}} &= \frac{2 \di \tilde{G} \zeta \sqrt{\alpha}\mathcal{S}_{-} m \nu \sqrt{x}}{R c^2} \sqrt{\frac{8\pi}{3}} \,\hat{\Psi}^{\ell \mathrm{m}} \e^{-\di \mathrm{m} \psi}\,.
\end{align}
\end{subequations}
By performing a PN expansion of the complex exponential, we will show that the phase redefinition successfully removes the logarithmic terms arising from scalar tail effects at the $\text{N}^{1.5}\text{LO}$ (corresponding to mass-scalar-dipole interaction, see Sec.~\ref{sec:radiative_moments}) from the $\psi^{\ell \mathrm{m}}$. If one were to compute $h^{\ell \mathrm{m}}$ at the $\text{N}^{1.5}\text{LO}$ (while we will limit here to the NLO), the phase redefinition~\eqref{eq:phase_redefinition} would similarly remove the logarithmic terms arising from the tail effects at the $\text{N}^{1.5}\text{LO}$ (corresponding to the mass-quadrupole interaction).

In order to derive the full waveform amplitude to the $\text{N}^{1.5}\text{LO}$, one has to know all modes $h^{\ell \mathrm{m}}$ with $2 \leq\ell \leq 4 $ and $\lvert \mathrm{m} \rvert \leq \ell$ up to the NLO, and all modes $\psi^{\ell \mathrm{m}}$ with $0 \leq\ell \leq 4 $ and $\lvert \mathrm{m} \rvert \leq \ell$ up to the $\text{N}^{1.5}\text{LO}$. In Table~\ref{tab:radiative_moments_for_waveform}, we summarize again the radiative moments needed to derive the full waveform amplitude to the $\text{N}^{1.5}\text{LO}$, along with the orders at which they must be obtained to reach that accuracy. 
\begin{table*}[ht]
    \centering
    \renewcommand{\arraystretch}{2}
    \begin{tabular}{|c||c|c|c|c|c||c|c|c|}
        \hline
       Moments & $\quad  {\mathcal{U}}^s \quad $ & $\quad{\mathcal{U}^s_i}\quad$ & $\quad{\mathcal{U}^s_{ij}}\quad$ & $\quad{\mathcal{U}^s_{ijk}}\quad$ & $\quad{\mathcal{U}^s_{ijkl}}\quad$ &$\quad{\mathcal{U}_{ij}}\quad$ & $\quad{\mathcal{V}_{ij}}/{\mathcal{U}_{ijk}} \quad$ & $\quad{\mathcal{V}_{ijk}}/{\mathcal{U}_{ijkl}}\quad$ \\
        \hline
       Required precision & NNLO & $\text{N}^{1.5}\text{LO}$ & NLO & $\text{N}^{0.5}\text{LO}$ & LO & NLO & $\text{N}^{0.5}\text{LO}$ & LO \\
        \hline
    \end{tabular}
    \caption{Required radiative moments to derive the full waveform amplitude to $\text{N}^{1.5}\text{LO}$.}
    \label{tab:radiative_moments_for_waveform}
\end{table*}
In particular, we need to calculate three additional radiative moments compared to those required for the NNLO flux, namely ${\mathcal{U}^s_{ijkl}}$, $\mathcal{V}_{ijk}$ and ${\mathcal{U}_{ijkl}}$ to LO. 

%------------------------------------------------------------------------
\subsection{Radiative multipole moments}\label{sec:radiative_moments}
%------------------------------------------------------------------------

Based on the PN-MPM formalism adapted to ST theories, the mass-type and current-type radiative moments can be expressed in terms of two sets of multipole moments, the sources moments $\{I_L, J_L\}$ and gauge ones $\{W_L, X_L, Y_L,Z_L\}$. These two sets are physically equivalent to the set of canonical moments $\{M_L, S_L\}$. However, as we anticipated in Sec.~\ref{sec:asymptotic_waveform}, the moment required at the highest order to compute the gravitational flux is the mass quadrupole $I_{ij}$, needed to the NLO. At this precision, the gauge moments do not contribute, so the radiative moments depend only on the source ones. \newline

Additionally, it is known in GR that the radiative moments acquire some non-linear and non-local-in-time contributions, notably the tail and memory terms.  Memory effects contribute only to the mass-type multipole moments $U_L$, with $\ell$ even. These contributions are semi-hereditary \footnote{Following the terminology of Ref.~\cite{ Blanchet:1992br}, we may refer to these terms as “semi-hereditary”, as they are given by some time anti-derivative of local terms; thus applying multiple time derivatives to a semi-hereditary term reduces it to be local.}. Since the energy flux involves time derivatives of the radiative moments, there is no hereditary effect in the energy flux associated to the memory. Note however that this is not the case for the angular momentum flux, see e.g. Ref.~\cite{Trestini:2024mfs}. Moreover, although memory terms become instantaneous once differentiated, their contribution to the energy flux arises only at very high PN orders, well beyond the accuracy considered in the present work.
By contrast, memory effects must be treated with care when computing the waveform. Even though memory contributions to the radiative moments formally appear at higher PN orders than those listed in Tables~\ref{tab:radiative_moments_for_flux} and~\ref{tab:radiative_moments_for_waveform}, they can lead to lower-order contributions in some waveform modes due to PN order promotions. This point will be addressed in detail in the next section. 
Tails effects are hereditary effects, hence are non-local-in-time, as well as their iterated time-derivatives. They do contribute to the mass-type and current-type radiative moments, but beyond the orders considered in Table~\ref{tab:radiative_moments_for_flux}, so can be neglected when computing the gravitational flux to the NLO.  As a consequence, the radiative moments relate to the source moments only by the simple relations
\begin{align}
U_L = I_L^{(\ell)} \quad \text{and} \quad V_L = J_L^{(\ell)}\,,
\end{align}
where the upper index $(\ell)$ refer to the $\ell^{\text{th}}$ time derivative. Similarly, it was shown in~\cite{Lang:2014osa} that the scalar radiative moments also acquire some tail contributions. It is notably the case for the scalar dipole moment $U_i^s$ that differs from its once-differentiated source moment by the relation~\cite{Bernard:2022noq}
\begin{align}\label{eq:dipole_moment_tail}
\mathcal{U}_i^s &= \overset{(1)}{I_i^s} + \frac{2G \ADM}{\phi_0 c^3} \int_{-\infty}^U \dd V \overset{(3)}{I_i^s}(V) \left[ \ln{\left(\frac{U-V}{2b_0} \right)} + 1 \right] \,.
\end{align}
 where $\ADM$ is the ADM mass, and $b_0$ is the time scale associated with the transformation to radiative coordinates.  The rest of the scalar moments needed to derive the NNLO scalar flux and $\text{N}^{1.5}\text{LO}$ scalar waveform simply satisfy
\begin{align}
U_L^s = {(I_L^s)}^{(\ell)}\,.
\end{align}
The general expression of the source moments in terms of the parameters of the source, specifically in terms of the PN expansion of the scalar and tensor pseudo-tensors~\eqref{tau}, along with the methodology to compute these moments, is provided in Sec.~\ref{sec:source_moments}.

%------------------------------------------------------------------------
\subsection{Memory contributions to $\mathrm{m}=0$ gravitational modes}
\label{sec:memory}
%------------------------------------------------------------------------

In the previous section, we have discussed how non-linear memory effects do not contribute to the gravitational part of the energy flux. However, they do impact the gravitational waveform. To illustrate this, we consider the quadrupole moment, whose memory contributions are given by~\cite{Bernard:2022noq} 
\begin{equation}\label{eq:U_ij_mem}
\mathcal{U}_{ij}^{(\text{mem})} = \frac{G(3+2 \omega_0)}{3c^3}\Bigg( \int_{-\infty}^U \!\dd V \overset{(2)}{I^s_{\langle i}}(V)\overset{(2)}{I^s_{j\rangle}}(V) -I^s_{\langle i}\overset{(3)}{I^s_{j\rangle}}- \overset{(1)}{I^s_{\langle i}}\overset{(2)}{I^s_{j\rangle}} -\frac{1}{2}I^s \overset{(3)}{I^s_{ij}} \Bigg) + \mathcal{O}\left(\frac{1}{c^5};\frac{\etidal}{c^5}\right) \,.
\end{equation}
These contributions formally appear at $1.5$PN order. When computing them, we will typically encounter two types of integrals: oscillatory integrals of the form
\begin{equation}\label{eq:AC}
\int_{-\infty}^{U} \dd \tau \, x^p(\tau) e^{-\di n \phi(\tau)} = \di \frac{\alpha \tilde{G} m}{n c^3}e^{-\di n \phi}x^{p-3/2}\,,
\end{equation}
and non-oscillatory integrals as
\begin{equation}\label{eq:DC}
\int_{-\infty}^{U} \dd \tau \, x^p(\tau) = \int_0^{x(U)} \dd y \, \frac{y^p}{\dot{x}(y)} \,.
\end{equation}
After integration, the oscillatory contributions coming from~\eqref{eq:AC} yield 1.5PN correction to the waveform and can therefore be neglected in this work. By contrast, for the non-oscillatory integrals, we need the time evolution of the gauge-invariant parameter $x$, which is governed at leading order in the point-particle approximation by
\begin{equation}\label{eq:xdot_0PN}
\dot{x} =\frac{8 c^3 \zeta \mathcal{S}_{-}{}^2 \nu x^4}{3 \alpha \tilde{G} m} \sim \mathcal{O}\left(\frac{1}{c^{5}}\right) \,,
\end{equation}
where $x\sim \mathcal{O}(c^{-2})$.  As a result, after integration, the non-oscillatory terms, although formally entering at 1.5PN order, effectively contribute at Newtonian order in the waveform.  Similarly, the $\text{N}^{1.5}\text{LO}$ tidal corrections to these non-oscillatory terms get promoted to the LO in the waveform. Computing the memory modes up to NLO will therefore require the knowledge of $\dot{x}$ also at NLO.  

We recall that memory effects only affect the mass-type radiative moments $U_L$, and consequently the gravitational modes, with $\ell$ even. Furthermore, because the motion is quasi-circular, non-oscillatory memory effects only contribute to the $m=0$ gravitational modes. Instead of computing these effects directly from the mass-type radiative moments, which would require us to compute them up to the $n+1.5$PN order to remain consistent (with $n$ denoting the order typically required, as summarized in Table~\ref{tab:radiative_moments_for_waveform}), we proceed similarly as in Ref.~\cite{Favata:2008yd} and compute the ${\rm m}=0$ modes via the angular integral: 
\begin{equation}\label{eq:hlmmem_generic}
h_{\ell \mathrm{m}}^\text{mem} = -\frac{16\pi G}{R \phi_0 c^4}\sqrt{\frac{(\ell-2)!}{(\ell+2)!}}\int^{U}_{-\infty} \dd t \int \dd \Omega \,\frac{\dd \mathcal{F}}{\dd\Omega} (\Omega) Y_{\ell \mathrm{m}}^{*}(\Omega) = \int_{-\infty}^{U} \dd t \,\dot{h}_{\ell \mathrm{m}}^\text{mem}\,.
\end{equation}
In ST theories, $\dot{h}_{\ell \mathrm{m}}^\text{mem}$ is made of two components, associated respectively to the gravitational and scalar fluxes, as :
\begin{equation}
\dot{h}_{\ell \mathrm{m}}^\text{mem}= \dot{h}_{\ell \mathrm{m}}^\text{mem, g} + \dot{h}_{\ell \mathrm{m}}^\text{mem, s} \,.
\end{equation}
The first component on the right hand side is the same as in GR and reads~\cite{Favata:2008yd}
\begin{equation}\label{eq:hlmmemdotgen_grav}
\dot{h}_{\ell \mathrm{m}}^\text{mem, g} = -\frac{R}{c}\sqrt{\frac{(\ell-2)!}{(\ell+2)!}}\sum_{\ell'=2}^\infty\sum_{\ell''=2}^\infty \sum_{\mathrm{m}'=-\ell'}^{\ell'}\sum_{\mathrm{m}''=-\ell''}^{\ell''} (-1)^{\mathrm{m}+\mathrm{m}''} G^{2, -2,0}_{\ell' ,\ell'', \ell, \mathrm{m}',-\mathrm{m}'',-\mathrm{m}}\,\dot{h}_{\ell' \mathrm{m}'}\dot{h}^{*}_{\ell'' \mathrm{m}''}\,,
\end{equation}
where the value of the angular integral $G^{s, s', s''}_{\ell, \ell', \ell'', \mathrm{m}, \mathrm{m}', \mathrm{m}''}$ is given in \textit{e.g.} Appendix A of~\cite{Favata:2008yd}. The second component is obtained by combining the definitions given in Eqs.~\eqref{eq:fluxes_scalar},~\eqref{eq:psi_spheri_harmonic}, and~\eqref{eq:hlmmem_generic}, and one finds
\begin{equation}\label{eq:hlmmemdotgen_scal}
\dot{h}_{\ell \mathrm{m}}^\text{mem, s} =-\frac{R (3+2 \omega_0)}{c}\sqrt{\frac{(\ell-2)!}{(\ell+2)!}}\sum_{\ell'=0}^\infty\sum_{\ell''=0}^\infty \sum_{\mathrm{m}'=-\ell'}^{\ell'}\sum_{\mathrm{m}''=-\ell''}^{\ell''} (-1)^{\mathrm{m}+\mathrm{m}''} G^{0, \,0,\, 0}_{\ell', \ell'', \ell, \mathrm{m}',-\mathrm{m}'',-\mathrm{m}}\,\dot{\psi}_{\ell' \mathrm{m}'}\dot{\psi}^{*}_{\ell'' \mathrm{m}''}\,.
\end{equation}
To our knowledge, this expression provides the first explicit formula for the scalar contribution to the gravitational-wave memory in scalar–tensor theories. After an explicit computation, the resulting gravitational memory modes are displayed in Eqs.~\eqref{Hlm_m0}. 

%------------------------------------------------------------------------
%------------------------------------------------------------------------
\section{Intermediate quantities}
\label{sec:intermediate_quantities}

%------------------------------------------------------------------------
%------------------------------------------------------------------------

To derive the NNLO flux and orbital phase, as well as the $\text{N}^{1.5}\text{LO}$ waveform, it is necessary to first compute a set of intermediate quantities. This section is dedicated to the derivation of these key building blocks. The most relevant results will be made available upon request to interested readers.

%------------------------------------------------------------------------
\subsection{Dissipative $\text{N}^{1.5}\text{LO}$ tidal terms in the equations of motion}
%------------------------------------------------------------------------

In a previous work~\cite{Bernard:2023eul}, we have derived the NNLO tidal corrections to the conservative equations of motion, focusing exclusively on the even PN contributions. However, to treat the radiative sector, we must also account for the dissipative $\text{N}^{1.5}\text{LO}$ tidal terms in the equations of motion. These contributions are important for two purposes: to compute the scalar dipolar moment, which contributes to the scalar energy flux and the mode $\hat{\Psi}^{11}$; and to determine the $\text{N}^{1.5}\text{LO}$ correction to the center-of-mass position, which is required to express the source moments in the center-of-mass frame. 
To derive the $\text{N}^{1.5}\text{LO}$ acceleration for body A in harmonic coordinates, we derive the geodesic equations by varying the matter action~\eqref{eq:Smatter+tidal} w.r.t the position of body A $y_A^{\mu}$. Since we only need the $\text{N}^{1.5}\text{LO}$ terms, it suffices to consider only the scalar dipolar part in the tidal action~\eqref{eq:Stidal}. Then, we perform a (3+1) splitting, as well as a PN expansion of that geodesic equation, which can be recasted into the following form~\cite{Blanchet:2013haa}
\be\label{eq:dPdtF}
\frac{\dd P^i}{\dd t} = F^i\,,
\ee
where $P^i=P^i_\text{pp}+P^i_\text{tidal}$ and $F^i=F^i_\text{pp}+F^i_\text{tidal}$ now read in ST theories:
\bse\label{eq:EoM_P_F}
\begin{align}
P^i_\text{pp} = & \, m \Biggl[v{}^{i} + \frac{1}{c^2}\biggl(\bigl(\psi_{(0)} (1 - 2 s) + \frac{1}{2} v^2 + 3 V\bigr) v{}^{i} - 4 V^{i}\biggr) \Biggr] + \calO\left(\frac{1}{c^4}\right)\,,\\
P^i_\text{tidal} = & \, \frac{\lambda^{(0)}}{c^4} \Biggl[2 (\partial_{a}\psi_{(0)} \partial^{a}\psi_{(0)}) v{}^{i} - 4 \biggl((v{}^{a} \partial_{a}\psi_{(0)}) + 
\partial_t \psi_{(0)}\biggr) \partial^{i}\psi_{(0)}\Biggr]+ \calO\left(\frac{\etidal}{c^4}\right)\,,\\
F^i_\text{pp} =& \, m \Biggl\{- (1 - 2 s) \partial^{i}\psi_{(0)} + \partial^{i}V + \frac1{c^2}\biggl[\biggl(\frac{(1 -  \zeta) (-4 \bar{\delta} -  \frac{32 \bar{\beta}\, \bar{\delta}}{\bar{\gamma}^2}) }{\zeta (2 + \bar{\gamma})^2}\psi_{(0)} + (1 - 2 s) \bigl(\frac{1}{2} v^2 + V\bigr)\biggr) \partial^{i}\psi_{(0)} \nn \\
& \quad + \bigl(\psi_{(0)} (1 - 2 s) + \frac{3}{2} v^2 -  V\bigr) \partial^{i}V - 4 v{}^{a} \partial^{i}V_{a}\biggl]\Biggr\}+ \calO\left(\frac{1}{c^4}\right)\,,\\
F^i_\text{tidal} = & \, - \frac{4 \lambda^{(0)}}{c^2} \partial^{a}\psi_{(0)} 
\partial^{i}\partial_{a}\psi_{(0)} + \frac{1}{c^4}\Biggl[4 \phi_{0}{} 
\lambda^{(1)} \biggl((\partial_{a}\psi_{(0)} \partial^{a}\psi_{(0)}) 
\partial^{i}\psi_{(0)} + 2 \psi_{(0)} \partial^{a}\psi_{(0)} 
\partial^{i}\partial_{a}\psi_{(0)}\biggr) \nn \\
& \quad - 2 \lambda^{(0)} \biggl(2 
(v{}^{a} \partial_{a}\psi_{(0)}) \partial_t \partial^{i}\psi_{(0)} - 
5 (\partial_{a}\psi_{(0)} \partial^{a}\psi_{(0)}) \partial^{i}\psi_{(0)} + 
4 \lambda_1 (\partial_{a}\psi_{(0)} \partial^{a}\psi_{(0)}) \partial^{i}\psi_{(0)} \nn \\
& \quad + \frac{4 (1 -  \zeta) \lambda_1 (\partial_{a}\psi_{(0)} 
\partial^{a}\psi_{(0)}) \partial^{i}\psi_{(0)}}{\zeta} - 3 
(\partial_{a}\psi_{(0)} \partial^{a}\psi_{(0)}) \partial^{i}V  + 2 
(v{}^{a} \partial_{a}\psi_{(0)}) v{}^{a} 
\partial^{i}\partial_{a}\psi_{(0)} \nn \\
& \quad + 2 v{}^{a} \partial_t \psi_{(0)} \
\partial^{i}\partial_{a}\psi_{(0)} - 10 \psi_{(0)} \partial^{a}\psi_{(0)} 
\partial^{i}\partial_{a}\psi_{(0)}  + 8 \lambda_1 \psi_{(0)} \partial^{a}\psi_{(0)} \partial^{i}\partial_{a}\psi_{(0)} \nn \\
& \quad+ \frac{8 (1 -  \zeta) \lambda_1 
\psi_{(0)} \partial^{a}\psi_{(0)} \partial^{i}\partial_{a}\psi_{(0)}}{\zeta} -  v^2 \partial^{a}\psi_{(0)} 
\partial^{i}\partial_{a}\psi_{(0)}  - 6 V \partial^{a}\psi_{(0)} 
\partial^{i}\partial_{a}\psi_{(0)}\biggr)\Biggr] \nn \\
& \quad+ \calO\left(\frac{\etidal}{c^4}\right)\,.
\end{align}
\ese
To derive the equations of motion for body $A$, one should  replace the mass $m$ by $m_A$, the velocity $v^i$ by $v^i_A$, the sensitivity $s$ by $s_A$, the parameter $\bar{\delta}$ by $\bar{\delta}_A$, the tidal deformability coefficient $\lambda^{(0)}$ by $\lambda_A^{(0)}$, and so on. Additionally, the potentials and their derivatives should be evaluated at the field point of body A using the Hadamard regularization~\cite{Blanchet:2000nu} (\textit{e.g.} $\partial^iV$ becomes $(\partial^iV)_A$).  From Eqs.~\eqref{eq:EoM_P_F}, we see that the potentials $V$ and $\psi_{(0)}$ are needed to $\text{N}^{1.5}\text{LO}$. These potentials are computed by inverting the flat spacetime wave equations~\eqref{eq:PotentialWaveEq} using the usual retarded d’Alembertian integral, written as an infinite PN-expansion of Poisson-like integrals, namely

\be\label{eq:invDalemvertian}
P(\mathbf{x},t) = \Box^{-1}_\text{ret} S(\mathbf{x},t) = -\frac{1}{4\pi} \sum_{k=0}^\infty \frac{(-)^k}{k!c^k} \frac{\dd^k}{\dd t^k}\underset{B=0}{\text{PF}}\int \dd^3\mathbf{x}' \left(\frac{r'}{r_0}\right)^{B}\vert \mathbf{x}-\mathbf{x}'\vert^{k-1} S(\mathbf{x}',t)\,.
\ee
Here, PF denotes the finite part regularization when $B\rightarrow 0$, where $B$ is a complex number. This regularization, based on analytical continuation, depends on the arbitrary constant $r_0$, and is introduced to cure IR divergences at spatial infinity~\cite{Blanchet:2005ft}~\footnote{Note that due to the low PN order at which the potentials are required for tidal effects, we have checked that Hadamard Regularisation is sufficient and there is no need to resort to Dimentional Regularisation~\cite{Blanchet:2003gy} in all our calculations.}.  The series~\eqref{eq:invDalemvertian} scales as $1/c$ rather than $1/c^2$, leading to the appearance of odd-PN contributions in the potentials, and consequently in the acceleration. After injecting the potentials and their derivatives in $P^i$ and $F^i$, and time-differentiating $P^i$, we order-reduce all the resulting accelerations which only enter to LO. Finally, we obtain the $\text{N}^{1.5}\text{LO}$ acceleration for body 1 in harmonic coordinates, whose NLO part is in agreement with our previous result~\cite{Bernard:2023eul}.  Here, we choose to display only the $\text{N}^{1.5}\text{LO}$ contribution
\be\label{eq:EoM_dissipative_pp_tidal}
\bm{a}^{\text{N}^{1.5}\text{LO}} = \bm{a}_{\text{pp}}^{\text{1.5PN}}+\bm{a}_{\text{tidal}}^{\text{N}^{1.5}\text{LO}}\,.
\ee
with
\begin{subequations}\label{eq:EoM_dissipative}
\begin{align}
\bm{a}_{\text{pp}}^{\text{1.5PN}} &= \frac{4 \alpha^2  \tilde{G}^2 m{}^2  \nu }{3 c^3 r^3} \mathcal{S}_{-}{}^2\zeta\bigl(3 \bm{n} (n v) -  \bm{v}\bigr) \\
\bm{a}_{\text{tidal}}^{\text{N}^{1.5}\text{LO}}& =  \, \frac{8 \alpha^2  \tilde{G}^2 m }{3 c^5 r^5}\frac{\zeta^2\mathcal{S}_{-}{}}{\bar{\gamma}(1 -  \zeta)} \Biggl\{\bm{v} \biggl[(n v)^2 \bigl((-45 \mathcal{S}_{+}{} - 45 \mathcal{S}_{-}{} \delta) \lambda_{-}^{(0)} + (-45 \mathcal{S}_{-}{} - 45 \mathcal{S}_{+}{} \delta) \lambda_{+}^{(0)}\bigr) \nn \\
& \hspace{-0.5cm} + v^2 \bigl((9 \mathcal{S}_{+}{} + 9 \mathcal{S}_{-}{} \delta) \lambda_{-}^{(0)} + (9 \mathcal{S}_{-}{} + 9 \mathcal{S}_{+}{} \delta) \lambda_{+}^{(0)}\bigr)\biggr] + \bm{n} \Bigl[(n v)v^2 \bigl((-45 \mathcal{S}_{+}{} - 45 \mathcal{S}_{-}{} \delta) \lambda_{-}^{(0)} + (-45 \mathcal{S}_{-}{} - 45 \mathcal{S}_{+}{} \delta) \lambda_{+}^{(0)}\bigr) \nn \\
& \hspace{-0.5cm} + (n v)^3 \bigl((105 \mathcal{S}_{+}{} + 105 \mathcal{S}_{-}{} \delta) \lambda_{-}^{(0)} + (105 \mathcal{S}_{-}{} + 105 \mathcal{S}_{+}{} \delta) \lambda_{+}^{(0)}\bigr)\Bigr]  + \frac{\alpha \tilde{G} m}{r} \biggl[\bm{n} (n v) \biggl((33 \mathcal{S}_{+}{} + 33 \mathcal{S}_{-}{} \delta) \lambda_{-}^{(0)} \nn \\
& \hspace{-0.5cm} + (33 \mathcal{S}_{-}{} + 33 \mathcal{S}_{+}{} \delta - 24 \bar{\gamma} \mathcal{S}_{-}{} \nu) \lambda_{+}^{(0)}\biggr) + \bm{v} \biggl((-7 \mathcal{S}_{+}{} - 7 \mathcal{S}_{-}{} \delta) \lambda_{-}^{(0)} + (-7 \mathcal{S}_{-}{} - 7 \mathcal{S}_{+}{} \delta + 4 \bar{\gamma} \mathcal{S}_{-}{} \nu) \lambda_{+}^{(0)}\biggr)\biggr]\Biggr\}  \,.
\end{align}
\end{subequations}
For the point-particle part $\bm{a}_{\text{pp}}^{\text{1.5PN}}$, we retrieve the result given in (A1.a) of~\cite{Bernard:2022noq}.

%------------------------------------------------------------------------
\subsection{Dissipative $\text{N}^{1.5}\text{LO}$ tidal terms in center-of-mass position}
%------------------------------------------------------------------------

While the center-of-mass position $G^i$ has already been computed in the conservative sector up to the NNLO and is displayed in Eq.~(45) of Ref.~\cite{Bernard:2023eul}, we now focus on deriving its dissipative contributions. In principle, $G^i$ can be computed by successively integrating the flux-balance equations for the linear momentum $P^i$ and for $G^i$ itself. These flux-balance equations take the form~\cite{Trestini:2024mfs}
\begin{align}\label{eq:flux_balance}
	\frac{\dd P^i}{\dd t} &= - \calF^i_{\bm{P}}-\calF^i_{s,\bm{P}} \,,\\
	\frac{\dd G^i}{\dd t} &= P^i - \calF^i_{\bm{G}} -\calF^i_{s,\bm{G}}\,,
\end{align}
where $\calF^i_{\bm{P}}$ and $\calF^i_{\bm{G}}$ denote the tensorial fluxes of linear momentum and of center-of-mass position, respectively, while $\calF^i_{s,\bm{P}}$ and $\calF^i_{s,\bm{G}}$ are their scalar counterparts. The explicit expressions for these tensorial fluxes, written in terms of the radiative moments, are the same as in GR and are given in Eqs.~(4.12b)–(4.15b) of Ref.~\cite{Compere:2019gft}. The corresponding scalar fluxes are defined in Eqs.~(5.5) of Ref.~\cite{Trestini:2024mfs}. Using these formulas, we find that $G^i$ acquires a dissipative contribution (or Schott term) at the $\text{N}^{1.5}\text{LO}$ relevant to this work, which we split as
\begin{equation}\label{eq:Gi_dissipative_1}
\left(G^i\right)^{\text{N}^{1.5}\text{LO}}=\left(G^i\right)_{\text{pp}}^{\text{1.5PN}}+\left(G^i\right)_{\text{tidal}}^{\text{N}^{1.5}\text{LO}}\,,
\end{equation}
with
\begin{subequations}\label{eq:Gi_dissipative_2}
\begin{align}
\left(G^i\right)_{\text{pp}}^{\text{1.5PN}} = & - \frac{\alpha \tilde{G} \zeta \mathcal{S}_{-}{}}{3 c^3} \Bigl[m_{2}{} (- \mathcal{S}_{-}{} + \mathcal{S}_{+}{}) + m_{1}{} (\mathcal{S}_{-}{} + \mathcal{S}_{+}{})\Bigr] (m_{1}{} v_{1}{}^{i} -  m_{2}{} v_{2}{}^{i})\,,   \\
%%%%%%%%%%%%%%%%%%%%%%%%%%%%%%%%%%%%%%%%%%%%%%%%%%%%%%
\left(G^i\right)_{\text{tidal}}^{\text{N}^{1.5}\text{LO}} = & \frac{2 \alpha^2 \tilde{G}^2 \zeta^2 (2 + \bar{\gamma}) }{3 c^5 (-1 + \zeta) r_{12}{}^3} \Bigl[m_{2}{} (- \mathcal{S}_{-}{} + \mathcal{S}_{+}{}) + m_{1}{} (\mathcal{S}_{-}{} + \mathcal{S}_{+}{})\Bigr] \Bigl[m_{2}{} (\mathcal{S}_{-}{} -  \mathcal{S}_{+}{}) \lambda_1^{(0)} \nn \\
&+ m_{1}{} (\mathcal{S}_{-}{} + \mathcal{S}_{+}{}) \lambda_2^{(0)}\Bigr] \Bigl[3 n_{12}{}^{i} \bigl((n_{12} v_{1}) -  (n_{12} v_{2})\bigr) -  v_{1}{}^{i} + v_{2}{}^{i}\Bigr] \,.
\end{align}
\end{subequations}

We now turn to the dissipative contributions to the transformation to the center-of-mass frame. In particular, we want to determine the dissipative corrections to the harmonic coordinates $\bm{y_A}$ and $\bm{v_A}$, expressed in terms of the CoM coordinates $\bm{x}=r\bm{n}$ and $\bm{v}$. In the conservative sector, this transformation is obtained by imposing the condition $G^i=0$, and solving it for $y_1^i$, from which the remaining coordinates follow straightforwardly. However, it was shown in GR~\cite{Blanchet:2018yqa,Blanchet:2024loi,Blanchet:2026suq} that in the radiative sector, the CoM condition is in fact $G^i+\Gamma^i=0$, where $\Gamma^i$ is a nonlocal-in-time term due to the recoil. In scalar-tensor theories,  an additional semi-hereditary scalar term arises, and the condition becomes~\cite{Trestini:2024mfs} 
\begin{align}
G^i+\Gamma^i+\Gamma^i_s=0  \,.
\end{align}
The quantities $\Gamma^i$ and $\Gamma^i_s$ are defined in Eqs.~(5.10) of ~\cite{Trestini:2024mfs}. Since $\Gamma^i$ contributes only starting at $3.5$PN order and $\Gamma^i_s$ at $2.5$PN order, both terms can be safely neglected at the NNLO. 
We now provide the ${\text{N}^{1.5}\text{LO}}$ contributions to the positions and the velocities in the CoM frame, which we decompose as
\begin{subequations}\label{eq:yAi_vAi_CM}
\begin{align}
\left(y_A^i\right)^{\text{N}^{1.5}\text{LO}}&=\left(y_A^i\right)_{\text{pp}}^{1.5\text{PN}}+\left(y_A^i\right)_{\text{tidal}}^{\text{N}^{1.5}\text{LO}}\,,\\
\left(v_A^i\right)^{\text{N}^{1.5}\text{LO}}&=\left(v_A^i\right)_{\text{pp}}^{1.5\text{PN}}+\left(v_A^i\right)_{\text{tidal}}^{\text{N}^{1.5}\text{LO}}\,.
\end{align}
\end{subequations}
with $A=1,2$. The point-particle approximation pieces can be found in \textit{e.g} Eqs.~(A2) of Ref.~\cite{Bernard:2022noq} and the tidal pieces read 
\begin{subequations}\label{eq:y1i_v1i_CM_dissipative}
\begin{align}
\left(y_1^i\right)_{\text{tidal}}^{\text{N}^{1.5}\text{LO}} &=  \frac{\alpha^2 \tilde{G}^2 m}{3 c^5  r^3}\frac{\zeta}{(1 -  \zeta) \bar{\gamma} (2 + \bar{\gamma})} \bigl(3 n^{i} (n v) -  v^{i}\bigr) \Bigl[\Bigl(8 \bar{\delta}^+ + 8 \bar{\delta}^- \delta - 4 \bigl(4 \bar{\delta}^+ + \bar{\gamma} (2 + \bar{\gamma})\bigr) \nu \Bigr) \lambda_{-}^{(0)} \nn \\
& + (8 \bar{\delta}^- + 8 \bar{\delta}^+ \delta - 16 \bar{\delta}^- \nu) \lambda_{+}^{(0)}\Bigr] \,,\\
%%%%%%%%%%
\left(v_1^i\right)_{\text{tidal}}^{\text{N}^{1.5}\text{LO}} &=  \frac{\alpha^2 \tilde{G}^2 m}{c^5  r^4} \frac{\zeta}{(1 -  \zeta) \bar{\gamma} (2 + \bar{\gamma})} \Bigl\{(n v) v^{i} \Bigl[\Bigl(16 \bar{\delta}^+ + 16 \bar{\delta}^- \delta + \bigl(-32 \bar{\delta}^+ - 8 \bar{\gamma} (2 + \bar{\gamma})\bigr) \nu \Bigr) \lambda_{-}^{(0)} \nn  \\
& + (16 \bar{\delta}^- + 16 \bar{\delta}^+ \delta - 32 \bar{\delta}^- \nu) \lambda_{+}^{(0)}\Bigr]   + \frac{\alpha \tilde{G} m}{r} n^{i} \Bigl[\Bigl(- \frac{16}{3} \bar{\delta}^+ -  \frac{16}{3} \bar{\delta}^- \delta + \bigl(\frac{32}{3} \bar{\delta}^+ + \frac{8}{3} \bar{\gamma} (2 + \bar{\gamma})\bigr) \nu \Bigr) \lambda_{-}^{(0)} \nn \\
&  + \Bigl(- \frac{16}{3} \bar{\delta}^- + \frac{16}{3} \bar{\delta}^- (2 + \bar{\gamma}) \nu + \delta \Bigl(- \frac{16}{3} \bar{\delta}^+ + \bigl(\frac{16}{3} \bar{\delta}^+ \bar{\gamma} + \frac{4}{3} \bar{\gamma}^2 (2 + \bar{\gamma})\bigr) \nu \Bigr)\Bigr) \lambda_{+}^{(0)}\Bigr] \nn \\
&  + n^{i} \Bigl[v^2 \Bigl(\Bigl(8 \bar{\delta}^+ + 8 \bar{\delta}^- \delta + \bigl(-16 \bar{\delta}^+ - 4 \bar{\gamma} (2 + \bar{\gamma})\bigr) \nu \Bigr) \lambda_{-}^{(0)} + (8 \bar{\delta}^- + 8 \bar{\delta}^+ \delta - 16 \bar{\delta}^- \nu) \lambda_{+}^{(0)}\Bigr)\nn \\
&  + (n v)^2 \Bigl(\Bigl(-40 \bar{\delta}^+ - 40 \bar{\delta}^- \delta + \bigl(80 \bar{\delta}^+ + 20 \bar{\gamma} (2 + \bar{\gamma})\bigr) \nu \Bigr) \lambda_{-}^{(0)} + (-40 \bar{\delta}^- - 40 \bar{\delta}^+ \delta + 80 \bar{\delta}^- \nu) \lambda_{+}^{(0)}\Bigr)\Bigr]\Bigr\}\,.
\end{align}
\end{subequations}
while $\left(y_2^i\right)_{\text{tidal}}^{\text{N}^{1.5}\text{LO}}$ and $\left(v_2^i\right)_{\text{tidal}}^{\text{N}^{1.5}\text{LO}}$ are given by $1 \leftrightarrow 2$. 

%------------------------------------------------------------------------
\subsection{Source multipole moments}
\label{sec:source_moments}
%------------------------------------------------------------------------

The first step to compute the radiative multipole moments consists in deriving their source counterparts at the same PN orders as the ones displayed in Tables~\ref{tab:radiative_moments_for_flux} and \ref{tab:radiative_moments_for_waveform}. The procedure to obtain their expressions in terms of the source, consisting in a matching of the MPM expansion of the fields to their PN expansions in the near zone, can be found in~\cite{Blanchet:1998in} in GR, and it has been adapted to ST theories in~\cite{Bernard:2022noq}. There are two different sets of STF moments, the usual mass-type and current-type ones $(I_L,J_L)$ given for any $ \ell\ge 2$, and the scalar-type moments $I^s_L$ given for any $ \ell\ge 0$. They read~\cite{Bernard:2022noq}
\begin{subequations}\label{eq:momentsIJIs}
	\begin{align}
	I_L(u) &= \mathop{\mathrm{FP}}_{B=0}\int \ud^3 \mathbf{x} \,\tilde{r}^B \int_{-1}^1 \ud z \bigg[ \delta_\ell(z) \hat{x}_L \Sigma - \frac{4(2\ell +1)}{c^2(\ell+1)(2\ell+3)} \delta_{\ell+1}(z) \hat{x}_{iL} \Sigma^{(1)}_i \nn\\
	& \qquad \qquad \qquad \qquad \qquad \qquad \qquad \qquad + \frac{2(2\ell + 1)}{c^4 (\ell+1)(\ell+2)(2\ell+5)} \delta_{\ell+2}(z) \hat{x}_{ijL} \Sigma^{(2)}_{ij}  \bigg] ( \mathbf{x}, u+z r/c) \,,\\
	J_L(u) &= \mathop{\mathrm{FP}}_{B=0} \int \ud^3 \mathbf{x} \,\tilde{r}^B \int_{-1}^1 \ud z \,\varepsilon_{ab\langle i_\ell} \bigg[ \delta_\ell(z) \hat{x}_{L-1\rangle a} \Sigma_b -\frac{2\ell+1}{c^2(\ell+2)(2\ell+3)} \delta_{\ell+1}(z) \hat{x}_{L-1\rangle ac} \Sigma^{(1)}_{bc}\bigg]( \mathbf{x}, u + z r/c)\,,\\
    I^s_L(u) &=\mathop{\mathrm{FP}}_{B=0} \int \ud^3 \mathbf{x}\,\tilde{r}^B  \int_{-1}^1 \ud z \,\delta_\ell(z) \,\hat{x}_L \Sigma^s ( \mathbf{x}, u+z r/c)\,.
	\end{align}
\end{subequations}
A regulator $\tilde{r}^B=\left(r/r_0\right)^B$, together with the finite part (FP) operator when $B\rightarrow 0$, is used to cure the IR divergences, since the PN-expanded integrands are expected to diverge at spatial infinity.  We have introduced the ST source terms $\Sigma \equiv (\bar\tau^{00} + \bar\tau^{ii})/c^2$, $\Sigma_i \equiv \bar\tau^{0i}/c$, $\Sigma_{ij} \equiv \bar\tau^{ij}$ and $\Sigma^s \equiv  \bar\tau_s/c^2$,  where the overbar on the pseudo stress-energy tensor components indicates that their PN expansion is considered.  The integrations over $z$ are transformed into infinite PN series~\cite{Poujade:2001ie}
\begin{equation}\label{eq:intdeltal}
\int^1_{-1} \ud z~ \delta_\ell(z) \,\Sigma( \mathbf{x}, u+z r/c) = \sum_{k=0}^{+\infty}\,\frac{(2\ell+1)!!}{(2k)!!(2\ell+2k+1)!!}
	\,\left(\frac{r}{c}\right)^{2k} \!\Sigma^{(2k)}(\mathbf{x},u)\,,
\end{equation}
and lead to time derivatives of the source terms. In practice, we start by performing the PN expansion of the pseudo stress-energy tensors $\tau^{\mu \nu}$ and $\tau^s$ defined by Eqs.~\eqref{tau} at the orders that we need, decomposing them into their spatial and temporal indices using the parametrization of the scalar and tensor fields~\eqref{metric_decomp}. Next, we insert such expansions into the definitions of the sources moments~\eqref{eq:momentsIJIs}, resulting into spatial integrals involving both compact and non-compact support terms. The former ones are either linear in the three dimensional Dirac distribution, $ \delta^{(3)}_A \equiv \delta^{(3)} (\mathbf{x} - \bm{y}_A) $, or involve higher spatial and/or time derivatives of expressions linear in $\delta^{(3)}_A$. As a consequence, we handle these integrations by parts before performing the integration at the location of body A.  This evaluation requires a regularization scheme to handle the infinite self fields of the point particles, which in general is the dimensional regularization~\cite{Damour:2001bu,Blanchet:2003gy}. But at the NNLO, it is the equivalent and simpler to use the Hadamard “partie finie” regularization instead~\cite{Blanchet:2005ft}. The non-compact support terms consist in products of the potentials, that are integrated over all space. Methods to evaluate these integrals are explained in \textit{e.g.}~\cite{Blanchet:2000nu}. The resulting source moments are general functions of the harmonic coordinates, and we reduce them to the CoM frame using Eqs.~\eqref{eq:y1i_v1i_CM_dissipative} and Eqs.(43)-(44) in~\cite{Bernard:2023eul}. Since the expressions for the source moments are very long, even in the CoM frame, and constitute only an intermediate result, we relegated them to the Mathematica supplementary file~\cite{SuppMaterial}. 

Finally, we specialize the source moments to quasi-circular orbits, following a similar strategy as in Sec. \ref{sec:dynamics_circ}. Since we are working in the radiative sector, radiation-reaction effects must be taken into account. In scalar–tensor theories, these effects enter already at 1.5PN order, in contrast to GR where they start at 2.5PN order. Accordingly, the relative velocity can be decomposed as $\bm{v}= \omega r \bm{\lambda} + \dot{r} \bm{n}$, where $\dot{r}\equiv\bm{n}\cdot\bm{v}$. In the point-particle approximation, $\dot{r}$ is formally of 1.5PN order and therefore contributes to the source moments at the $\text{N}^{1.5}\text{LO}$ if one includes the tidal effects. As a consequence, only its LO expression is required here. Using the relation between the orbital separation and the PN parameter, $r=\frac{\tilde{G}\alpha m}{c^2 \gamma}$, together with the relation $\gamma(x)$ obtained at the NNLO in Sec.~\ref{sec:dynamics_circ} and the LO evolution equation for $x$ derived from the flux-balance equation \eqref{eq:flux_balance_eq}, see Eq.~\eqref{eq:xdot_0PN}, we obtain the following expression for $\dot{r}$ at LO  
\begin{align}\label{eq:rdot}
\dot{r} &=- \frac{8}{3} c\, \zeta \mathcal{S}_{-}{}^2 \nu x^2 + \frac{16 \,c\, \zeta x^5}{9 (1 -  \zeta) \bar{\gamma}^2} \Bigl\{16 \zeta \bar{\gamma}^2 \mathcal{S}_{-}{}^2 \nu \tilde{\lambda}_{+}^{(0)} +  \sqrt{\alpha} \Bigl[-3 \mathcal{S}_{-}{} \delta (4 \zeta \mathcal{S}_{-}{} \mathcal{S}_{+}{} \tilde{\Lambda}_{-}^{(0)} -  \bar{\gamma} \tilde{\Lambda}_{+}^{(0)} + 4 \zeta \mathcal{S}_{-}{}^2 \tilde{\Lambda}_{+}^{(0)}) \nn \\
& \quad - 24 \zeta \bar{\gamma} \mathcal{S}_{-}{}^2 \nu (\mathcal{S}_{-}{} \tilde{\Lambda}_{-}^{(0)} + \mathcal{S}_{+}{} \tilde{\Lambda}_{+}^{(0)}) - 3 \mathcal{S}_{-}{} (- \bar{\gamma} \tilde{\Lambda}_{-}^{(0)} + 4 \zeta \mathcal{S}_{-}{}^2 \tilde{\Lambda}_{-}^{(0)} + 4 \zeta \mathcal{S}_{-}{} \mathcal{S}_{+}{} \tilde{\Lambda}_{+}^{(0)})\Bigr]\Bigr\} \ +\ \mathcal{O}\left( \frac{1}{c^2} ; \frac{\epsilon_{\text{tidal}}}{c^2} \right) \,.
\end{align}
After expressing the source moments entirely in terms of the parameter $\gamma$, inserting  $\gamma$ and $\dot{r}$ as functions of the gauge invariant parameter $x$, and truncating at the correct order, we have reduced our expressions for the source moments to the case of quasi-circular orbits. The corresponding lengthy expressions are displayed in Appendix~\ref{App:source_moments_circ}. 

%------------------------------------------------------------------------
%------------------------------------------------------------------------
\section{Fluxes and orbital phase}
\label{sec:fluxes_phases}
%------------------------------------------------------------------------
%------------------------------------------------------------------------

We have now all the ingredients to compute the radiated scalar and gravitational energy fluxes and orbital phase up to the NNLO. These quantities are derived within the quasi-circular orbit approximation and written in terms of the frequency-dependent parameter~$x$. The complete results presented in this section are directly available in the supplementary \textit{Mathematica} file~\cite{SuppMaterial}. 

%------------------------------------------------------------------------
\subsection{Scalar and gravitational fluxes}
%------------------------------------------------------------------------

First, we derive the instantaneous gravitational and scalar fluxes in the CoM frame, discarding tail contributions. We decompose these fluxes as 
\begin{subequations}\label{eq:Fginst_Fsinst}
\begin{align}
\mathcal{F}^g_{\text{inst}}&=\left(\mathcal{F}^g_{\text{inst}}\right)_{\text{pp}}+\left(\mathcal{F}^g_{\text{inst}}\right)_{\text{tidal}}\,,\\
\mathcal{F}^s_{\text{inst}}&=\left(\mathcal{F}^s_{\text{inst}}\right)_{\text{pp}}+\left(\mathcal{F}^s_{\text{inst}}\right)_{\text{tidal}}\,.
\end{align}
\end{subequations}
Substituting our expressions for the radiative moments (which relate to the source ones only by multiple time-differentiations) into Eq.~\eqref{eq:fluxes_1PN} yields 
\begin{subequations}
\begin{align}
\left(\mathcal{F}^g_{\text{inst}}\right)_{\text{tidal}}&= \frac{\alpha^4 \tilde{G}^4 m^4 \nu^2 \zeta (2 + \bar{\gamma}) }{c^5 (1 -  \zeta) r^7} \Biggl\{\frac{1}{c^2}\Biggl[\frac{896}{15} (n v)^2 \lambda_{+}^{(0)} -  \frac{256}{5} v^2 \lambda_{+}^{(0)}\Biggr] + \frac{1}{c^4}\Biggl[\frac{\alpha^2 \tilde{G}^2 m^2}{r^2} (- \frac{64}{21} + \frac{256}{21} \nu) \lambda_{+}^{(0)} \nn \\
& \hspace{-0.5cm}+ \bigl(- \frac{32}{35} (211 + 112 \bar{\gamma}) + 384 \nu \bigr) (n v)^4 \lambda_{+}^{(0)} + \bigl(\frac{32}{105} (705 + 392 \bar{\gamma}) - 448 \nu \bigr) (n v)^2 v^2 \lambda_{+}^{(0)} \nn \\
& \hspace{-0.5cm}+ \bigl(- \frac{64}{15} (11 + 6 \bar{\gamma}) + \frac{768}{7} \nu \bigr) v^4 \lambda_{+}^{(0)} + \frac{\alpha \tilde{G} m}{r} \biggl[\frac{(1 -  \zeta)}{\zeta} \Bigl((n v)^2 (80 \delta \lambda_1 \Lambda_{-}^{(0)} - 80 \lambda_1 \Lambda_{+}^{(0)}) + v^2 (-64 \delta \lambda_1 \Lambda_{-}^{(0)} + 64 \lambda_1 \Lambda_{+}^{(0)})\Bigr) \nn \\
&\hspace{-0.5cm}+ v^2 \Bigl(\bigl(\frac{256 \bar{\beta}^-}{\bar{\gamma}} -  \frac{16 (80 \bar{\beta}^+ + 14 \bar{\gamma} + 15 \bar{\gamma}^2) \delta}{5 \bar{\gamma}}\bigr) \lambda_{-}^{(0)} + \delta (80 - 64 \lambda_1) \Lambda_{-}^{(0)} + 32 \phi_{0}{} \delta \Lambda_{-}^{(1)} \nn \\
&\hspace{-0.5cm}+ \bigl(\frac{16 (-80 \bar{\beta}^+ + 161 \bar{\gamma} + 32 \bar{\beta}^+ \bar{\gamma} + 63 \bar{\gamma}^2)}{5 \bar{\gamma}} -  \frac{256 \bar{\beta}^- (-5 + 2 \bar{\gamma}) \delta}{5 \bar{\gamma}} -  \frac{64}{7} \nu \bigr) \lambda_{+}^{(0)} + (-80 + 64 \lambda_1) \Lambda_{+}^{(0)} - 32 \phi_{0}{} \Lambda_{+}^{(1)}\Bigr) \nn \\
&\hspace{-0.5cm}+ (n v)^2 \Bigl(\bigl(- \frac{320 \bar{\beta}^-}{\bar{\gamma}} + \frac{4 (1200 \bar{\beta}^+ + 218 \bar{\gamma} + 225 \bar{\gamma}^2) \delta}{15 \bar{\gamma}}\bigr) \lambda_{-}^{(0)} + \delta (-100 + 80 \lambda_1) \Lambda_{-}^{(0)} - 40 \phi_{0}{} \delta \Lambda_{-}^{(1)} \nn \\
& \hspace{-0.5cm} + \bigl(- \frac{4 (-8400 \bar{\beta}^+ + 15562 \bar{\gamma} + 2688 \bar{\beta}^+ \bar{\gamma} + 6055 \bar{\gamma}^2)}{105 \bar{\gamma}} + \frac{64 \bar{\beta}^- (-25 + 8 \bar{\gamma}) \delta}{5 \bar{\gamma}} + \frac{928}{35} \nu \bigr) \lambda_{+}^{(0)} + (100 - 80 \lambda_1) \Lambda_{+}^{(0)} \nn \\
& \hspace{-0.5cm} + 40 \phi_{0}{} \Lambda_{+}^{(1)}\Bigr)\biggr]\Biggr] +\calO\left(\frac{\etidal}{c^{3}} \right)  \Biggr\}\\
%%%%%%%%%%%%%%%%%%%%%%%%%%%%%%%%%%%%%%%
\left(\mathcal{F}^s_{\text{inst}}\right)_{\text{tidal}}&= \frac{\alpha^3 \tilde{G}^3 m^3 \zeta }{c^3 (1 -  \zeta) \bar{\gamma}^2 r^6} \Biggl\{\frac{( s_1 -  s_2)}{c^2} \Biggl[(n v)^2 \bigl(24 \delta \nu (4 \zeta \mathcal{S}_{-}{} \mathcal{S}_{+}{} \Lambda_{-}^{(0)} -  \bar{\gamma} \Lambda_{+}^{(0)} + 4 \zeta \mathcal{S}_{-}{}^2 \Lambda_{+}^{(0)}) \nn \\
&- 24 \nu (\bar{\gamma} \Lambda_{-}^{(0)}  - 4 \zeta \mathcal{S}_{-}{}^2 \Lambda_{-}^{(0)} - 4 \zeta \mathcal{S}_{-}{} \mathcal{S}_{+}{} \Lambda_{+}^{(0)})\bigr)  + \frac{\alpha \tilde{G} m }{r} \Bigl(\frac{16}{3} \delta \nu (4 \zeta \mathcal{S}_{-}{} \mathcal{S}_{+}{} \Lambda_{-}^{(0)} -  \bar{\gamma} \Lambda_{+}^{(0)} + 4 \zeta \mathcal{S}_{-}{}^2 \Lambda_{+}^{(0)}) \nn \\
& -  \frac{64}{3} \zeta \bar{\gamma} \mathcal{S}_{-}{} \nu^2 (\mathcal{S}_{-}{} \Lambda_{-}^{(0)} + \mathcal{S}_{+}{} \Lambda_{+}^{(0)}) -  \frac{16}{3} \nu (\bar{\gamma} \Lambda_{-}^{(0)} - 4 \zeta \mathcal{S}_{-}{}^2 \Lambda_{-}^{(0)} - 4 \zeta \mathcal{S}_{-}{} \mathcal{S}_{+}{} \Lambda_{+}^{(0)})\Bigr) + v^2 \bigl(-8 \delta \nu (4 \zeta \mathcal{S}_{-}{} \mathcal{S}_{+}{} \Lambda_{-}^{(0)} \nn \\
& -  \bar{\gamma} \Lambda_{+}^{(0)} + 4 \zeta \mathcal{S}_{-}{}^2 \Lambda_{+}^{(0)}) + 8 \nu (\bar{\gamma} \Lambda_{-}^{(0)} - 4 \zeta \mathcal{S}_{-}{}^2 \Lambda_{-}^{(0)} - 4 \zeta \mathcal{S}_{-}{} \mathcal{S}_{+}{} \Lambda_{+}^{(0)})\bigr)\Biggr] + \frac{1}{c^4}\Bigl[\cdots\Bigr] +  \frac{1}{c^5}\Bigl[\cdots\Bigr] \nn \\
& + \frac{1}{c^6}\Biggl[\frac{6 \alpha \tilde{G} m \zeta \bar{\gamma}^2  \mathcal{S}_{-}{}^2 \nu^2}{r^3} \bigl(-16 c^4\mu_{+}^{(0)} + c^6 (c_{+}^{(0)} - 4 \nu_{+}^{(0)})\bigr)+(\cdots)\Biggr] +\calO\left(\frac{\etidal}{c^{5}}\right) \Biggr\}\,. 
\end{align}
\end{subequations}
Next, we specialize to the case of quasi-circular orbits, which permits us to explicitely compute the tail integral~\eqref{eq:dipole_moment_tail} which contributes to the radiative scalar dipole moment.   Tail integrals are computed under the assumption of an adiabatic inspiral, meaning that the orbital elements (orbital separation, phase and frequency) evolve due to radiation reaction on a timescale much longer than the orbital period. To describe the radiation reaction scale, we introduce a dimensionless adiabatic parameter $\xi = \dot{\omega}/\omega^2$, which scales as $\xi\sim \mathcal{O}\left(c^{-5}\right)$ in GR and $\xi\sim \mathcal{O}\left(c^{-3}\right)$ in ST theories. Since we only need to derive the tail integral~\eqref{eq:dipole_moment_tail} at LO, the adiabatic approximation holds and the orbital elements can be approximated as constant inside the integral. Furthermore, although tail integrals are formally hereditary and depend on the entire past history of the source, it was shown in Ref.~\cite{Blanchet:1993ec} that they are negligible compared to recent past contributions.  See Ref.~\cite{Arun:2004ff} for a detailed explanation on how to compute the tail integrals. While still accounting for radiation-reaction contributions, we get the full gravitational and scalar fluxes, which we write as
\begin{subequations}\label{eq:Fg_Fs}
\begin{align}
\mathcal{F}^g&=\mathcal{F}^g_{\text{pp}}+\mathcal{F}^g_{\text{tidal}}\,,\\
\mathcal{F}^s&=\mathcal{F}^s_{\text{pp}}+\mathcal{F}^s_{\text{tidal}}\,,
\end{align}
\end{subequations}
with
\begin{subequations}
\begin{align}
\mathcal{F}^g_{\text{tidal}}=&- \frac{512 c^5 (2 + \bar{\gamma}) \nu^2 x^8}{15 \alpha \tilde{G} (1 -  \zeta) \bar{\gamma} } \biggl\{\zeta \bar{\gamma} \tilde{\lambda}_{+}^{(0)} + x \Bigl[- \frac{23}{4} \zeta \bar{\gamma} \nu \tilde{\lambda}_{+}^{(0)} + \frac{1}{16} \delta (80 \bar{\beta}^+ \zeta \tilde{\lambda}_{-}^{(0)} + 15 \zeta \bar{\gamma} \tilde{\lambda}_{-}^{(0)} + 15 \zeta \bar{\gamma}^2 \tilde{\lambda}_{-}^{(0)} \nn \\
&- 25 \zeta \bar{\gamma} \tilde{\Lambda}_{-}^{(0)} + 20 \bar{\gamma} \lambda_1 \tilde{\Lambda}_{-}^{(0)} - 10 \zeta \bar{\gamma} \phi_{0}{} \tilde{\Lambda}_{-}^{(1)} - 80 \bar{\beta}^- \zeta \tilde{\lambda}_{+}^{(0)} - 32 \bar{\beta}^- \zeta \bar{\gamma} \tilde{\lambda}_{+}^{(0)}) + \frac{1}{224} (-1120 \bar{\beta}^- \zeta \tilde{\lambda}_{-}^{(0)} \nn \\
&+ 1120 \bar{\beta}^+ \zeta \tilde{\lambda}_{+}^{(0)} - 393 \zeta \bar{\gamma} \tilde{\lambda}_{+}^{(0)} + 448 \bar{\beta}^+ \zeta \bar{\gamma} \tilde{\lambda}_{+}^{(0)} - 210 \zeta \bar{\gamma}^2 \tilde{\lambda}_{+}^{(0)} + 350 \zeta \bar{\gamma} \tilde{\Lambda}_{+}^{(0)} - 280 \bar{\gamma} \lambda_1 \tilde{\Lambda}_{+}^{(0)} \nn \\
& + 140 \zeta \bar{\gamma} \phi_{0}{} \tilde{\Lambda}_{+}^{(1)})\Bigr]+\calO\left(\frac{\etidal}{c^{3}}\right)\biggr\} \\
%%%%%%%%%%%%%%%%%%%%%%%%%%%%%%%%%%%%%%%
\mathcal{F}^s_{\text{tidal}}=& - \frac{128 c^5 \zeta \mathcal{S}_{-}{} \nu x^7}{9 \alpha \tilde{G} (-1 + \zeta) \bar{\gamma}^2 } \biggl\{\zeta \bar{\gamma}^2 \mathcal{S}_{-}{} \nu \tilde{\lambda}_{+}^{(0)} + \sqrt{\alpha} \bigl(\frac{3}{16} \delta (4 \zeta \mathcal{S}_{-}{} \mathcal{S}_{+}{} \tilde{\Lambda}_{-}^{(0)} -  \bar{\gamma} \tilde{\Lambda}_{+}^{(0)} + 4 \zeta \mathcal{S}_{-}{}^2 \tilde{\Lambda}_{+}^{(0)}) \nn \\
& + \frac{3}{2} \zeta \bar{\gamma} \mathcal{S}_{-}{} \nu (\mathcal{S}_{-}{} \tilde{\Lambda}_{-}^{(0)} + \mathcal{S}_{+}{} \tilde{\Lambda}_{+}^{(0)}) -  \frac{3}{16} (\bar{\gamma} \tilde{\Lambda}_{-}^{(0)} - 4 \zeta \mathcal{S}_{-}{}^2 \tilde{\Lambda}_{-}^{(0)} - 4 \zeta \mathcal{S}_{-}{} \mathcal{S}_{+}{} \tilde{\Lambda}_{+}^{(0)})\bigr)  \nn \\
& + x \Bigl[\cdots\Bigr] + (2 + \bar{\gamma}) \pi x^{3/2} \Bigl[- \frac{1}{2} \zeta \bar{\gamma}^2 \mathcal{S}_{-}{} \nu \tilde{\lambda}_{+}^{(0)} + \sqrt{\alpha}  \bigl(\frac{3}{16} \delta (4 \zeta \mathcal{S}_{-}{} \mathcal{S}_{+}{} \tilde{\Lambda}_{-}^{(0)} -  \bar{\gamma} \tilde{\Lambda}_{+}^{(0)} + 4 \zeta \mathcal{S}_{-}{}^2 \tilde{\Lambda}_{+}^{(0)}) \nn \\
& -  \frac{3}{16} (\bar{\gamma} \tilde{\Lambda}_{-}^{(0)} - 4 \zeta \mathcal{S}_{-}{}^2 \tilde{\Lambda}_{-}^{(0)} - 4 \zeta \mathcal{S}_{-}{} \mathcal{S}_{+}{} \tilde{\Lambda}_{+}^{(0)})\bigr)\Bigr] + x^2\Bigl[\frac{9}{64} \zeta \bar{\gamma}^2 \mathcal{S}_{-}{} \nu  \bigl(\tilde{c}_{+}^{(0)} - 4 (4 \tilde{\mu}_{+}^{(0)} + \tilde{\nu}_{+}^{(0)})\bigr) + (\cdots)\Bigl] \nn \\
& +\calO\left(\frac{\etidal}{c^{5}}\right)\biggr\} \,. 
\end{align}
\end{subequations}
We have verified that the point-particle part of the gravitational and scalar fluxes up to 2PN order agrees with results from~\cite{Sennett:2016klh,Bernard:2022noq}. 

%------------------------------------------------------------------------
\subsection{Orbital phase in the time domain}\label{eq:orbital_phase_TD}
%------------------------------------------------------------------------

The orbital phase can be obtained from the flux balance equation for the energy
\begin{align}\label{eq:flux_balance_eq}
\frac{\dd E}{\dd t} = - \mathcal{F}
\end{align}
where $\mathcal{F}$ is the total flux, including both scalar and gravitational contributions,  and $E$ represents the binding energy. It differs from the conserved energy due to Schott terms, which contribute starting at the 3PN order, analogous to the terms appearing at the 4PN order in GR \cite{Trestini:2025nzr}. Since we are only concerned with the energy at the NNLO, we neglect these Schott terms and use the same notation for both the conserved and binding energy.  Up to a constant, it satisfies
\begin{align}\label{eq:phase_TD}
\phi = -\frac{c^3}{\alpha \tilde{G} m}\int \dd x \, x^{3/2}\,\frac{\dd E/ \dd x}{\mathcal{F}(x)}\,.
\end{align}
By substituting our results for the energy and the flux, along with their point-particle counterparts taken from Refs.~\cite{Bernard:2018ivi} and~\cite{Bernard:2022noq}, into Eq.~\eqref{eq:phase_TD}, re-expanding the ratio $(\dd E/ \dd x)/\mathcal{F}$ 
and after truncating at the NNLO, we get an analytical expression for the time domain orbital phase up to the NNLO. Special care must be taken when re-expanding this ratio. Indeed, as highlighted in~\cite{Sennett:2016klh}, the radiated flux contains both scalar and gravitational components, giving rise to two different regimes: a dipolar-driven (DD) regime, where the energy flux is dominated by dipolar emission which vanishes when $s_1 = s_2$, and a quadrupolar-driven (QD) regime, where the flux is dominated by quadrupolar emission. 
In the following, we focus exclusively in the DD-driven regime as the corrections to tidal effects due to the presence of the scalar field in ST theories are expected to be small and hardly observable in the QD regime~\cite{Bernard:2019yfz}.

Focusing on the DD regime, the derivation of the full phase $\phi = \phi_{\text{pp}}+\phi_{\text{tidal}}$ is straightforward and yields the following NNLO tidal correction 
\begin{align}\label{eq:phase_TD_value_DD}
\phi_{\text{tidal}} & = - \frac{4 x^{3/2}}{(-1 + \zeta) \zeta \bar{\gamma}^2 \mathcal{S}_{-}{}^3 \nu^2} \Bigl\{\zeta \bar{\gamma}^2 \mathcal{S}_{-}{} \nu \tilde{\lambda}_{+}^{(0)} + \sqrt{\alpha} \bigl(\frac{1}{8} \delta (-4 \zeta \mathcal{S}_{-}{} \mathcal{S}_{+}{} \tilde{\Lambda}_{-}^{(0)} + \bar{\gamma} \tilde{\Lambda}_{+}^{(0)} - 4 \zeta \mathcal{S}_{-}{}^2 \tilde{\Lambda}_{+}^{(0)}) \nn \\
& -  \zeta \bar{\gamma} \mathcal{S}_{-}{} \nu (\mathcal{S}_{-}{} \tilde{\Lambda}_{-}^{(0)} + \mathcal{S}_{+}{} \tilde{\Lambda}_{+}^{(0)}) + \frac{1}{8} (\bar{\gamma} \tilde{\Lambda}_{-}^{(0)} - 4 \zeta \mathcal{S}_{-}{}^2 \tilde{\Lambda}_{-}^{(0)} - 4 \zeta \mathcal{S}_{-}{} \mathcal{S}_{+}{} \tilde{\Lambda}_{+}^{(0)})\bigr) \nn \\
& +x\Bigl[ \cdots \Bigr] + \sqrt{\alpha} (2 + \bar{\gamma}) \pi x^{3/2} \Bigl[\frac{1}{16} \delta (4 \zeta \mathcal{S}_{-}{} \mathcal{S}_{+}{} \tilde{\Lambda}_{-}^{(0)} -  \bar{\gamma} \tilde{\Lambda}_{+}^{(0)} + 4 \zeta \mathcal{S}_{-}{}^2 \tilde{\Lambda}_{+}^{(0)}) + \zeta \bar{\gamma} \mathcal{S}_{-}{} \nu (\mathcal{S}_{-}{} \tilde{\Lambda}_{-}^{(0)} + \mathcal{S}_{+}{} \tilde{\Lambda}_{+}^{(0)}) \nn \\
& + \frac{1}{16} (- \bar{\gamma} \tilde{\Lambda}_{-}^{(0)} + 4 \zeta \mathcal{S}_{-}{}^2 \tilde{\Lambda}_{-}^{(0)} + 4 \zeta \mathcal{S}_{-}{} \mathcal{S}_{+}{} \tilde{\Lambda}_{+}^{(0)})\Bigr] + x^2 \Bigl[-  \frac{45}{112} \zeta \bar{\gamma}^2 \mathcal{S}_{-}{} \nu  \bigl(\tilde{c}_{+}^{(0)} - 4 (4 \tilde{\mu}_{+}^{(0)} + \tilde{\mu}_{+}^{(0)}) \bigr)+ (\cdots)\Bigr] \nn \\
& + \calO\left(\frac{\etidal}{c^{5}}\right)\Bigr\}\,.
\end{align}
The corresponding point-particle contribution $\phi_{\text{pp}}$ agrees with the 2PN result displayed \textit{e.g.} in Eq.~(5.7) of~\cite{Bernard:2022noq}. 

%------------------------------------------------------------------------
\subsection{Orbital phase in the stationary phase approximation}
%------------------------------------------------------------------------

For data analysis purposes, we also present the phase within the stationary phase approximation (SPA)~\cite{Tichy:1999pv}. This technique consists in performing the Fourier transform of the gravitational waveform. We proceed similarly as in Ref.~\cite{Sennett:2016klh} and write the amplitude modes in the time domain as
\begin{equation}
h_{\ell \mathrm{m}}(t) = \mathcal{A}_{\ell \mathrm{m}}(t)\mathrm{e}^{- \mathrm{i}\mathrm{m} \phi(t)}\,,
\end{equation}
where $\mathcal{A}_{\ell \mathrm{m}}(t)$ and $\phi(t)$ are the time-domain amplitude and orbital phase, respectively. Following the convention of Ref.~\cite{Sennett:2016klh}, the Fourier transform of the waveform modes reads
\begin{equation}\label{eqref:fourier_transform}
h_{\ell \mathrm{m}}(f) = \int_{-\infty}^{\infty} \ud t \,\mathcal{A}_{\ell \mathrm{m}}(t)\mathrm{e}^{- \mathrm{i}(\mathrm{m} \phi(t)-2\pi f t)}\,,
\end{equation}
where $f$ is the Fourier frequency and we denote $\Phi_{\ell \mathrm{m}}(t) = \mathrm{m} \phi(t)-2\pi f t$. In the SPA, for each Fourier frequency, we identify a stationary time $t_f^{(\mathrm{m})}$ such that the phase $\Phi_{\ell \mathrm{m}}$ becomes stationary at that time. This is defined by the condition $\mathrm{m} \omega(t_f^{(\mathrm{m})})=2\pi f$, where $\mathrm{m}$ is the harmonic index associated with the different modes of the signal. Thus, the Fourier transform~\eqref{eqref:fourier_transform} becomes
\begin{equation}
h_{\ell \mathrm{m}}^{\text{SPA}}(f) = \mathcal{A}_{\ell \mathrm{m}}(f)\mathrm{e}^{- \mathrm{i}\Phi_{\ell \mathrm{m}}(f)-\mathrm{i} \pi/4 }\,,
\end{equation}
where the SPA phase is given by 
\begin{equation}
\Phi_{\ell \mathrm{m}}(f) = \mathrm{m} \phi(t_f^{(m)}) - 2\pi f t_f^{(m)}\,.
\end{equation}
Next, we define the time-domain parameter $v=x^{1/2}$, where $x=\left(\tilde{G}\alpha m \omega/c^3\right)^{2/3}$, is related to the instantaneous orbital frequency $\omega(t)$ defined at each time $t$. We also define the frequency-domain parameter $v_f=(\pi \tilde{G}\alpha \,m f/c^3)^{1/3}$ associated to the Fourier frequency $f$, which can take all values within the frequency band. At the stationary time $t_f^{(\mathrm{m})}$, the time-domain parameter $v$ is linked to frequency-domain parameter $v_f$ via
\begin{align}
v (t_f^{(\mathrm{m})})=\left(\frac{2}{\mathrm{m}}\right)^{1/3}v_f\,.
\end{align}
This ensures that the waveform at some frequency $f$ corresponds to a state of the binary at the stationary point $t_f^{(\mathrm{m})}$, where the Fourier phase of the waveform, $\Phi_{\ell \mathrm{m}}$, is stationary with respect to the time $t_f^{(\mathrm{m})}$. Then, the SPA phase can be rewritten as
\begin{equation}\label{eq:phase_SPA}
\Phi_{\ell \mathrm{m}}(f) = \left. \mathrm{m} \left( \phi(v) - \frac{c^3}{\tilde{G}\alpha m}v^3t(v) \right)\right|_{v=\left(2/\mathrm{m}\right)^{1/3}v_f}
\end{equation}
where $v=v(t_f^{(\mathrm{m})})$. The functions $\phi(v)$ and $t(v)$ are computed directly from the flux balance equation~\eqref{eq:flux_balance_eq}, and they satisfy the following relations
\begin{subequations}
\label{eq:SPA_phi_t}
\begin{align}
\phi(v) &= \phi(v_0) - \frac{c^3}{\alpha \tilde{G} m}\int^v_{v_0} \dd v \,v^3 \frac{\dd E/ \dd v}{\mathcal{F}(v)} \\
t(v) &= t(v_0)- \int^v_{v_0} \dd v \, \frac{\dd E/ \dd v}{\mathcal{F}(v)}
\end{align} 
\end{subequations}
where $\phi(v_0)$ and $t(v_0)$ are constants evaluated at some reference point $v_0$. We have explicitly computed Eq.~\eqref{eq:phase_SPA} within the DD regime and we decompose the Fourier domain phase as
\begin{align}
\Phi^{\ell \mathrm{m}}(f) = \Phi^{\ell \mathrm{m}}_{\text{pp}}(f) + \Phi^{\ell \mathrm{m}}_{\text{tidal}}(f) + \mathrm{m} \phi_0 - 2 \pi f \,t_0 \,.
\end{align}
The constants $\phi_0$ and $t_0$ result from the sum of $\phi(v_0)$ and $t(v_0)$, respectively, and come from the lower boundaries of the integrals~\eqref{eq:SPA_phi_t}. Finally, the NNLO tidal correction to the phase reads
\begin{align}\label{eq:phase_SPA_value}
\Phi^{\ell \mathrm{m}}_{\text{tidal}}(f) &= \frac{4 {\rm m} v^3}{(-1 + \zeta) \zeta \bar{\gamma}^2 \mathcal{S}_{-}{}^3 \nu^2} \biggl\{
\bigl(-1 + 3 \log(v)\bigr) \Bigl(\zeta \bar{\gamma}^2 \mathcal{S}_{-}{} \nu \tilde{\lambda}_{+}^{(0)} + \sqrt{\alpha} \bigl(\frac{1}{8} \delta (-4 \zeta \mathcal{S}_{-}{} \mathcal{S}_{+}{} \tilde{\Lambda}_{-}^{(0)} + \bar{\gamma} \tilde{\Lambda}_{+}^{(0)} - 4 \zeta \mathcal{S}_{-}{}^2 \tilde{\Lambda}_{+}^{(0)}) \nn \\
& -  \zeta \bar{\gamma} \mathcal{S}_{-}{} \nu (\mathcal{S}_{-}{} \tilde{\Lambda}_{-}^{(0)} + \mathcal{S}_{+}{} \tilde{\Lambda}_{+}^{(0)}) + \frac{1}{8} (\bar{\gamma} \tilde{\Lambda}_{-}^{(0)} - 4 \zeta \mathcal{S}_{-}{}^2 \tilde{\Lambda}_{-}^{(0)} - 4 \zeta \mathcal{S}_{-}{} \mathcal{S}_{+}{} \tilde{\Lambda}_{+}^{(0)})\bigr)\Bigr) + v^2 \Bigl[\cdots\Bigr] \nn \\
& +\sqrt{\alpha} (2 + \bar{\gamma}) \pi v^3 \Bigl[\frac{1}{16} \delta (4 \zeta \mathcal{S}_{-}{} \mathcal{S}_{+}{} \tilde{\Lambda}_{-}^{(0)} -  \bar{\gamma} \tilde{\Lambda}_{+}^{(0)} + 4 \zeta \mathcal{S}_{-}{}^2 \tilde{\Lambda}_{+}^{(0)}) + \zeta \bar{\gamma} \mathcal{S}_{-}{} \nu (\mathcal{S}_{-}{} \tilde{\Lambda}_{-}^{(0)} + \mathcal{S}_{+}{} \tilde{\Lambda}_{+}^{(0)}) + \frac{1}{16} (- \bar{\gamma} \tilde{\Lambda}_{-}^{(0)} \nn \\
& + 4 \zeta \mathcal{S}_{-}{}^2 \tilde{\Lambda}_{-}^{(0)} + 4 \zeta \mathcal{S}_{-}{} \mathcal{S}_{+}{} \tilde{\Lambda}_{+}^{(0)})\Bigr] + v^4 \Bigl[-  \frac{135}{448} \zeta \bar{\gamma}^2 \mathcal{S}_{-}{} \nu \bigl(\tilde{c}_{+}^{(0)} - 4 (4 \tilde{\mu}_{+}^{(0)} + \tilde{\nu}_{+}^{(0)})\bigr) + (\cdots)\Bigr] + \calO\left(\frac{\etidal}{c^{5}}\right)\biggr\}  \,.
\end{align} 
The 2PN point-particle counterpart to Eq.~\eqref{eq:phase_SPA_value} agrees with Eq.~(77) of~\cite{Sennett:2016klh} and the LO dipolar tidal correction matches Eq.~(14) of~\cite{Bernard:2019yfz}. For a detailed comparison with the results of Ref.~\cite{Creci:2024wfu}, we refer the reader to Appendix~\ref{eq:comparison_Creci}. 

%------------------------------------------------------------------------
%------------------------------------------------------------------------
\section{Waveform amplitude modes}
\label{sec:amplitudemodes}
%------------------------------------------------------------------------
%------------------------------------------------------------------------

Finally, we present the complete set of waveform amplitude modes at $\text{N}^{1.5}\text{LO}$, including the gravitational memory (${\rm m}=0$) modes. The derivation follows the general procedure described in Secs.~\ref{sec:asymptotic_waveform} and \ref{sec:memory}. As discussed in Sec.~\ref{sec:asymptotic_waveform}, achieving this accuracy requires computing all gravitational modes up to NLO order, while scalar modes must be obtained up to the $\text{N}^{1.5}\text{LO}$, since the scalar waveform enters at -$0.5$PN w.r.t the gravitational waveform. The derivation of the memory modes specifically required the computation of $\dot{x}$, the time derivative of the PN parameter $x$, given at LO in Eq.~\eqref{eq:xdot_0PN}, up to the NLO, which directly follows from the flux balance equation~\eqref{eq:flux_balance_eq}. We provide such a result for $\dot{x}$ up to the NNLO in the Supplemental Material~\cite{SuppMaterial}. The resulting waveform modes are presented below.
 
%------------------------------------------------------------------------
\subsection{The gravitational modes}
%------------------------------------------------------------------------

The gravitational modes are decomposed as $\hat{H}_{\ell {\rm m}}=\hat{H}^\text{pp}_{\ell {\rm m}}+\hat{H}^\text{tidal}_{\ell {\rm m}}$. The point-particle parts $\hat{H}^\text{pp}_{\ell {\rm m}}$ were derived up to 2PN order in~\cite{Sennett:2016klh}. Here, we present the non-vanishing tidal corrections to the gravitational modes up to NLO:
\begin{subequations} \label{Hlm}
\begin{align}
\hat{H}^\text{tidal}_{22}=& - \frac{16 x^3}{3 (1 -  \zeta) \bar{\gamma}} \Bigl\{\zeta \bar{\gamma} \tilde{\lambda}_{+}^{(0)} + x \Bigl[\frac{209}{168} \zeta \bar{\gamma} \nu \tilde{\lambda}_{+}^{(0)} + \frac{5}{48} \delta (48 \bar{\beta}^+ \zeta \tilde{\lambda}_{-}^{(0)} + 9 \zeta \bar{\gamma} \tilde{\lambda}_{-}^{(0)} + 9 \zeta \bar{\gamma}^2 \tilde{\lambda}_{-}^{(0)} - 15 \zeta \bar{\gamma} \tilde{\Lambda}_{-}^{(0)} \nn \\
& + 12 \bar{\gamma} \lambda_1 \tilde{\Lambda}_{-}^{(0)}  - 6 \zeta \bar{\gamma} \phi_{0}{} \tilde{\Lambda}_{-}^{(1)} - 48 \bar{\beta}^- \zeta \tilde{\lambda}_{+}^{(0)} - 32 \bar{\beta}^- \zeta \bar{\gamma} \tilde{\lambda}_{+}^{(0)}) + \frac{1}{336} (-1680 \bar{\beta}^- \zeta \tilde{\lambda}_{-}^{(0)} + 1680 \bar{\beta}^+ \zeta \tilde{\lambda}_{+}^{(0)} \nn \\
&- 431 \zeta \bar{\gamma} \tilde{\lambda}_{+}^{(0)}  + 1120 \bar{\beta}^+ \zeta \bar{\gamma} \tilde{\lambda}_{+}^{(0)} - 91 \zeta \bar{\gamma}^2 \tilde{\lambda}_{+}^{(0)} + 525 \zeta \bar{\gamma} \tilde{\Lambda}_{+}^{(0)} - 420 \bar{\gamma} \lambda_1 \tilde{\Lambda}_{+}^{(0)} + 210 \zeta \bar{\gamma} \phi_{0}{} \tilde{\Lambda}_{+}^{(1)})\Bigr]\Bigr\}\,, \\
%%%%%%%%%%%%%%%%%%%%%%%%%%%%%%%%
\hat{H}^\text{tidal}_{21}=& \frac{8i \zeta \delta x^{7/2} \tilde{\lambda}_{+}^{(0)}}{3 (-1 + \zeta)}\,, \\
\hat{H}^\text{tidal}_{33}=& - \frac{3i \sqrt{30} \zeta \delta x^{7/2} \tilde{\lambda}_{+}^{(0)}}{\sqrt{7} (-1 + \zeta)}\,, \\
\hat{H}^\text{tidal}_{32}=& - \frac{32 \sqrt{5} \zeta (-1 + 3 \nu) x^4 \tilde{\lambda}_{+}^{(0)}}{9 \sqrt{7} (-1 + \zeta)} \,, \\
\hat{H}^\text{tidal}_{31}=& \frac{i \sqrt{2} \zeta \delta x^{7/2} \tilde{\lambda}_{+}^{(0)}}{3 \sqrt{7} (-1 + \zeta)}\,, \\
\hat{H}^\text{tidal}_{44}= &\frac{256 \sqrt{5} \zeta (-1 + 3 \nu) x^4 \tilde{\lambda}_{+}^{(0)}}{27 \sqrt{7} (-1 + \zeta)}\,, \\
\hat{H}^\text{tidal}_{42}=& - \frac{32 \sqrt{5} \zeta (-1 + 3 \nu) x^4 \tilde{\lambda}_{+}^{(0)}}{189 (-1 + \zeta)}\,. 
\end{align}
\end{subequations}
For the sake of simplicity we have omitted the remainder $\calO\left(\etidal/c^{3}\right)$. We have also computed for the first time the ${\rm m}=0$ modes in the DD regime at the NLO, therefore we display both the point-particle and tidal parts 
\begin{subequations} \label{Hlm_m0}
\begin{align}
\hat{H}^\text{pp}_{20}=& \frac{1}{4 \sqrt{6}}\Bigl\{1 + \Bigl[- \frac{3}{7} + \frac{2}{3} \bar{\beta}^+ -  \frac{2}{3} \bar{\gamma} -  \frac{204}{35 \zeta \mathcal{S}_{-}{}^2} -  \frac{3 \bar{\gamma}}{\zeta \mathcal{S}_{-}{}^2} + (- \frac{2}{3} \bar{\beta}^- -  \frac{9 \mathcal{S}_{+}{}}{28 \mathcal{S}_{-}{}}) \delta -  \frac{61}{84} \nu \Bigr] x\Bigr\} + \calO\left(\frac{1}{c^{3}}\right)\,, \\
%%%%%%%%%%%%%%%%%%%%%%%%%%%%%%%%%%%%%%%
\hat{H}^\text{tidal}_{20}=& \frac{x^3}{12600 \sqrt{6} \alpha (-1 + \zeta)^2 \zeta \bar{\gamma}^3 \mathcal{S}_{-}{}^3 \nu} \Bigl\{-8400 \alpha (-1 + \zeta) \zeta^2 \bar{\gamma}^3 \mathcal{S}_{-}{}^3 \nu \tilde{\lambda}_{+}^{(0)} \nn \\
&+ 12600 \alpha^{3/2} (-1 + \zeta) \zeta^2 \bar{\gamma}^2 \mathcal{S}_{-}{}^3 \nu (\mathcal{S}_{-}{} \tilde{\Lambda}_{-}^{(0)} + \mathcal{S}_{+}{} \tilde{\Lambda}_{+}^{(0)}) + x \Bigl[\cdots\Bigr] \Bigr\} + \calO\left(\frac{\etidal}{c^{3}}\right) \,, \\
%%%%%%%%%%%%%%%%%%%%%%%%%%%%%%%%%%%%%%%
\hat{H}^\text{pp}_{40}=& \frac{x}{1680 \sqrt{2} \zeta \mathcal{S}_{-}{}^2}\Bigl\{-8 - 15 \zeta \mathcal{S}_{-}{}^2 + 15 \zeta \mathcal{S}_{-}{} \mathcal{S}_{+}{} \delta + 30 \zeta \mathcal{S}_{-}{}^2 \nu \Bigr\} + \calO\left(\frac{1}{c^{3}}\right) \,, \\
%%%%%%%%%%%%%%%%%%%%%%%%%%%%%%%%%%%%%%%
\hat{H}^\text{tidal}_{40}=& \frac{x^4 }{12600 \sqrt{2} \alpha^2 (-1 + \zeta) \zeta \bar{\gamma}^2 \mathcal{S}_{-}{}^3 \nu \Bigl(-8 + 15 \zeta \mathcal{S}_{-}{} \bigl(\mathcal{S}_{+}{} \delta + \mathcal{S}_{-}{} (-1 + 2 \nu)\bigr)\Bigr)}\Bigl\{\cdots\Bigr\} + \calO\left(\frac{\etidal}{c^{3}}\right) \,. 
\end{align}
\end{subequations}
We do not display the complete and explicit expressions for $\hat{H}^\text{tidal}_{20}$ and $\hat{H}^\text{tidal}_{40}$ due to their length but all the gravitational modes are directly available in~\cite{SuppMaterial}. Note that the mode $\hat{H}^\text{pp}_{20}$ agrees at Newtonian order with the result found in~\cite{Sennett:2016klh} in the DD regime. All the gravitational modes with negative $m$ can be deduced from these expressions using the relation $\hat{H}^{\ell, -{\rm m}} = (-)^\ell (\hat{H}^{\ell {\rm m}})^{*}$. 

%------------------------------------------------------------------------
\subsection{The scalar modes}
%------------------------------------------------------------------------

The scalar modes are decomposed as $\hat{\Psi}_{\ell {\rm m}}=\hat{\Psi}^\text{pp}_{\ell {\rm m}}+\hat{\Psi}^\text{tidal}_{\ell {\rm m}}$. The point-particle parts $\hat{\Psi}^\text{pp}_{\ell {\rm m}}$ were derived up to 1.5PN order in~\cite{Bernard:2022noq}, and the associated tidal corrections up to $\text{N}^{1.5}\text{LO}$ read:
\begin{subequations} \label{Psilm}
\begin{align}
\hat{\Psi}^\text{tidal}_{00}=&- \frac{i \sqrt{2} x^{7/2}}{\sqrt{3} (-1 + \zeta) \bar{\gamma}^2 \mathcal{S}_{-}{} \nu} \Bigl\{-8 \zeta (-4 \bar{\beta}^- \bar{\gamma} \mathcal{S}_{-}{} + 16 \bar{\beta}^- \zeta \mathcal{S}_{-}{}^3 + 4 \bar{\beta}^+ \bar{\gamma} \mathcal{S}_{+}{} -  \bar{\gamma}^2 \mathcal{S}_{+}{} - 16 \bar{\beta}^- \zeta \mathcal{S}_{-}{} \mathcal{S}_{+}{}^2) \nu \tilde{\lambda}_{+}^{(0)} \nn \\
& + \frac{\nu}{\sqrt{\alpha}} \Bigl[-12 \bar{\gamma}^2 \lambda_1 \tilde{\lambda}_{+}^{(0)} + 3 \zeta \bar{\gamma}^2 (5 \tilde{\lambda}_{+}^{(0)} + 2 \phi_{0}{} \tilde{\lambda}_{+}^{(1)})\Bigr] +\sqrt{\alpha}  \Bigl[\nu (24 \bar{\beta}^- \tilde{\Lambda}_{-}^{(0)} + 4 \zeta \bar{\gamma} \mathcal{S}_{-}{} \mathcal{S}_{+}{} \tilde{\Lambda}_{-}^{(0)} - 24 \bar{\beta}^+ \tilde{\Lambda}_{+}^{(0)} - 5 \bar{\gamma}^2 \tilde{\Lambda}_{+}^{(0)} \nn \\
& + 4 \zeta \bar{\gamma} \mathcal{S}_{-}{}^2 \tilde{\Lambda}_{+}^{(0)}) - 4 \zeta \bar{\gamma} \mathcal{S}_{-}{} \delta \nu (\mathcal{S}_{-}{} \tilde{\Lambda}_{-}^{(0)} + \mathcal{S}_{+}{} \tilde{\Lambda}_{+}^{(0)})\Bigr]+ x \Bigl[\cdots\Bigr]\Bigr\}\,, \\
%%%%%%%%%%%%%%%%%%%%%%%%%
\hat{\Psi}^\text{tidal}_{11}=&\frac{8 x^3}{3 (-1 + \zeta) \bar{\gamma}^2 \mathcal{S}_{-}{} \nu} \Bigl\{\zeta \bar{\gamma}^2 \mathcal{S}_{-}{} \nu \tilde{\lambda}_{+}^{(0)} + \sqrt{\alpha} \Bigl[- \frac{3}{8} \delta (4 \zeta \mathcal{S}_{-}{} \mathcal{S}_{+}{} \tilde{\Lambda}_{-}^{(0)} -  \bar{\gamma} \tilde{\Lambda}_{+}^{(0)} + 4 \zeta \mathcal{S}_{-}{}^2 \tilde{\Lambda}_{+}^{(0)}) + \frac{3}{8} (\bar{\gamma} \tilde{\Lambda}_{-}^{(0)}  - 4 \zeta \mathcal{S}_{-}{}^2 \tilde{\Lambda}_{-}^{(0)} \nn \\
&- 4 \zeta \mathcal{S}_{-}{} \mathcal{S}_{+}{} \tilde{\Lambda}_{+}^{(0)})\Bigr] + x \Bigl[\cdots\Bigr] + x^{3/2} \Bigl[\cdots\Bigr] \Bigr\}\,, \\
%%%%%%%%%%%%%%%%%%%%%%%%%
\hat{\Psi}^\text{tidal}_{22}=&- \frac{4i x^{7/2}}{3 \sqrt{5} (1 -  \zeta) \bar{\gamma}^2 \mathcal{S}_{-}{}} \Bigl\{\zeta \bar{\gamma}^2 (\mathcal{S}_{+}{} -  \mathcal{S}_{-}{} \delta) \tilde{\lambda}_{+}^{(0)} + \sqrt{\alpha}  \Bigl[- \frac{3}{2} (16 \zeta \mathcal{S}_{-}{} \mathcal{S}_{+}{} \tilde{\Lambda}_{-}^{(0)} + 2 \zeta \bar{\gamma} \mathcal{S}_{-}{} \mathcal{S}_{+}{} \tilde{\Lambda}_{-}^{(0)} - 4 \bar{\gamma} \tilde{\Lambda}_{+}^{(0)} -  \bar{\gamma}^2 \tilde{\Lambda}_{+}^{(0)} \nn \\
& + 16 \zeta \mathcal{S}_{-}{}^2 \tilde{\Lambda}_{+}^{(0)} + 2 \zeta \bar{\gamma} \mathcal{S}_{-}{}^2 \tilde{\Lambda}_{+}^{(0)}) + 3 \zeta \bar{\gamma} \mathcal{S}_{-}{} \delta (\mathcal{S}_{-}{} \tilde{\Lambda}_{-}^{(0)} + \mathcal{S}_{+}{} \tilde{\Lambda}_{+}^{(0)})\Bigr] + x \Bigl[\cdots\Bigr]\Bigr\}\,, \\
%%%%%%%%%%%%%%%%%%%%%%%%%
\hat{\Psi}^\text{tidal}_{33}=& \frac{2 \sqrt{6} x^4}{\sqrt{35} (1 -  \zeta) \bar{\gamma}^2 \mathcal{S}_{-}{}} \Bigl\{\zeta \bar{\gamma}^2 (\mathcal{S}_{-}{} -  \mathcal{S}_{+}{} \delta - 2 \mathcal{S}_{-}{} \nu) \tilde{\lambda}_{+}^{(0)}  - \frac{\sqrt{\alpha}}{2} \Bigl[\frac{1}{4} \delta \Bigl(-108 \zeta \mathcal{S}_{-}{} \mathcal{S}_{+}{} \tilde{\Lambda}_{-}^{(0)} - 28 \zeta \bar{\gamma} \mathcal{S}_{-}{} \mathcal{S}_{+}{} \tilde{\Lambda}_{-}^{(0)} \nn \\
& + 27 \bar{\gamma} \tilde{\Lambda}_{+}^{(0)} + 14 \bar{\gamma}^2 \tilde{\Lambda}_{+}^{(0)} - 108 \zeta \mathcal{S}_{-}{}^2 \tilde{\Lambda}_{+}^{(0)} - 28 \zeta \bar{\gamma} \mathcal{S}_{-}{}^2 \tilde{\Lambda}_{+}^{(0)}\Bigr) - 14 \zeta \bar{\gamma} \mathcal{S}_{-}{} \nu \Bigl(\mathcal{S}_{-}{} \tilde{\Lambda}_{-}^{(0)} + \mathcal{S}_{+}{} \tilde{\Lambda}_{+}^{(0)}\Bigr) \nn \\
& + \frac{1}{4} \Bigl(-27 \bar{\gamma} \tilde{\Lambda}_{-}^{(0)} + 108 \zeta \mathcal{S}_{-}{}^2 \tilde{\Lambda}_{-}^{(0)} + 28 \zeta \bar{\gamma} \mathcal{S}_{-}{}^2 \tilde{\Lambda}_{-}^{(0)} + 108 \zeta \mathcal{S}_{-}{} \mathcal{S}_{+}{} \tilde{\Lambda}_{+}^{(0)} + 28 \zeta \bar{\gamma} \mathcal{S}_{-}{} \mathcal{S}_{+}{} \tilde{\Lambda}_{+}^{(0)}\Bigr)\Bigr]\Bigr\}\,, \\
%%%%%%%%%%%%%%%%%%%%%%%%%
\hat{\Psi}^\text{tidal}_{31}=& \frac{6 \sqrt{2} x^4}{5\sqrt{7} (1 -  \zeta) \bar{\gamma}^2 \mathcal{S}_{-}{}} \Bigl\{\zeta \bar{\gamma}^2 (\mathcal{S}_{-}{} -  \mathcal{S}_{+}{} \delta - 2 \mathcal{S}_{-}{} \nu) \tilde{\lambda}_{+}^{(0)}  - \frac{\sqrt{\alpha}}{2} \Bigl[\frac{1}{12} \delta \Bigl(12 \zeta \mathcal{S}_{-}{} \mathcal{S}_{+}{} \tilde{\Lambda}_{-}^{(0)} + 28 \zeta \bar{\gamma} \mathcal{S}_{-}{} \mathcal{S}_{+}{} \tilde{\Lambda}_{-}^{(0)} - 3 \bar{\gamma} \tilde{\Lambda}_{+}^{(0)} \nn \\
& - 14 \bar{\gamma}^2 \tilde{\Lambda}_{+}^{(0)} + 12 \zeta \mathcal{S}_{-}{}^2 \tilde{\Lambda}_{+}^{(0)} + 28 \zeta \bar{\gamma} \mathcal{S}_{-}{}^2 \tilde{\Lambda}_{+}^{(0)}\Bigr) + \frac{14}{3} \zeta \bar{\gamma} \mathcal{S}_{-}{} \nu \Bigl(\mathcal{S}_{-}{} \tilde{\Lambda}_{-}^{(0)} + \mathcal{S}_{+}{} \tilde{\Lambda}_{+}^{(0)}\Bigr) + \frac{1}{12} \Bigl(3 \bar{\gamma} \tilde{\Lambda}_{-}^{(0)} - 12 \zeta \mathcal{S}_{-}{}^2 \tilde{\Lambda}_{-}^{(0)} \nn \\
& - 28 \zeta \bar{\gamma} \mathcal{S}_{-}{}^2 \tilde{\Lambda}_{-}^{(0)} - 12 \zeta \mathcal{S}_{-}{} \mathcal{S}_{+}{} \tilde{\Lambda}_{+}^{(0)} - 28 \zeta \bar{\gamma} \mathcal{S}_{-}{} \mathcal{S}_{+}{} \tilde{\Lambda}_{+}^{(0)}\Bigr)\Bigr]\Bigr\}\,, \\
%%%%%%%%%%%%%%%%%%%%%%%%%
\hat{\Psi}^\text{tidal}_{44}=& \frac{122i x^{9/2}}{9 \sqrt{105} (1 -  \zeta) \bar{\gamma}^2 \mathcal{S}_{-}{}} \Bigl\{\zeta \bar{\gamma}^2 (\mathcal{S}_{+}{} -  \mathcal{S}_{-}{} \delta - 3 \mathcal{S}_{+}{} \nu + \mathcal{S}_{-}{} \delta \nu) \tilde{\lambda}_{+}^{(0)} - \frac{\sqrt{\alpha}}{2} \Bigl[\frac{3}{61} \Bigl(512 \zeta \mathcal{S}_{-}{} \mathcal{S}_{+}{} \tilde{\Lambda}_{-}^{(0)} \nn \\
& + 130 \zeta \bar{\gamma} \mathcal{S}_{-}{} \mathcal{S}_{+}{} \tilde{\Lambda}_{-}^{(0)} - 128 \bar{\gamma} \tilde{\Lambda}_{+}^{(0)} - 65 \bar{\gamma}^2 \tilde{\Lambda}_{+}^{(0)} + 512 \zeta \mathcal{S}_{-}{}^2 \tilde{\Lambda}_{+}^{(0)} + 130 \zeta \bar{\gamma} \mathcal{S}_{-}{}^2 \tilde{\Lambda}_{+}^{(0)}\Bigr) -  \frac{3}{61} \nu \Bigl(1024 \zeta \mathcal{S}_{-}{} \mathcal{S}_{+}{} \tilde{\Lambda}_{-}^{(0)} \nn \\
& + 390 \zeta \bar{\gamma} \mathcal{S}_{-}{} \mathcal{S}_{+}{} \tilde{\Lambda}_{-}^{(0)} - 256 \bar{\gamma} \tilde{\Lambda}_{+}^{(0)} - 195 \bar{\gamma}^2 \tilde{\Lambda}_{+}^{(0)} + 1024 \zeta \mathcal{S}_{-}{}^2 \tilde{\Lambda}_{+}^{(0)}  + 390 \zeta \bar{\gamma} \mathcal{S}_{-}{}^2 \tilde{\Lambda}_{+}^{(0)}\Bigr) \nn \\
&+ \frac{390}{61} \zeta \bar{\gamma} \mathcal{S}_{-}{} \delta \nu \Bigl(\mathcal{S}_{-}{} \tilde{\Lambda}_{-}^{(0)} + \mathcal{S}_{+}{} \tilde{\Lambda}_{+}^{(0)}\Bigr) -  \frac{6}{61} \delta \Bigl(-64 \bar{\gamma} \tilde{\Lambda}_{-}^{(0)} + 256 \zeta \mathcal{S}_{-}{}^2 \tilde{\Lambda}_{-}^{(0)} + 65 \zeta \bar{\gamma} \mathcal{S}_{-}{}^2 \tilde{\Lambda}_{-}^{(0)} \nn \\
& + 256 \zeta \mathcal{S}_{-}{} \mathcal{S}_{+}{} \tilde{\Lambda}_{+}^{(0)} + 65 \zeta \bar{\gamma} \mathcal{S}_{-}{} \mathcal{S}_{+}{} \tilde{\Lambda}_{+}^{(0)}\Bigr)\Bigr]\Bigr\}\,, \\
%%%%%%%%%%%%%%%%%%%%%%%%%
\hat{\Psi}^\text{tidal}_{42}=& \frac{4i \sqrt{5} x^{9/2}}{9 \sqrt{3} (1 -  \zeta) \bar{\gamma}^2 \mathcal{S}_{-}{}} \Bigl\{\zeta \bar{\gamma}^2 (\mathcal{S}_{+}{} -  \mathcal{S}_{-}{} \delta - 3 \mathcal{S}_{+}{} \nu + \mathcal{S}_{-}{} \delta \nu) \tilde{\lambda}_{+}^{(0)} - \frac{\sqrt{\alpha}}{2}\Bigl[ - \frac{3}{35} \Bigl(32 \zeta \mathcal{S}_{-}{} \mathcal{S}_{+}{} \tilde{\Lambda}_{-}^{(0)} \nn \\
& + 34 \zeta \bar{\gamma} \mathcal{S}_{-}{} \mathcal{S}_{+}{} \tilde{\Lambda}_{-}^{(0)} - 8 \bar{\gamma} \tilde{\Lambda}_{+}^{(0)} - 17 \bar{\gamma}^2 \tilde{\Lambda}_{+}^{(0)} + 32 \zeta \mathcal{S}_{-}{}^2 \tilde{\Lambda}_{+}^{(0)} + 34 \zeta \bar{\gamma} \mathcal{S}_{-}{}^2 \tilde{\Lambda}_{+}^{(0)}\Bigr) + \frac{3}{35} \nu \Bigl(64 \zeta \mathcal{S}_{-}{} \mathcal{S}_{+}{} \tilde{\Lambda}_{-}^{(0)} \nn \\
& + 102 \zeta \bar{\gamma} \mathcal{S}_{-}{} \mathcal{S}_{+}{} \tilde{\Lambda}_{-}^{(0)} - 16 \bar{\gamma} \tilde{\Lambda}_{+}^{(0)} - 51 \bar{\gamma}^2 \tilde{\Lambda}_{+}^{(0)} + 64 \zeta \mathcal{S}_{-}{}^2 \tilde{\Lambda}_{+}^{(0)} + 102 \zeta \bar{\gamma} \mathcal{S}_{-}{}^2 \tilde{\Lambda}_{+}^{(0)}\Bigr) -  \frac{102}{35} \zeta \bar{\gamma} \mathcal{S}_{-}{} \delta \nu \Bigl(\mathcal{S}_{-}{} \tilde{\Lambda}_{-}^{(0)} + \mathcal{S}_{+}{} \tilde{\Lambda}_{+}^{(0)}\Bigr) \nn \\
& + \frac{6}{35} \delta \Bigl(-4 \bar{\gamma} \tilde{\Lambda}_{-}^{(0)} + 16 \zeta \mathcal{S}_{-}{}^2 \tilde{\Lambda}_{-}^{(0)} + 17 \zeta \bar{\gamma} \mathcal{S}_{-}{}^2 \tilde{\Lambda}_{-}^{(0)} + 16 \zeta \mathcal{S}_{-}{} \mathcal{S}_{+}{} \tilde{\Lambda}_{+}^{(0)} + 17 \zeta \bar{\gamma} \mathcal{S}_{-}{} \mathcal{S}_{+}{} \tilde{\Lambda}_{+}^{(0)}\Bigr)\Bigr]\Bigr\}\,, 
\end{align}
\end{subequations}
As previously, we have omitted here the remainder $\calO\left(\etidal/c^{5}\right)$. We also do not display the full explicit expressions for $\hat{\Psi}^\text{tidal}_{00}$, $\hat{\Psi}^\text{tidal}_{11}$ and $\hat{\Psi}^\text{tidal}_{22}$ due to their length but all the scalar modes are directly available in details in the supplementary file~\cite{SuppMaterial}. As usual, scalar modes with negative $m$ can be deduced using the relation $\Psi^{\ell, -{\rm m}} = (-)^\ell (\Psi^{\ell {\rm m}})^{*}$. As noted in~\cite{Bernard:2022noq}, the Newtonian contribution to $\psi_{00}$ is an $x$-independent constant, implying that the leading mode in the flux is in fact $\psi_{11}$. This remains true when tidal effects are included, as their contribution to $\psi_{00}$ only enters at the NLO.

As a final remark, we notice that the $\text{N}^{1.5}\text{LO}$ scalar waveform only contains contributions from the dipolar scalar tidal effects. Quadrupolar scalar, tensorial and scalar-tensorial tidal contributions will start contributing only at the NNLO, thus requiring the knowledge of the scalar waveform at half a PN order beyond the $\text{N}^{1.5}\text{LO}$ to which we are working. 

%------------------------------------------------------------------------
%------------------------------------------------------------------------
\section{Conclusion}
%------------------------------------------------------------------------
%------------------------------------------------------------------------

We have extended the modeling of gravitational waveforms from inspiralling, non-spinning compact binaries in ST theories by incorporating adiabatic tidal effects to the radiative sector. We have used the PN-MPM formalism adapted to ST theories, and worked within the adiabatic approximation for the description of tides, including the scalar tides. As our main results, we have computed the tidal corrections to the total flux up to the NNLO, and derived the corresponding NNLO correction to the GW phasing in the dipolar-driven regime. At this level of accuracy, we had to account for the contributions of the scalar, tensorial, and mixed scalar-tensorial tidal deformabilities. We have also computed the full wavefoorm, including both gravitational and scalar modes, to $\text{N}^{1.5}\text{LO}$ in the same regime. Only dipolar tidal corrections contribute to the waveform modes at this precision. In order to capture quadrupolar tidal effects in the waveform, specifically the second-order scalar as well as the leading-order tensorial and mixed scalar-tensorial contributions, one needs to extend the PN computations up to NNLO. This is left for future work.

In this work, we have focused on non-spinning, quasi-circular systems, and worked in the adiabatic approximations for the tides and scalarization. Future prospects to extend this work would be to go beyond the adiabatic approximations by considering dynamical tides and scalarization. Another aspect would be to consider spinning objects, and in particular tidal-induced spin contributions. Indeed, such effects should be important in other classes of ST theories that include higher curvature corrections and that differ from GR for black holes, such as Einstein-scalar-Gauss-Bonnet gravity,  where the dynamics of spinning binary black holes has recently been investigated within the EFT framework~\cite{Almeida:2024cqz} .
Finally, we emphasize that taking into account scalar tidal effects is crucial when devising tests of gravity with gravitational waves, as their dipolar nature starting formally at 3PN order presents them a clear signature compared to the traditional tides in GR~\cite{Bernard:2019yfz,Bernard:2025dyh}. Hence, we encourage people working on fundamental physics aspects of the science exploitation of future gravitational wave detectors, such as LISA and Einstein Telescope, to take into account these effects when devising fully agnostic or theory-specific tests. 

%------------------------------------------------------------------------
\acknowledgments
%------------------------------------------------------------------------

L.~B. and E.~D. thank Iris van Gemeren for insightful conversations. L.~B. and E.~D. acknowledges financial support by the ANR PRoGRAM project, grant ANR-21-CE31-0003-001.
L.~B. acknowledges financial support from the EU Horizon 2020 Research and Innovation Programme under the Marie Sklodowska-Curie Grant Agreement no. 101007855.
L. B. is also grateful for the hospitality of Perimeter Institute where part of this work was carried out. Research at Perimeter Institute is supported in part by the Government of Canada through the Department of Innovation, Science and Economic Development Canada and by the Province of Ontario through the Ministry of Colleges and Universities. This research was also supported in part by the Simons Foundation through the Simons Foundation Emmy Noether Fellows Program at Perimeter Institute. 
E.D. is grateful for the hospitality of the Max Planck Institute for Gravitational Physics (Albert Einstein Institute, AEI) in Potsdam, where part of this work was carried out. E. D. acknowledges financial support from the SALTO exchange program between the Centre National de la Recherche Scientifique (CNRS) and the Max-Planck-Gesellschaft (MPG). E.D. is also supported by IBS under the project code IBS-R018-D3.

%------------------------------------------------------------------------
%------------------------------------------------------------------------
\appendix
%------------------------------------------------------------------------
%------------------------------------------------------------------------

%------------------------------------------------------------------------
%------------------------------------------------------------------------
\section{Source moments for quasi-circular orbits}
\label{App:source_moments_circ}
%------------------------------------------------------------------------
%------------------------------------------------------------------------

We present here the tidal corrections to the source multipole moments relevant for the computation of the 1PN gravitational and scalar energy fluxes~\eqref{eq:fluxes_1PN}, specializing to the case of quasi-circular orbits. The corresponding  expressions in the point-particle approximation can be found in~\cite{Bernard:2022noq}. Due to their length, the complete expressions for the scalar monopole $ I^s$, dipole $I^s_i$ and quadrupole $I^s_{ij}$ are not displayed explicitly. However, all source moments, given both in the center-of-mass frame and for quasi-circular orbits, will be made available upon request to interested readers.

\begin{subequations}
\begin{align}\label{eq:source_moments_circ}  
\left(I^s\right)^\text{tidal} & =  \frac{16 \zeta m \nu x^3}{3 (-1 + \zeta)^2 \bar{\gamma}^2} \Bigl\{x\Bigl[\sqrt{\alpha} \zeta \Bigl(4 \bar{\beta}^- \bar{\gamma} \mathcal{S}_{-}{} - 16 \bar{\beta}^- \zeta \mathcal{S}_{-}{}^3 - 4 \bar{\beta}^+ \bar{\gamma} \mathcal{S}_{+}{} + \bar{\gamma}^2 \mathcal{S}_{+}{} + 16 \bar{\beta}^- \zeta \mathcal{S}_{-}{} \mathcal{S}_{+}{}^2\Bigr) \tilde{\lambda}_{+}^{(0)} + \frac{3}{8} \bar{\gamma}^2 (5 \zeta - 4 \lambda_1) \tilde{\lambda}_{+}^{(0)} \nn \\
&+ \frac{3}{4} \zeta \bar{\gamma}^2 \phi_{0}{} \tilde{\lambda}_{+}^{(1)} + \alpha \Bigl(\frac{1}{8} (24 \bar{\beta}^- \tilde{\Lambda}_{-}^{(0)} + 4 \zeta \bar{\gamma} \mathcal{S}_{-}{} \mathcal{S}_{+}{} \tilde{\Lambda}_{-}^{(0)} - 24 \bar{\beta}^+ \tilde{\Lambda}_{+}^{(0)} - 5 \bar{\gamma}^2 \tilde{\Lambda}_{+}^{(0)} + 4 \zeta \bar{\gamma} \mathcal{S}_{-}{}^2 \tilde{\Lambda}_{+}^{(0)}) \nn \\
&-  \frac{1}{2} \zeta \bar{\gamma} \mathcal{S}_{-}{} \delta (\mathcal{S}_{-}{} \tilde{\Lambda}_{-}^{(0)} + \mathcal{S}_{+}{} \tilde{\Lambda}_{+}^{(0)})\Bigr)\Bigr] + x^2\Bigl[ \cdots \Bigr] +\mathcal{O}\left(\frac{\epsilon_{\text{tidal}}}{c^5} \right)\Bigr\}  \,, \nn \\
%%%%%%%%%%%%%%%%%%%%%%%%%%%%%%%%%%%%%%%%%%%%%%%%%%%%%%%%%%%%%
\left(I^s_i\right)^\text{tidal} & = \frac{16 \alpha \tilde{G} m^2 \zeta  x^2}{3 c^2 (-1 + \zeta)^2 \bar{\gamma}^2}  \biggl\{n^{i} \Bigl[\sqrt{\alpha} \zeta \bar{\gamma}^2 \mathcal{S}_{-}{} \nu \tilde{\lambda}_{+}^{(0)} + \alpha \Bigl(- \frac{3}{8} (\bar{\gamma} \tilde{\Lambda}_{-}^{(0)} + 4 \zeta \mathcal{S}_{+}{}^2 \tilde{\Lambda}_{-}^{(0)} + 4 \zeta \mathcal{S}_{-}{} \mathcal{S}_{+}{} \tilde{\Lambda}_{+}^{(0)}) \nn \\
& -  \frac{3}{8} \delta (4 \zeta \mathcal{S}_{-}{} \mathcal{S}_{+}{} \tilde{\Lambda}_{-}^{(0)} + \bar{\gamma} \tilde{\Lambda}_{+}^{(0)} + 4 \zeta \mathcal{S}_{+}{}^2 \tilde{\Lambda}_{+}^{(0)})\Bigr)\Bigr] + x \, n^i \Bigl[ \cdots \Bigr] + x^{3/2} \lambda^{i} \Bigl[\sqrt{\alpha} \Bigl(\frac{1}{6} \zeta \bar{\gamma}^2 \mathcal{S}_{-}{} (\bar{\gamma}  + 4 \zeta \mathcal{S}_{+}{}^2) \nu  \nn \\
&+ \frac{1}{3} \zeta \bar{\gamma}^2 \mathcal{S}_{+}{} (\bar{\gamma} + 2 \zeta \mathcal{S}_{+}{}^2) \delta \nu -  \frac{2}{3} \zeta \bar{\gamma}^2 \mathcal{S}_{-}{} (\bar{\gamma} + 2 \zeta \mathcal{S}_{+}{}^2) \nu^2\Bigr) \tilde{\lambda}_{+}^{(0)} + \alpha \Bigl(\frac{1}{4} (\bar{\gamma} + 2 \zeta \mathcal{S}_{+}{}^2) \delta \nu (4 \zeta \mathcal{S}_{-}{} \mathcal{S}_{+}{} \tilde{\Lambda}_{-}^{(0)} + \bar{\gamma} \tilde{\Lambda}_{+}^{(0)} \nn \\
&+ 4 \zeta \mathcal{S}_{+}{}^2 \tilde{\Lambda}_{+}^{(0)})  + \frac{1}{16} (- \bar{\gamma}^2 \tilde{\Lambda}_{-}^{(0)} - 16 \zeta \bar{\gamma} \mathcal{S}_{+}{}^2 \tilde{\Lambda}_{-}^{(0)} - 32 \zeta^2 \mathcal{S}_{+}{}^4 \tilde{\Lambda}_{-}^{(0)} - 8 \zeta \bar{\gamma} \mathcal{S}_{-}{} \mathcal{S}_{+}{} \tilde{\Lambda}_{+}^{(0)} - 32 \zeta^2 \mathcal{S}_{-}{} \mathcal{S}_{+}{}^3 \tilde{\Lambda}_{+}^{(0)}) \nn \\
& + \frac{1}{4} \nu (\bar{\gamma}^2 \tilde{\Lambda}_{-}^{(0)} + 14 \zeta \bar{\gamma} \mathcal{S}_{+}{}^2 \tilde{\Lambda}_{-}^{(0)}  + 24 \zeta^2 \mathcal{S}_{+}{}^4 \tilde{\Lambda}_{-}^{(0)} + 8 \zeta \bar{\gamma} \mathcal{S}_{-}{} \mathcal{S}_{+}{} \tilde{\Lambda}_{+}^{(0)} + 24 \zeta^2 \mathcal{S}_{-}{} \mathcal{S}_{+}{}^3 \tilde{\Lambda}_{+}^{(0)}) + \frac{1}{16} \delta (-8 \zeta \bar{\gamma} \mathcal{S}_{-}{} \mathcal{S}_{+}{} \tilde{\Lambda}_{-}^{(0)} \nn \\
& - 32 \zeta^2 \mathcal{S}_{-}{} \mathcal{S}_{+}{}^3 \tilde{\Lambda}_{-}^{(0)} -  \bar{\gamma}^2 \tilde{\Lambda}_{+}^{(0)}  - 16 \zeta \bar{\gamma} \mathcal{S}_{+}{}^2 \tilde{\Lambda}_{+}^{(0)} - 32 \zeta^2 \mathcal{S}_{+}{}^4 \tilde{\Lambda}_{+}^{(0)})\Bigr)\Bigr] + x^2 \, n^{i} \Bigl[-  \frac{9}{32} \sqrt{\alpha} \zeta \bar{\gamma}^2 \mathcal{S}_{-}{} \nu   \bigl(\tilde{c}_{+}^{(0)} - 4 (4 \tilde{\mu}_{+}^{(0)} + \tilde{\nu}_{+}^{(0)})\bigr) \nn \\
& + (\cdots) \Bigr]+\mathcal{O}\left(\frac{\epsilon_{\text{tidal}}}{c^5} \right)\biggr\} \,, \nn \\
%%%%%%%%%%%%%%%%%%%%%%%%%%%%%%%%%%%%%%%%%%%%%%%%%%%%%%%%%%%%%
\left(I^s_{ij}\right)^\text{tidal} & =  \frac{4 \alpha^2 \tilde{G}^2 m^3 \nu \zeta x }{3 c^4 (-1 + \zeta)^2 \bar{\gamma}^2}\Bigl\{ n^{\langle i} n^{j \rangle}\Bigl[\sqrt{\alpha} \zeta \bar{\gamma}^2 (4 \mathcal{S}_{+}{} - 4 \mathcal{S}_{-}{} \delta) \tilde{\lambda}_{+}^{(0)} + 6 \alpha (-4 \zeta \mathcal{S}_{-}{} \mathcal{S}_{+}{} \tilde{\Lambda}_{-}^{(0)} + \bar{\gamma} \tilde{\Lambda}_{+}^{(0)} - 4 \zeta \mathcal{S}_{-}{}^2 \tilde{\Lambda}_{+}^{(0)})\Bigr] \nn \\
& + x \Bigl[ n^{\langle i} n^{j\rangle}\Bigl( \cdots \Bigr)+\lambda^{\langle i} \lambda^{j \rangle}\Bigl( \cdots \Bigr) \Bigr]+\mathcal{O}\left(\frac{\epsilon_{\text{tidal}}}{c^3} \right)\Bigr\} \,, \nn \\
%%%%%%%%%%%%%%%%%%%%%%%%%%%%%%%%%%%%%%%%%%%%%%%%%%%%%%%%%%%%%
\left(I^s_{ijk}\right)^\text{tidal} & = - \frac{16 \alpha^3 \tilde{G}^3 m^4 \nu \zeta  }{c^6 (-1 + \zeta) \bar{\gamma}^2 (2 + \bar{\gamma})}  \biggl\{n^{\langle i} n^{j} n^{k \rangle}\biggl[\frac{\zeta \bar{\gamma}^2 (\mathcal{S}_{-}{} -  \mathcal{S}_{+}{} \delta - 2 \mathcal{S}_{-}{} \nu) \tilde{\lambda}_{+}^{(0)}}{\sqrt{\alpha}} -  \frac{3}{4} \delta (-4 \zeta \mathcal{S}_{-}{} \mathcal{S}_{+}{} \tilde{\Lambda}_{-}^{(0)} + \bar{\gamma} \tilde{\Lambda}_{+}^{(0)} - 4 \zeta \mathcal{S}_{-}{}^2 \tilde{\Lambda}_{+}^{(0)}) \nn \\
& + \frac{3}{4} (\bar{\gamma} \tilde{\Lambda}_{-}^{(0)} - 4 \zeta \mathcal{S}_{-}{}^2 \tilde{\Lambda}_{-}^{(0)} - 4 \zeta \mathcal{S}_{-}{} \mathcal{S}_{+}{} \tilde{\Lambda}_{+}^{(0)})\biggr]+\mathcal{O}\left(\frac{\epsilon_{\text{tidal}}}{c} \right)\biggr\} \,, \nn \\
%%%%%%%%%%%%%%%%%%%%%%%%%%%%%%%%%%%%%%%%%%%%%%%%%%%%%%%%%%%%%
\left(I_{ij}\right)^\text{tidal} & = - \frac{16 \alpha^2 \tilde{G}^2 m^3 \nu x^3}{3 c^4 (1 -  \zeta) \bar{\gamma}} \biggl\{\zeta \bar{\gamma}\, n^{\langle i} n^{j\rangle} \tilde{\lambda}_{+}^{(0)} + x \Bigl[\zeta \bar{\gamma} \, \lambda^{\langle i} \lambda^{j \rangle} \Bigl(\frac{22}{21} - \frac{22}{7} \nu \Bigr)\tilde{\lambda}_{+}^{(0)} + n^{\langle i} n^{j \rangle} \Bigl(- \frac{319}{168} \zeta \bar{\gamma} \nu \tilde{\lambda}_{+}^{(0)} \nn \\
& + \frac{5}{48} \delta (48 \bar{\beta}^+ \zeta \tilde{\lambda}_{-}^{(0)} + 9 \zeta \bar{\gamma} \tilde{\lambda}_{-}^{(0)} + 9 \zeta \bar{\gamma}^2 \tilde{\lambda}_{-}^{(0)} - 15 \zeta \bar{\gamma} \tilde{\Lambda}_{-}^{(0)} + 12 \bar{\gamma} \lambda_1 \tilde{\Lambda}_{-}^{(0)} - 6 \zeta \bar{\gamma} \phi_{0}{} \tilde{\Lambda}_{-}^{(1)} - 48 \bar{\beta}^- \zeta \tilde{\lambda}_{+}^{(0)} - 32 \bar{\beta}^- \zeta \bar{\gamma} \tilde{\lambda}_{+}^{(0)}) \nn \\
& + \frac{1}{336} (-1680 \bar{\beta}^- \zeta \tilde{\lambda}_{-}^{(0)} + 1680 \bar{\beta}^+ \zeta \tilde{\lambda}_{+}^{(0)} - 79 \zeta \bar{\gamma} \tilde{\lambda}_{+}^{(0)} + 1120 \bar{\beta}^+ \zeta \bar{\gamma} \tilde{\lambda}_{+}^{(0)} - 91 \zeta \bar{\gamma}^2 \tilde{\lambda}_{+}^{(0)} + 525 \zeta \bar{\gamma} \tilde{\Lambda}_{+}^{(0)} - 420 \bar{\gamma} \lambda_1 \tilde{\Lambda}_{+}^{(0)} \nn \\
& + 210 \zeta \bar{\gamma} \phi_{0}{} \tilde{\Lambda}_{+}^{(1)})\Bigr)\Bigr]+\mathcal{O}\left(\frac{\epsilon_{\text{tidal}}}{c^3} \right)\biggr\}\,, \nn \\
%%%%%%%%%%%%%%%%%%%%%%%%%%%%%%%%%%%%%%%%%%%%%%%%%%%%%%%%%%%%%
\left(I_{ijk}\right)^\text{tidal} &=  - \frac{8 \alpha^3  \tilde{G}^3 m^4 \delta \nu}{c^6 }\frac{\zeta}{(-1 + \zeta)} \biggl\{n^{\langle i} n^{j} n^{k \rangle} \tilde{\lambda}_{+}^{(0)} +\mathcal{O}\left(\frac{\epsilon_{\text{tidal}}}{c} \right) \biggr\}\,, \nn \\
%%%%%%%%%%%%%%%%%%%%%%%%%%%%%%%%%%%%%%%%%%%%%%%%%%%%%%%%%%%%%
\left(J_{ij}\right)^\text{tidal} &=  - \frac{8 \alpha^2  \tilde{G}^2 m^3 \delta \nu x^{3/2}}{c^3}\frac{\zeta}{ (-1 + \zeta)} \biggl\{\ell^{\langle i} n^{j \rangle} \, \tilde{\lambda}_{+}^{(0)}+\mathcal{O}\left(\frac{\epsilon_{\text{tidal}}}{c} \right) \biggr\}\,.
\end{align}
\end{subequations}

%------------------------------------------------------------------------
%------------------------------------------------------------------------
\section{Comparisons with Ref.~\cite{Creci:2024wfu}}\label{eq:comparison_Creci}
%------------------------------------------------------------------------
%------------------------------------------------------------------------

In Ref.~\cite{Creci:2024wfu}, the tidal corrections to both the dynamical and radiative sectors in ST theories were recently derived at LO in the various types of tidal deformability coefficients. Within the context of our work, this corresponds to contributions that are linear in the zeroth-order coefficients, namely $(\lambda_A^{(0)},\mu_A^{(0)},\nu_A^{(0)},c_A^{(0)})$.

A direct comparison between their results and ours is not straightforward due to differences in the treatment of matter coupling. In our approach, the matter fields $\mathfrak{m}$ are coupled to the Jordan-frame metric $g_{\mu \nu}$, so that the tidal deformability parameters entering our tidal action~\eqref{eq:Stidal} correspond to physical, observable quantities. In contrast, in Ref.~\cite{Creci:2024wfu}, the following conformal transformation is performed:
\begin{align}
g_{\mu \nu}^{(J)} = A(\hat{\varphi})g_{\mu \nu}^{(E)}
\end{align}
where $A(\hat{\varphi})$ is the conformal factor, $g_{\mu \nu}^{(J)}$ denotes the Jordan-frame metric and $g_{\mu \nu}^{(E)}$ the Einstein-frame metric; and the matter fields are coupled to the Einstein-frame metric. As a results, tidal corrections derived in Ref.~\cite{Creci:2024wfu} depend on tidal deformabilities which are Einstein-frame quantities. 

Therefore, to perform a consistent comparison between our results and those of Ref.~\cite{Creci:2024wfu}, it is necessary to apply the appropriate transformation rules that relate the physical Jordan-frame tidal deformabilities to their Einstein-frame counterparts. We find the following transformation rules
%
%After such a transformation, the action within the Einstein frame takes the form 
%
%\begin{align}
%S_{ST}^{(E)} = \int \dd^4x \frac{c^3\sqrt{-g}}{16 \pi G}\left[ R-2 \, g^{\mu \nu}\, \partial_{\mu}\hat{\varphi}\,\partial_{\nu}\hat{\varphi}\right] + S_{\text{matter}}\left[\mathfrak{m}, A(\hat{\varphi})g_{\mu \nu}\right]\,,
%\end{align}
%
%Since they couple the matter fields with the Einstein-frame metric, mapping our tidal deformability coefficients to those used in~\cite{Creci:2024wfu} introduces a mixing between the different types of coefficients. Such a mapping is summarized in Table~\ref{tab:mapping_TLN}.
%
\hspace{0.5cm}
\begin{center}
\begin{tabular}{|c|c|}
	\hline 
	\textbf{Our notation} & \textbf{Notation from Ref.~\cite{Creci:2024wfu}} \\[5pt]
	\hline 
	\hline  
	 $\lambda_A^{(0)}$ & $-\frac{\Delta_0}{A_0^3} \, \lambda^{\ell=1}_S$  \\[10pt]
	 $\mu_A^{(0)}$ & $-\frac{\Delta_0}{A_0}\,\left[\lambda^{\ell=2}_{S}+ c^2 \left(\frac{\partial ln A}{\partial \hat{\varphi}}(\hat{\varphi}_0)\right)^2 \lambda^{\ell=2}_{T} +2 c^2\left(\frac{\partial ln A}{\partial \hat{\varphi}}(\hat{\varphi}_0)\right)\lambda^{\ell=2}_{ST}  \right]$ \\[10pt]
     $\nu_A^{(0)}$ & $-A_0\sqrt{\Delta_0}\,\left[\lambda^{\ell=2}_{ST}+\left(\frac{\partial ln A}{\partial \hat{\varphi}}(\hat{\varphi}_0)\right)\lambda^{\ell=2}_{T} \right] $ \\[10pt]
	 $c_A^{(0)}$ & $A_0^3\, \lambda^{\ell=2}_T$ \\[8pt]
\hline
\end{tabular}
\captionof{table}{Translation map for tidal deformability coefficients. \label{tab:mapping_TLN}}
\end{center}
Mapping our Jordan-frame tidal deformabilities to those in the Einstein frame (using the notation from Ref.~\cite{Creci:2024wfu}) results in a mixing between the different types of tidal deformability coefficients. This mixing is a key difference from the transformation rules presented in Eqs.~(A61)-(A63) of Ref.~\cite{Creci:2023cfx}, where no such mixing occurs. The reason for this discrepancy lies in the definition of the tidal field. In the present work, the tidal field $G^{\mu\nu}$ is defined in terms of the Riemann tensor (see Eq.~\eqref{eq:tensor_field}), which does transform under the conformal transformation~\eqref{eq:conformal_metric}. In contrast, in Ref.~\cite{Creci:2023cfx}, the tidal field is defined using the Weyl tensor, which, when having one raised index, is invariant under a conformal transformation.
However, as we are working at the leading order, such different definitions do not impact the dynamics.
Hence, it does not affect the final results but we should keep in mind that different transformation rules hold depending on our choice for the tidal field.  Using these relations, we are in full agreement with the results of Ref.~\cite{Creci:2024wfu} in both the dynamical and radiative sectors. Note that for the gravitational flux, the comparison is restricted to the leading-order dipolar scalar tidal contribution. Indeed, our work only requires precision up to the NLO, a level at which quadrupolar tidal deformations do not yet appear.

%------------------------------------------------------------------------
%------------------------------------------------------------------------
\bibliography{DB25}

%------------------------------------------------------------------------
%------------------------------------------------------------------------

\end{document}